\documentclass[a4paper,11pt]{article}
\pdfoutput=1
\usepackage{jinstpub}
\usepackage[T1]{fontenc}
\usepackage{caption}
\usepackage{subcaption}
\usepackage{bm}
\usepackage{upgreek}
\usepackage{mathtools}
\usepackage{booktabs}
\usepackage{boldline}
\usepackage{multirow}
\usepackage{tikz-network}
\usepackage{nicematrix}
\newtheorem{theorem}{Theorem}[section]
\newcommand{\bigzero}{\mbox{\normalfont\Large\bfseries 0}}
\newcommand{\rvline}{\hspace*{-\arraycolsep}\vline\hspace*{-\arraycolsep}}

\newcommand\coolrightbrace[2]{%
  \left.\vphantom{\begin{matrix} #1 \end{matrix}}\right\}#2}

\title{\boldmath Kinematic and vertex fitting package for the $\text{CMD-}3$
  experiment} \author[a,b]{S.S. Gribanov,} \author[a,b]{A.S. Popov}
\affiliation[a]{Budker Institute of Nuclear Physics, SB RAS,\\Prospekt Akademika
  Lavrent’yeva, 11, Novosibirsk, 630090 Russia} \affiliation[b]{Novosibirsk
  State University, Physics Department,\\Pirogova, 2, Novosibirsk, 630090
  Russia} \emailAdd{S.S.Gribanov@inp.nsk.su} \emailAdd{aspopov1@inp.nsk.su} \abstract{In this article, we
  discuss a new software package of kinematic and vertex fitting for the $\text{CMD-}3$ experiment
  at the VEPP-$2000$ electron-positron collider. The authors describe in detail
  the fitting algorithm, parametrization of four-momenta and
  trajectories of various particles and present the results of testing the
  fitting pakage using events of Monte Carlo simulation of various $e^+e^-$
  annihilation processes. The authors also provide several examples of the
  fitting package validation using Gaussian simulation. Although the package discussed
  in this article is
  intended for the $\text{CMD-}3$ experiment, it can also be used in other experiments.
  The authors consider the described package as their first step towards a more
  universal and rigorous kinematic and vertex fitting package that can be used in
  future $e^+e^-$ experiments, such as the Super Charm-Tau factory.}

\begin{document}
\maketitle
\flushbottom

\section{Introduction \label{sec:intro}}
Kinematic and vertex fitting is widely used data analysis technique in
particle physics experiments~\cite{Bauer2000,Prokofiev:2004,Prokofiev:2005,Liang_2010,Jorg2018,Goepfert,Smith2019:nucphys,HULSBERGEN2005566,KROHN2020164269,Radkhorrami2021}.
This technique can be used in
order to separate events corresponding to different kinematic hypotheses
and to reconstruct
interaction and decay vertices. In addition, the use of this technique often leads
to significant improvements in resolution of some quantities, such as various
invariant masses. This paper describes kinematic and vertex fitting package
developed for the $\text{CMD-}3$ experiment~\cite{GRANCAGNOLO2010114,Khazin:2008,AkhmetshinBGO209,Razuvaev:2017,Anisenkov_2017,ANISENKOV2013463} located at the
$\text{VEPP-}2000$ electron-positron
collider~\cite{Danilov:860802,BERKAEV2012303} with a maximum center-of-mass
energy of $2$~GeV.
The discussed package is currently
actively used in the study of a number of processes with the $\text{CMD-}3$
detector. This package can also be used in similar experiments. To do
this, it is necessary to make a number of experiment-dependent changes related
to the format of input data and probably to some particle parametrizations.
The authors believe that it is quite easy to do, since the discussed package
is divided into detector-dependent and detector-independent parts.

In general, any kinematic and vertex fitting algorithm consists in constrained
minimization of the chi-square function:
\begin{equation}
  \label{eq:chi-square-0}
  \begin{split}
    \chi^2(\bm{x}) &= \bm{\Delta x}^{\intercal}\hat{\tilde{C}}^{-1}\bm{\Delta x}, \\
    \bm{\Delta x} &= \bm{x} - \bm{\tilde{x}}
  \end{split}
\end{equation}
where $\bm{x}\in\mathbb{R}^{n}$ is the vector of the optimization variables
corresponding to measurable parameters, i.e. such parameters that can be
measured in an experiment. The natural number $n$ is the number of such parameters, the
vector $\bm{\tilde{x}}$ is the vector of these parameters, measured in a single event.
The matrix $\hat{\tilde{C}}$ is the corresponding covariance matrix, obtained in the same
event. The parameters measured by a detector are usually referred to one particle
or another. Moreover, usually the parameters corresponding to different
particles are measured independently. With these assumptions in mind, the
chi-square from equation~\eqref{eq:chi-square-0} can be written as
\begin{equation}
  \label{eq:chi-square-1}
  \begin{split}
    \chi^2(\bm{x}) &= \sum\limits^{N_p}_{i = 1}\left.\bm{\Delta x^{(i)}}\right.^{\intercal}\left.\hat{\tilde{C}}^{(i)}\right.^{-1}\bm{\Delta x^{(i)}},\\
    \bm{\Delta x^{(i)}} &= \bm{x^{(i)}} - \bm{\tilde{x}^{(i)}},\\
    \bm{x} &= \bm{x^{(1)}}\oplus\bm{x^{(2)}}\oplus\ldots\oplus\bm{x^{(N_p)}},\\
    \bm{\tilde{x}} &= \bm{\tilde{x}^{(1)}}\oplus\bm{\tilde{x}^{(2)}}\oplus\ldots\oplus\bm{\tilde{x}^{(N_p)}},\\
    \hat{\tilde{C}} &= \hat{\tilde{C}}^{(1)}\oplus\hat{\tilde{C}}^{(2)}\oplus\ldots\oplus\hat{\tilde{C}}^{(N_p)},
  \end{split}
\end{equation}
where $i$ is the particle index, $N_p$ is the number of particles in a hypothesis,
$\bm{x^{(i)}}$ is the vector of the optimization variables corresponding to
measurable parameters of the $i$-th particle. The vector $\bm{\tilde{x}^{(i)}}$ is the
vector of these parameters, measured in a single event. $\hat{\tilde{C}}^{(i)}$ is the
corresponding covariance matrix. The notation $\bm{x^{(i)}}\oplus\bm{x^{(j)}}$
means the concatenation of the vectors $\bm{x^{(i)}}$ and $\bm{x^{(j)}}$, while the notation
$\hat{\tilde{C}}^{(i)}\oplus\hat{\tilde{C}}^{(j)}$ means the direct sum of the matrices
$\hat{\tilde{C}}^{(i)}$ and $\hat{\tilde{C}}^{(j)}$. The parameters measured by
a detector can belong not only to particles, but also, for example, to the
$e^+e^-$ interaction vertex. The
contribution of these parameters can also be included in the
chi-square given by equation~\ref{eq:chi-square-1}.

In the case of kinematic and vertex fitting, equality constraints are commonly
used. These constraints have the following form:
\begin{equation}
  \label{eq:equlity-constraints}
  \begin{split}
    f_k(\bm{y}) &= 0,\;k=1,2,\ldots, m,\\
    \bm{y} &= \bm{x}\oplus\bm{a},\\
  \end{split}
\end{equation}
where $k$ is the constraint index, $f_k$ is the $k$-th constraint function, $m$ is the number
of constraints in a hypothesis, $\bm{a}\in\mathbb{R}^{l}$ is the vector of non-measurable
optimization parameters, $l$ is the number of these parameters.
In this article we mean that non-measurable parameters
are those parameters that cannot be measured experimentally, but can be
obtained as result of fitting. There is a number of such parameters: momentum
components of a lost particle, time parameter of a charged particle trajectory,
vertex coordinates, and some other parameters. Some examples of
non-measurable parameters are given in sections~\ref{sec:vertices} and
\ref{sec:particles}.

Taking into account the chi-square function~\eqref{eq:chi-square-0}
or~\eqref{eq:chi-square-1} and the equality
constraints~\eqref{eq:equlity-constraints}, one can formulate the following
problem for finding the conditional extremum of the chi-square function:
\begin{equation}
  \label{eq:opt-problem}
  \begin{split}
    \begin{cases}
    \chi^2(\bm{x}) \to \text{extremum},\\
    f_k(\bm{y}) = 0,\;k=1,2,\ldots, m.
    \end{cases}
  \end{split}
\end{equation}
It is important to note that the initial values of the measurable parameters
$\bm{x}$ should be set equal to their measured values $\bm{\tilde{x}}$. In this
case, the minimization algorithm starts from the lowest point of the chi-square
paraboloid, which increases the probability that the fit will converge to a
global minimum. The criterion that the found extremum is a local minimum and not
a local maximum is discussed in section~\ref{sec:minimization-algorithm}.

\section{Minimization algorithm \label{sec:minimization-algorithm}}
In this work the well-known method of Lagrange
multipliers~\cite{Ioffe1979,Bertsekas1996,Herman2018} is used in
order to reduce the constrained least-squares optimization problem to the unconstrained
one. The method of Lagrange multipliers is based on the concept of the Lagrange
function. In the context of the problem~\eqref{eq:opt-problem} this function has
the following form:
\begin{equation}
  \label{eq:lagrange-function}
  \mathcal{L}(\bm{y}, \bm{\lambda}) = \chi^2(\bm{x}) + \sum\limits^{m}_{k = 1}\lambda_k f_k(\bm{y}),
  \;\bm{\lambda} = \begin{bmatrix}\lambda_1\\ \lambda_2 \\ \vdots \\ \lambda_m\end{bmatrix},
\end{equation}
where variables $\lambda_1,\lambda_2,\ldots,\lambda_m$ are the Lagrange
multipliers. Taking into the account the Lagrange function, the necessary
condition for the existence of an extremum in the problem~\eqref{eq:opt-problem}
can be formulated using the following theorem~\cite{Ioffe1979}.
\begin{theorem}[necessary condition]
  \label{th:lagrange-principle}
  If $\bm{y^{\prime}}\in\Omega\equiv\{\bm{y}\in\mathbb{R}^{n+l}| f_k(\bm{y}) = 0,\;k=1,2,\ldots,m\}$
  is a conditional extremum point of the
  problem~\eqref{eq:opt-problem}, then there is a nonzero vector
  $\bm{\lambda^{\prime}}\in\mathbb{R}^{m}$ such that
  \begin{equation}
    \bm{\nabla}_{\bm{y}}\mathcal{L}(\bm{y}, \bm{\lambda^{\prime}})|_{\bm{y}=\bm{y^{\prime}}} = \bm{0}.
  \end{equation}
\end{theorem}

In order to determine whether the found extremum is a local minimum or a local
maximum, it is necessary to formulate sufficient conditions for an
extremum~\cite{Ioffe1979}.
\begin{theorem}[sufficient conditions]
  Let $\bm{y}^{\prime}\in\Omega$ and there is a vector $\bm{\lambda^{\prime}}$, such
  that
  $\bm{\nabla}_{\bm{y}}\mathcal{L}(\bm{y},\bm{\lambda^{\prime}})|_{\bm{y}=\bm{y^{\prime}}}=\bm{0}$.
  If for any vector $\bm{dy}\in\mathbb{R}^{n+l},\;\bm{dy}\neq\bm{0}$ that
  satisfies the conditions
  $\bm{\nabla}_{\bm{y}}f_k(\bm{y})|_{\bm{y}=\bm{y^{\prime}}}\cdot\bm{dy} = 0,\;k=1,2,\ldots,m$,
  the following inequality holds
  \begin{equation}
    \label{eq:min-cond}
    \sum\limits_{i=1}^{n + l}\sum\limits_{j=1}^{n + l}\frac{\partial^2\mathcal{L}(\bm{y},\bm{\lambda^{\prime}})}{\partial y_i \partial y_j}\Big|_{\bm{y}=\bm{y^{\prime}}}{dy_i}{dy_j} > 0,
  \end{equation}
  then the strict local minimum of the function $\chi^2$ on the set $\Omega$ is reached at the point $\bm{y}=\bm{y^{\prime}}$. If the following inequality holds
  \begin{equation}
    \label{eq:max-cond}
    \sum\limits_{i=1}^{n + l}\sum\limits_{j=1}^{n + l}\frac{\partial^2\mathcal{L}(\bm{y},\bm{\lambda^{\prime}})}{\partial y_i \partial y_j}\Big|_{\bm{y}=\bm{y^{\prime}}}{dy_i}{dy_j} < 0,
  \end{equation}
  then the strict local maximum is reached at the point $\bm{y}=\bm{y^{\prime}}$.
\end{theorem}

Thus, in order to find the conditional minimum of the $\chi^2$
function~\eqref{eq:chi-square-1}, one should follow the algorithm below. First
of all, we need to compose the Lagrange function according
equation~\eqref{eq:lagrange-function}. Then we need to use the necessary
condition for the existence of an extremum (theorem~\ref{th:lagrange-principle}), i.e.
solve the following system of equations:
\begin{equation}
  \label{eq:system-of-lagrange-equations}
    \begin{cases}
      \bm{\nabla}_{\bm{y}}\mathcal{L}(\bm{y}, \bm{\lambda}) = \bm{0},\\
      f_k(\bm{y}) = 0,\;k=1,2,\ldots,m.
    \end{cases}
\end{equation}
Let us note that the system of equations~\eqref{eq:system-of-lagrange-equations}
can be rewritten in a form that is more convenient for further use:
\begin{equation}
  \label{eq:system-of-lagrange-equations-1}
  \begin{split}
    \bm{\nabla}_{\bm{q}}\mathcal{L}(\bm{y},\bm{\lambda}) &= \bm{0},\\
    \bm{q} &= \bm{y}\oplus\bm{\lambda},
  \end{split}
\end{equation}
where $\bm{q}$ is the vector of all parameters involved in the constrained
chi-square minimization, including the Lagrange multipliers. Suppose that as a
result of solving the system of
equations~\eqref{eq:system-of-lagrange-equations-1}, a pair of vectors is
obtained: $\bm{y}=\bm{y^{\prime}}$ and $\bm{\lambda}=\bm{\lambda^{\prime}}$.
This pair of vectors corresponds to a conditional local extremum of the
chi-square function. Finally, it is necessary to check whether this pair of
vectors corresponds to a local minimum or to a local maximum point. To do this,
we can use inequalities~\eqref{eq:min-cond} and \eqref{eq:max-cond}.

To solve the system of equations~\eqref{eq:system-of-lagrange-equations-1}, the
package discussed in this paper uses Newton's method~\cite{Gill1982,Izmailov2014,Vishnoi2021}.
This method is based on the following iterative approach:
\begin{equation}
  \label{eq:newton-method}
  \bm{q}_{s} = \bm{q}_{s - 1} - \hat{\mathcal{H}}^{-1}(\bm{q}_{s - 1})\bm{\nabla}_{\bm{q}}\mathcal{L}(\bm{y}_{s - 1}, \bm{\lambda}_{s - 1}),\;s=1,2,\ldots,
\end{equation}
where $\bm{q}_{s}=\bm{x}_{s}\oplus\bm{a}_{s}\oplus\bm{\lambda}_{s}$ is the
vector of optimization parameters at the $s$-th step of the iterative procedure,
$\bm{q}_{0}=\bm{x}_{0}\oplus\bm{a}_{0}\oplus\bm{\lambda}_{0}$ is the vector of their initial values,
$\hat{\mathcal{H}}(\bm{q}_{s-1})$ is the Hessian of the Lagrange function with respect to
the variables $\bm{q}$,
and $\bm{\nabla}_{\bm{q}}\mathcal{L}(\bm{y}_{s-1},\bm{\lambda}_{s-1})$ is its gradient
calculated at the point $\bm{q}_{s-1}$.
The vector $\bm{y}_{s-1}$ is the concatenation of the vectors of measurable $\bm{x}_{s-1}$
and non-measurable $\bm{a}_{s-1}$ parameters at the $(s-1)$-th step of the optimization algorithm:
$\bm{y}_{s-1}=\bm{x}_{s-1}\oplus\bm{a}_{s-1}$.
The initial values $\bm{x}_0$ of the measurable parameters $\bm{x}$ are set
equal to their measured values $\bm{\tilde{x}}$. The initial values
$\bm{\lambda}_0$ of the Lagrange multipliers are set to zero. The initial values
$\bm{a}_0$ of the non-measurable parameters can optionally be set to custom
values.

The Hessian $\hat{\mathcal{H}}(\bm{q}_{s-1})$ of the Lagrange function has the following form:
\begin{equation}
  \label{eq:hessian}
  \begin{split}
    \hat{\mathcal{H}}(\bm{q}_{s-1}) &=
    \begin{bmatrix}
      2\hat{M}+\hat{Q}_{s - 1} & \rvline & \hat{J}_{s - 1} \\
      \hline
      \hat{J}^{\intercal}_{s - 1} & \rvline & \bigzero_{m\times m}
    \end{bmatrix},\\
    \hat{M} &= \hat{\tilde{C}}^{-1}\oplus{\bigzero_{l\times l}},
  \end{split}
\end{equation}
where $\bigzero_{m\times m}$ and $\bigzero_{l\times l}$ are $m$ by $m$ and $l$
by $l$ zero matrices, respectively; $\hat{Q}_{s - 1}$ is the matrix containing the
Hessians of the constraint functions with respect to the $\bm{y}$ variables. Let
us denote the Hessian of the $k$-th constraint function calculated at the point
$\bm{y}_{s - 1}$ as $\hat{Q}^{(k)}_{s - 1}$. In these notations, the matrix
$\hat{Q}_{s - 1}$ has the following form:
\begin{equation}
  \label{eq:q-s-1}
    \hat{Q}_{s - 1} = \sum\limits^{m}_{k = 1}\lambda_{k, s - 1}\hat{Q}^{(k)}_{s - 1},
\end{equation}
where $\lambda_{k, s - 1}$ is the value of the $k$-th Lagrange multiplier at the $(s-1)$-th
step of the iterative procedure given by equation~\eqref{eq:newton-method}. The matrix $\hat{J}_{s - 1}$
in equation~\eqref{eq:hessian} is the Jacobian $\frac{\partial (f_1,\ldots,f_m)}{\partial (y_1,\ldots,y_{(n+l)})}$ of
the constraint functions with respect to the variables $\bm{y}$. This Jacobian is also calculated at the point $\bm{y}_{s - 1}$.

The constraints~\eqref{eq:equlity-constraints} do not explicitly depend on the parameters
of particles and vertices. The constraint functions usually explicitly depend only on parametrizations
of the particle four-momenta and trajectories (see section~\ref{sec:particles}), as well as the vertex
parametrization (see section~\ref{sec:vertices}).
In this case, it is convenient to use the chain rule to obtain partial derivatives of the constraints:
\begin{equation}
  \label{eq:chain-rule}
  \frac{\partial f_k(g_1(\bm{y}),\ldots,g_{N_g}(\bm{y}))}{\partial y_i} = \frac{\partial f_k}{\partial g_j}\frac{\partial g_j}{\partial y_i},
\end{equation}
where $g_j$ is a function of some parametrization, $N_g$ is the number of such functions, the repeated index $j$ is summed up.
Using a chain rule in the fitting package makes it possible to implement constraint, particle, and vertex classes
independently. Thus, for example, adding a new particle class to the package will not change the implementation of the
constraints, and vice versa.

In the case of the package discussed in this paper, the procedure given
by equation~\eqref{eq:newton-method} continues until one of the following
conditions is met:
\begin{enumerate}
  \item\label{item:cond-1} $|\chi^2(\bm{x}_s) - \chi^2(\bm{x}_{s - 1})|<\varepsilon$ and
    $\delta(\bm{y}_s)<\varepsilon$, where $\delta(\bm{y}_s) =\sum\limits_{k=1}^{m}|f_k(\bm{y}_s)|$ is the residual and $\varepsilon$ is
    some tolerance;
  \item\label{item:cond-2} $s>N_{\text{iter}}$, where $s$ is the step number of the iterative
    procedure given by equation~\eqref{eq:newton-method} and $N_{\text{iter}}$
    is the maximum number of iterations.
\end{enumerate}
If the first condition (\ref{item:cond-1}) is met, then the fit has converged to a local extremum.
If the second condition (\ref{item:cond-2}) is met, then the fit has not converged.
The default tolerance $\varepsilon$ and maximum number of iterations $N_{\text{iter}}$ used in the
fitting package are equal to $10^{-4}$ and $20$, respectively. These values can be easily adjusted if desired.
Although the default value for the maximum number of iterations is equal to $20$, the
optimization algorithm usually converges in fewer iterations. For example, in the case of fitting the events
of the $e^+e^-\rightarrow K_S K^{\pm}\pi^{\mp},\; K_S\rightarrow\pi^+\pi^-$ process under the corresponding signal
hypothesis (see sections~\ref{sec:hypotheses} and~\ref{sec:kskpi-kskpi}), the optimization algorithm most often
converges in three iterations. At the same time, in about $93\%$ of all events, the algorithm converges in
$2$--$6$ iterations. In the case of fitting the events of the
$e^+e^-\rightarrow\eta\pi^+\pi^-,\;\eta\rightarrow\gamma\gamma$ process (see section~\ref{sec:etapipi-2pi2gamma}),
the algorithm most often converges in two iterations. In about $98\%$ of the
$e^+e^-\rightarrow\eta\pi^+\pi^-,\;\eta\rightarrow\gamma\gamma$ events, the algorithm converges in $2$--$4$
iterations.

In the implementation of the optimization algorithm,
special attention is paid to periodic parameters. If, during
optimization, such a parameter goes beyond the period boundaries, then the
corresponding contribution to the chi-square becomes incorrect because the
measured value of this parameter is within the period. Therefore, after each
iteration, if the parameter goes beyond the period boundaries, then it is
returned back using a shift by an integer number of periods. Since it is
assumed that it does not matter for the constraint functions whether the
parameter is inside the period or not, the parameter values before and
after the shift are equivalent.

Another important feature of the fitting package is the ability to use upper
and lower limits for parameters. In the current version of the package, if
the parameter is out of range, it returns to its initial value, after which
the optimization algorithm continues.

The fitting package provides an interface for fixing and releasing parameters.
Fixing a parameter means that this parameter does not change during the
execution of the optimization algorithm. This can be achieved in the following
way. In the Hessian~\eqref{eq:hessian}, all off-diagonal elements of rows and
columns corresponding to a fixed parameter are set to zero. At the same time,
the diagonal element of the Hessian corresponding to this parameter must be set
equal to one. The component of the Lagrange function gradient corresponding to a
fixed parameter must be set to zero (see equation~\eqref{eq:newton-method}).
It is this approach to fixing parameters
that is used in the considered fitting package. However, this approach is not
unique. For instance, fixed parameters can be excluded
programmatically, i.e. one can consider the Hessian with respect to free
parameters only. Both approaches are equivalent, but the latter approach makes
it possible to reduce the size of the Hessian matrix. With the further
development of the fittings package, it is planned to switch to using this
mechanism for fixing parameters.

The fitting package also provides an interface for enabling and disabling
constraints. This feature is coming very useful when it is necessary to perform
a fit in several hypotheses (see section~\ref{sec:hypotheses}) that differ from
each other only in the set of constraints.

\section{Hypotheses \label{sec:hypotheses}}
Further, in this work the term hypothesis is often used. By hypothesis we mean
a set of all particles, vertices and constraints involved in the constrained
chi-square minimization. The hypothesis also depends on the state of particles
and vertices, since some parameters of these entities can be fixed or limited.
The fitting package discussed in this paper has special classes responsible for
describing these hypotheses. Further, for simplicity, we name the hypotheses
according to the particles they contain, for example, as
$e^+e^-\rightarrow\gamma\gamma\gamma$. One has to distinguish a hypothesis from
a process, since the same notation is used for both of them. The text of the
paper always notes whether the reader is dealing with the notation of a
hypothesis or a process.

The paper also often uses the terms signal and background hypothesis. A
hypothesis is called a signal hypothesis for some process if it describes this
process. Otherwise, the hypothesis is a background hypothesis. There may be
several signal hypotheses for some processes. Such hypotheses can differ from
each other, for example, by the set of constraints, as well as by the presence
or absence of lost particles. To show this, let us consider the
$e^+e^-\rightarrow\eta\pi^+\pi^-,\;\eta\rightarrow\gamma\gamma$ process. Since
the final state of this process contains two charged pions and two photons, then
hypothesis $e^+e^-\rightarrow\pi^+\pi^-\gamma\gamma$ is a signal hypothesis for
this process. Another signal hypothesis is
$e^+e^-\rightarrow\eta\pi^+\pi^-,\;\eta\rightarrow\gamma\gamma$. This hypothesis
takes into account the intermediate particle $\eta$. This particle can be taken
into account using the constraint on the two-photon invariant mass.

Some of the hypotheses discussed in the article contain
a particle denoted by the letter $X$. Throughout this paper the
designation $X$ is used for a neutral intermediate particle of unknown mass.
An example of such a hypothesis is
$e^+e^-\rightarrow X K^+\pi^-,\;X\rightarrow\pi^+\pi^-$, which is a signal
hypothesis for the $e^+e^-\rightarrow K_S K^+\pi^-,\;K_S\rightarrow\pi^+\pi^-$
process.

\section{Chi-square distribution \label{sec:chi-square-distribution}}
As a result of fitting the events of a certain process,
one can obtain the distribution of the chi-square given
by equation~\eqref{eq:chi-square-1}. The properties of
this distribution need to be known as the chi-square is often
used in selection criteria, and analysis of the chi-square
distribution is in some cases useful for validating kinematic
and vertex fitting packages. This raises the question of whether
this distribution is consistent with the chi-squared probability
density function. This probability density function (PDF) has the
following form:
\begin{equation}
  \label{eq:chi-square-pdf}
  f_{\chi^2}(t; \nu) =
  \begin{cases}
    \frac{1}{2^{\nu/2}\Gamma(\nu/2)}t^{\nu/2 - 1}e^{-t/2},\;t > 0;\\
    0,\;\text{otherwise},
  \end{cases}
\end{equation}
where the variable $\nu$ has the meaning of the number degrees of freedom ($\text{NDF}$),
and the variable $t$ has the meaning of the chi-square value.

This section is dedicated to answering the question posed in the previous paragraph.
Sections~\ref{sec:ndf-linear} and~\ref{sec:ndf-non-linear} discuss how linear and
nonlinear constraints affect the properties of the chi-square distribution. In
section~\ref{sec:chi2-non-gaussian}, we briefly discuss how this distribution is
affected by the non-Gaussian response of the detector. In
section~\ref{sec:gaussian-simulation}, we describe an approach to the verification
of the fitting package based on a Gaussian simulation technique.

\subsection{Chi-square in the case of linear constraints \label{sec:ndf-linear}}
Let us first consider the case where all constraints are linear. Suppose that
these constraints have the following form:
\begin{equation}
  \label{eq:linear-constraints}
  \hat{J}\bm{y} + \bm{v} = \bm{0},
\end{equation}
where $\hat{J}= \frac{\partial (f_1,\ldots,f_m)}{\partial (y_1,\ldots,y_{(n + l)})}$ is the Jacobian
of constraint functions, $\bm{v}$ is some vector, $\dim{\bm{v}}=m$. The Lagrange function
in this case has the form:
\begin{equation}
  \label{eq:lagrangian-linear-constraints}
  \begin{split}
    \mathcal{L}(\bm{y}, \bm{\lambda}) &=
    \left(\bm{x} - \bm{\tilde{x}}\right)^{\intercal}\hat{\tilde{C}}^{-1}\left(\bm{x} - \bm{\tilde{x}}\right) +
    \bm{\lambda}^{\intercal}\left(\hat{J}\bm{y} + \bm{v}\right),\\
    \bm{y} &= \bm{x}\oplus\bm{a}.\\
  \end{split}
\end{equation}
After substituting the Lagrange function~\eqref{eq:lagrangian-linear-constraints}
into equation~\eqref{eq:system-of-lagrange-equations-1}, this equation will be
reduced to the following system of linear equations\footnote{Thus, within the
framework of the iterative algorithm given by equation~\eqref{eq:newton-method},
the optimization problem~\eqref{eq:opt-problem} is solved in one iteration in
the case of linear constraints.}:
\begin{equation}
  \label{eq:linear-system-llc}
  \begin{bmatrix}
    2\hat{M} & \rvline & \hat{J}^{\intercal} \\
    \hline
    \hat{J} & \rvline & \bigzero_{m\times m}
  \end{bmatrix}\Delta\bm{q} = -\Delta\bm{g},
\end{equation}
where $\hat{M}=\hat{\tilde{C}}^{-1}\oplus\bigzero_{l\times l}$,
$\Delta\bm{q}=\Delta\bm{y}\oplus\Delta\bm{\lambda}$, $\Delta\bm{y} = \bm{y} -\bm{\tilde{y}}$,
$\Delta\bm{\lambda} = \bm{\lambda} - \bm{0}=\bm{\lambda}$.
$\bigzero_{m\times m}$ and $\bigzero_{l\times l}$ are zero $m$ by $m$ and $l$ by
$l$ zero matrices, respectively. The vector
$\bm{\tilde{y}}=\bm{\tilde{x}}\oplus\bm{\tilde{a}}$ has the meaning of the
vector of initial values of the parameters $\bm{y}$. As discussed above, the
vector $\bm{\tilde{x}}$ at the same time has the meaning of the measured values
of the parameters $\bm{x}$. The vector $\Delta\bm{g}$ can be written in the following
form:
\begin{equation}
  \Delta\bm{g}=\bm{\nabla}_{\bm{q}}\mathcal{L}(\bm{\tilde{y}}, \bm{0}) =
  \begin{bmatrix}
    0\\
    \vdots\\
    0\\
    f_1(\bm{\tilde{y}})\\
    f_2(\bm{\tilde{y}})\\
    \vdots\\
    f_m(\bm{\tilde{y}})
  \end{bmatrix}\equiv
  \begin{bmatrix}
    0\\
    \vdots\\
    0\\
    \Delta f_1\\
    \Delta f_2\\
    \vdots\\
    \Delta f_m
  \end{bmatrix},
\end{equation}
where $\Delta f_k =  f_k(\bm{\tilde{y}}) - f_k(\bm{y^{\prime}})=f_k(\bm{\tilde{y}})$,
$f_k(\bm{y^{\prime}}) = 0$ and $\bm{y^{\prime}}$
is a local extremum point.

Note that the matrix of the system of equations~\eqref{eq:linear-system-llc} is
non-singular only if $\text{rank}(\hat{J})=m$. If the last equality is not satisfied,
then linearly dependent rows and columns appear in the matrix of the system
of linear equations~\eqref{eq:linear-system-llc}, i.e. this matrix becomes
degenerate. Note that the condition $\text{rank}(\hat{J})=m$ must also hold in the
nonlinear case, otherwise the Hessian~\eqref{eq:hessian} is degenerate. Thus,
it makes sense to consider only functionally independent constraints ($\text{rank}(\hat{J})=m$).
In addition, it is easy to show that the inequality $n + l \geq m$ must also hold,
otherwise the constraints are functionally dependent:
$\text{rank}(\hat{J})\leq\min(n + l, m) = n + l < m$.

Let us represent the Jacobian as the concatenation of two matrices:
$\hat{J} = \begin{bmatrix} \hat{J}_{\bm{x}} & \hat{J}_{\bm{a}} \end{bmatrix}$. The matrix
$\hat{J}_{\bm{x}} = \frac{\partial (f_1,\ldots, f_m)}{\partial (x_1,\ldots,x_n)}$ is the
Jacobian of the constraint functions with respect to the parameters $\bm{x}$, and the matrix
$\hat{J}_{\bm{a}} = \frac{\partial (f_1,\ldots, f_m)}{\partial (a_1,\ldots,a_l)}$ is the
Jacobian of these functions with respect to the parameters $\bm{a}$. Taking into account this
form of the Jacobian, the system of equations~\eqref{eq:linear-system-llc} can be rewritten as
follows:
\begin{equation}
  \label{eq:linear-system-llc-2}
  \begin{bmatrix}
    2\hat{\tilde{C}}^{-1} & \rvline &
    \begin{matrix}
      \bigzero_{n\times l} & \hat{J}^{\intercal}_{\bm{x}}
    \end{matrix}\\
    \hline
    \begin{matrix}
      \bigzero_{l\times n} \\
      \hat{J}_{\bm{x}}
    \end{matrix} & \rvline &
    \begin{matrix}
      \bigzero_{l\times l} & \hat{J}^{\intercal}_{\bm{a}} \\
      \hat{J}_{\bm{a}} & \bigzero_{m\times m}
    \end{matrix}
  \end{bmatrix}\Delta\bm{q} = -\Delta\bm{g}.
\end{equation}
It can be seen from the last system of linear equations
that $\text{rank}(\hat{J}_{\bm{a}})=l$, otherwise the matrix of this system is degenerate. It is also seen that
the inequality $l\leq m$ must be satisfied, since $\min(m, l)\geq\text{rank}(\hat{J}_{\bm{a}}) = l$.

Equation~\eqref{eq:linear-system-llc-2} can be easily solved by inverting the corresponding block
matrix. To do this, one can use the well-known Forbenius formula~\cite{Bukov2006} for block matrix inversion:
\begin{equation}
  \label{eq:formenius-formula}
  \hat{A}^{-1} =
  \begin{bmatrix}
    \hat{B} & \hat{C} \\
    \hat{D} & \hat{F}
  \end{bmatrix}^{-1}=
  \begin{bmatrix}
    \hat{B}^{-1} + \hat{B}^{-1}\hat{C}\hat{G}^{-1}\hat{D}\hat{B}^{-1} & \rvline & - \hat{B}^{-1}\hat{C}\hat{G}^{-1} \\
    \hline
    - \hat{G}^{-1}\hat{D}\hat{B}^{-1} & \rvline & \hat{G}^{-1}
  \end{bmatrix},
\end{equation}
where $\hat{B}$ is a $n_1\times n_1$ non-singular matrix, $\hat{F}$ is a $n_2\times n_2$ matrix, and
$\hat{G}=\hat{F}-\hat{D}\hat{B}^{-1}\hat{C}$. Taking into account the last formula, one can obtain the
following result:
\begin{equation}
  \label{eq:applying-forbenius-formula}
  \begin{split}
  \begin{bmatrix}
    2\hat{\tilde{C}}^{-1} & \rvline &
    \begin{matrix}
      \bigzero_{n\times l} & \hat{J}^{\intercal}_{\bm{x}}
    \end{matrix}\\
    \hline
    \begin{matrix}
      \bigzero_{l\times n} \\
      \hat{J}_{\bm{x}}
    \end{matrix} & \rvline &
    \begin{matrix}
      \bigzero_{l\times l} & \hat{J}^{\intercal}_{\bm{a}} \\
      \hat{J}_{\bm{a}} & \bigzero_{m\times m}
    \end{matrix}
  \end{bmatrix}^{-1} &=
  \begin{bmatrix}
    2\hat{\tilde{C}}^{-1} & \rvline & \hat{K}^{\intercal} \\
    \hline
    \hat{K} & \rvline & \hat{S}
  \end{bmatrix}^{-1} \\
  &= \begin{bmatrix}
    \frac{1}{2}\hat{\tilde{C}} + \frac{1}{4}\hat{\tilde{C}}\hat{K}^{\intercal}\hat{G}^{-1}\hat{K}\hat{\tilde{C}} &
    \rvline &
    -\frac{1}{2}\hat{\tilde{C}}\hat{K}^{\intercal}\hat{G}^{-1} \\
    \hline
    -\frac{1}{2}\hat{G}^{-1}\hat{K}\hat{\tilde{C}} &
    \rvline &
    \hat{G}^{-1}
  \end{bmatrix},
  \end{split}
\end{equation}
where $\hat{S}$ is the following matrix:
\begin{equation}
  \hat{S} =
  \begin{bmatrix}
    \bigzero_{l\times l} & \hat{J}^{\intercal}_{\bm{a}} \\
    \hat{J}_{\bm{a}} & \bigzero_{m\times m}
  \end{bmatrix};
\end{equation}
$\hat{K}$ is the concatenation of matrices $\bigzero_{l\times n}$ and $\hat{J}_{\bm{x}}$:
\begin{equation}
  \hat{K} =
  \begin{bmatrix}
    \bigzero_{l\times n} \\
    \hat{J}_{\bm{x}}
    \end{bmatrix}.
\end{equation}
The matrix $\hat{G}$ in the case of equation~\eqref{eq:applying-forbenius-formula} is as follows:
\begin{equation}
  \hat{G} = \hat{S} - \frac{1}{2}\hat{K}\hat{\tilde{C}}\hat{K}^{\intercal}.
\end{equation}
Thus, the solution for the vector $\Delta\bm{q}$ has the form:
\begin{equation}
  \begin{split}
    \Delta\bm{q} &=
    \begin{bmatrix}
      \frac{1}{2}\hat{\tilde{C}} + \frac{1}{4}\hat{\tilde{C}}\hat{K}^{\intercal}\hat{G}^{-1}\hat{K}\hat{\tilde{C}} &
      \rvline &
      -\frac{1}{2}\hat{\tilde{C}}\hat{K}^{\intercal}\hat{G}^{-1} \\
      \hline
      -\frac{1}{2}\hat{G}^{-1}\hat{K}\hat{\tilde{C}} &
      \rvline &
      \hat{G}^{-1}
    \end{bmatrix}
    \begin{bmatrix}
      0 \\
      \vdots\\
      0\\
      \Delta f_1\\
      \vdots \\
      \Delta f_m
    \end{bmatrix}%
    \begin{matrix}
      \coolrightbrace{x \\ x \\ y\\ y}{&n+l}\\
      \coolrightbrace{y \\ y \\ x }{&m}
    \end{matrix}\\
    &=
    \begin{bmatrix}
      -\frac{1}{2}\hat{\tilde{C}}\hat{K}^{\intercal}\hat{G}^{-1}\\
      \hat{G}^{-1}
    \end{bmatrix}
    \begin{bmatrix}
      0 \\
      \vdots\\
      0\\
      \Delta f_1\\
      \vdots \\
      \Delta f_m
    \end{bmatrix}%
    \begin{matrix}
      \coolrightbrace{x \\ x \\ y\\ y}{&l}\\
      \coolrightbrace{y \\ y \\ x }{&m}
    \end{matrix} 
    = \hat{R}\hat{G}^{-1}\hat{P}\Delta\bm{f},
  \end{split}
\end{equation}
where matrix $\hat{R}$ is such that
\begin{equation}
  \hat{R} =
  \begin{bmatrix}
    -\frac{1}{2}\hat{\tilde{C}}\hat{K}^{\intercal} \\
    \hat{I}_{l + m}
  \end{bmatrix},
\end{equation}
matrix $\hat{P}$ has
the following form:
\begin{equation}
  \hat{P} =
  \begin{bmatrix}
    \bigzero_{l\times m}\\
    \hat{I}_{m}
    \end{bmatrix},
\end{equation}
matrix $\hat{I}_{k}$ is the $k\times k$ identity matrix,
$\Delta\bm{f}=\begin{bmatrix} \Delta f_1 & \Delta f_2 & \cdots & \Delta f_m\end{bmatrix}^{\intercal}$.
The chi-square at the local minimum point $\bm{x^{\prime}}$ has the following form:
\begin{equation}
  \label{eq:chi2-linear}
  \begin{split}
  \chi^2(\bm{x^{\prime}}) &= \left(\bm{x^{\prime}} - \bm{\tilde{x}}\right)^{\intercal}
  \hat{\tilde{C}}^{-1}\left(\bm{x^{\prime}} - \bm{\tilde{x}}\right) =
  \Delta\bm{x}^{\intercal}\hat{\tilde{C}}^{-1}\Delta\bm{x} = \Delta\bm{q}^{\intercal}\hat{Y}\Delta\bm{q} \\
  &= \Delta\bm{f}^{\intercal}\hat{P}^{\intercal}\hat{G}^{-1}\hat{R}^{\intercal}\hat{Y}\hat{R}\hat{G}^{-1}\hat{P}\Delta\bm{f} \\
  &= \Delta\bm{f}^{\intercal}\left(\frac{1}{4}\hat{P}^{\intercal}\hat{G}^{-1}\hat{K}\hat{\tilde{C}}\hat{K}^{\intercal}\hat{G}^{-1}\hat{P}\right)\Delta\bm{f}
  = \Delta\bm{f}^{\intercal}\hat{\mathcal{M}}\Delta\bm{f},
  \end{split}
\end{equation}
where $\hat{Y}=\hat{\tilde{C}}^{-1}\oplus\bigzero_{(l+m)\times(l+m)}$ and $\hat{\mathcal{M}}=\frac{1}{4}\hat{P}^{\intercal}\hat{G}^{-1}\hat{K}\hat{\tilde{C}}\hat{K}^{\intercal}\hat{G}^{-1}\hat{P}$.

\subsubsection{Case $l=0$ \label{sec:case-of-zero-l}}
Let us consider first the case where the number of non-measurable parameters is zero, $l=0$.
In this case $\hat{G}=-\frac{1}{2}\hat{J_{\bm{x}}}\hat{\tilde{C}}\hat{J_{\bm{x}}}^{\intercal}$. This matrix must
be non-singular, otherwise the system of equations~\eqref{eq:linear-system-llc-2} is unsolvable.

Consider the random vector $\Delta\bm{f}=\hat{J}_{\bm{x}}\bm{\tilde{x}}+\bm{v}$.
Since this vector is linearly related to the random vector
$\bm{\tilde{x}}\sim\mathcal{N}(\bm{\mu}_{\bm{x}},\hat{\tilde{C}})$, it is easy to show
that it is also normally distributed. The expectation $\text{E}[.]$ of this vector has the
following form: $\text{E}[\Delta\bm{f}] =
\hat{J}_{\bm{x}}\text{E}[\bm{\tilde{x}}]+\bm{v}=\hat{J}_{\bm{x}}\bm{\mu}_{\bm{x}}+\bm{v}\equiv\bm{0}$.
In the previous formula, we used the identity $\text{E}[\bm{\tilde{x}}]=\bm{\mu}_{\bm{x}}$.
The covariance matrix $\hat{\text{Cov}}[.]$ for this vector can also be easily obtained:
\begin{equation}
  \begin{split}
    \hat{\text{Cov}}[\Delta\bm{f}]_{ij} &= \text{E}[(\Delta\bm{f} - \bm{0})_i(\Delta\bm{f} - \bm{0})_j] =\\
    \text{E}[(\hat{J}_{\bm{x}}\bm{\tilde{x}} + \bm{v})_i(\hat{J}_{\bm{x}}\bm{\tilde{x}} + \bm{v})_j] &=
    (\hat{J}_{\bm{x}})_{ia}(\hat{J}_{\bm{x}})_{jb}\text{E}[(\bm{\tilde{x}})_a(\bm{\tilde{x}})_b] +
    (\bm{v})_i(\hat{J}_{\bm{x}}\bm{\mu}_{\bm{x}})_j = \\
    (\hat{J}_{\bm{x}})_{ia}(\hat{J}_{\bm{x}})_{jb}
    \text{E}[(\bm{\tilde{x}} - \bm{\mu}_{\bm{x}})_a(\bm{\tilde{x}} - \bm{\mu}_{\bm{x}})_b] &=
    \left(\hat{J}_{\bm{x}}\hat{\tilde{C}}\hat{J}^{\intercal}_{\bm{x}}\right)_{ij}.
  \end{split}
\end{equation}
In the last equation it is assumed that summation is carried out over the repeated indices $a$ and $b$. Indices $i$ and $j$ are
free matrix indices. Thus we got that $\Delta\bm{f}\sim\mathcal{N}(\bm{0}, \hat{J}_{\bm{x}}\hat{\tilde{C}}\hat{J}^{\intercal}_{\bm{x}})$.
Since the matrix $\hat{G}=-\frac{1}{2}\hat{J_{\bm{x}}}\hat{\tilde{C}}\hat{J_{\bm{x}}}^{\intercal}$ is non-singular, the matrix
$\hat{J_{\bm{x}}}\hat{\tilde{C}}\hat{J_{\bm{x}}}^{\intercal}$ is also non-singular. Therefore, the random variable
$\Delta\bm{f}^{\intercal}\left(\hat{J}_{\bm{x}}\hat{\tilde{C}}\hat{J}^{\intercal}_{\bm{x}}\right)^{-1}\Delta\bm{f}$ is distributed according to
the chi-square distribution with $\text{rank}\left(\hat{J}_{\bm{x}}\hat{\tilde{C}}\hat{J}^{\intercal}_{\bm{x}}\right) = m$ degrees of freedom.

On the other hand, according to equation~\eqref{eq:chi2-linear}, the following equality holds in the
case of $l=0$:
\begin{equation}
  \chi^2(\bm{x^{\prime}})=\Delta\bm{x}^{\intercal}\hat{\tilde{C}}^{-1}\Delta\bm{x} =
  \Delta\bm{f}^{\intercal}\hat{M}\Delta\bm{f} =
  \Delta\bm{f}^{\intercal}\left(\hat{J}_{\bm{x}}\hat{\tilde{C}}\hat{J}^{\intercal}_{\bm{x}}\right)^{-1}\Delta\bm{f},
\end{equation}
i.e. the random variable $\chi^2(\bm{x}^{\prime})=\Delta\bm{x}^{\intercal}\hat{\tilde{C}}^{-1}\Delta\bm{x}$ has the same distribution
as the random variable $\Delta\bm{f}^{\intercal}\left(\hat{J}_{\bm{x}}\hat{\tilde{C}}\hat{J}^{\intercal}_{\bm{x}}\right)^{-1}\Delta\bm{f}$.
Thus, in the case of linear constraints and with $l=0$, the random variable $\chi^2(\bm{x^{\prime}})$ is distributed according
to the probability density function~\eqref{eq:chi-square-pdf} with $m$ degrees of freedom.

\subsubsection{Case $l\neq 0$ \label{sec:case-of-nonzero-l}}
Now consider the case when the number of non-measurable parameters is non-zero.
The system of linear equations~\eqref{eq:linear-constraints} describing the constraints
can be rewritten as follows:
\begin{equation}
  \label{eq:linear-constraints-2}
  \hat{J}_{\bm{x}}\bm{x} + \hat{J}_{\bm{a}}\bm{a} + \bm{v} = \bm{0}.
\end{equation}
Since $\text{rank}(\hat{J})=m$, the latter system contains $m$ linearly independent equations.
Let us multiply this system of equations on the left by the matrix $\hat{J}^{\intercal}_{\bm{a}}$.
Since $\text{rank}\left(\hat{J}^{\intercal}_{\bm{a}}\hat{J}_{\bm{a}}\right)=\text{rank}(\hat{J}_{\bm{a}})=l$,
the $l\times l$ matrix $\hat{J}^{\intercal}_{\bm{a}}\hat{J}_{\bm{a}}$ is a full rank matrix,
i.e. this matrix is non-singular. Thus, non-measurable variables $\bm{a}$ can be expressed
from equation~\eqref{eq:linear-constraints-2} as follows:
\begin{equation}
  \label{eq:non-measurable-params}
  \bm{a} = -\left(\hat{J}^{\intercal}_{\bm{a}}\hat{J}_{\bm{a}}\right)^{-1}\hat{J}^{\intercal}_{\bm{a}}
  \left(\hat{J}_{\bm{x}}\bm{x} + \bm{v}\right).
\end{equation}
Further, non-measurable parameters can be excluded from equation~\eqref{eq:linear-constraints-2}
by substituting equation~\eqref{eq:non-measurable-params} back into equation~\eqref{eq:linear-constraints-2}.
As a result of this substitution, one can obtain the following system of equations:
\begin{equation}
  \label{eq:constraint-replacement}
  \left(\hat{I} - \hat{J}_{\bm{a}}\left(\hat{J}^{\intercal}_{\bm{a}}\hat{J}_{\bm{a}}\right)^{-1}\hat{J}^{\intercal}_{\bm{a}}\right)
  \left(\hat{J}_{\bm{x}}\bm{x} + \bm{v}\right) = \bm{0},
\end{equation}
where $\hat{I}$ is the identity matrix. The system of linear equations~\eqref{eq:constraint-replacement}
contains $m$ equations, but some of the equations in this system
are linearly dependent. Since this system was obtained from a system of $m$ linearly independent
equations~\eqref{eq:linear-constraints-2} by eliminating $l$ non-measurable parameters,
it contains only $m - l$ linearly independent equations. Thus the case $l\neq 0$
can be reduced to the case $l=0$. Therefore, by analogy with section~\ref{sec:case-of-nonzero-l},
one can conclude that
in the case of linear constraints and with $l\neq 0$, the random variable $\chi^2(\bm{x^{\prime}})$ is
distributed according to the probability density function~\eqref{eq:chi-square-pdf} with $m - l$
degrees of freedom.

\subsection{Chi-square in the case of nonlinear constraints \label{sec:ndf-non-linear}}
In the case of nonlinear constraints, the random vectors $\bm{x^{\prime}}$ and
$\Delta\bm{f}$ are expressed in terms of the random vector $\bm{\tilde{x}}$
nonlinearly. Therefore, in the case of nonlinear constraints, the distribution of
the random variable
$\chi^2(\bm{x^{\prime}}) = \left(\bm{x^{\prime}} - \bm{\tilde{x}}\right)^{\intercal}\hat{\tilde{C}}^{-1}\left(\bm{x^{\prime}} - \bm{\tilde{x}}\right)$
is not consistent with probability density function~\eqref{eq:chi-square-pdf}.

The kinematic and vertex fitting implies extensive use
of nonlinear constraints (see sections~\ref{sec:particles} and~\ref{sec:constraints}).
Thus, one should expect that the distribution of the random variable
$\chi^2(\bm{x^{\prime}})$ will not be consistent with the probability density
function~\eqref{eq:chi-square-pdf} even in the case of the Gaussian response of the
detector. However, in the case of many hypotheses, nonlinear effects do not lead to significant
distortion of the chi-square distribution. That is, despite the nonlinearity,
the distribution of the random variable $\chi^2(\bm{x^{\prime}})$ in these cases
is well described by the probability density function~\eqref{eq:chi-square-pdf}. At the same time,
in the case of some hypotheses, the nonlinear effects can be so large that the
distribution of the random variable $\chi^2(\bm{x^{\prime}})$ in these cases is not
described by the probability density function~\eqref{eq:chi-square-pdf}.
A detailed discussion of how constraint nonlinearity affects the $\chi^2(\bm{x^{\prime}})$
distribution is given in section~\ref{sec:gaussian-simulation-examples} with
examples of Gaussian simulation.

\subsection{Chi-square in the case of Monte Carlo simulation and experimental data \label{sec:chi2-non-gaussian}}
It often happens that the uncertainties of some measurable parameters are of a non-Gaussian
nature. The reason is the non-Gaussian response of the detector.
This behavior of parameter uncertainties can be observed not only in experimental data,
but also in Monte Carlo simulation\footnote{Further, by Monte Carlo simulation we understand the
simulation of events of any $e^+e^-$ annihilation process, taking into account
the response of a detector.} events. In section~\ref{sec:ndf-linear},
it is shown that the chi-square distribution is described by the probability density
function~\eqref{eq:chi-square-pdf} only if the parameter uncertainties are Gaussian.
Thus, we expect that the chi-square distribution obtained for experimental events or
Monte Carlo simulation events will not be consistent with the chi-squared probability
density function. Another effect that distorts the chi-square distribution in
the case of the $\text{CMD-}3$ experiment is the imperfect calibration of the
covariance matrix $\hat{\tilde{C}}$. These effects are well illustrated by Monte
Carlo simulation examples in section~\ref{eq:mc-examples}.

Since the Monte Carlo simulation does not perfectly describe the experimental data,
the distributions of the random variable $\chi^2(\bm{x}^{\prime})$ in the experiment
and simulation may be different. Suppose that in the analysis of some physical process,
a selection criterion for $\chi^2(\bm{x^{\prime}})$ is used. Since the distributions
of the random value $\chi^2(\bm{x^{\prime}})$ in experiment and simulation are different,
the detection efficiencies in simulation and experiment are also different.
The detection efficiency in the experiment is usually unknown, so the corresponding
efficiency from the simulation is usually used instead. Since these efficiencies are different,
this difference must be taken into account using efficiency corrections and finding the
corresponding systematic uncertainties. Examples of detection efficiency corrections related
to a chi-square selection criterion can be found, for example, in articles~\cite{Gribanov2020,Aulchenko2015}.

\subsection{Gaussian simulation \label{sec:gaussian-simulation}}
In sections~\ref{sec:ndf-non-linear} and~\ref{sec:chi2-non-gaussian}, it is discussed
that the nonlinearity of the constraints, the non-Gaussian response of the
detector and the imperfect calibration of the covariance matrix $\hat{\tilde{C}}$ can
lead to distortion of the chi-square distribution. However, possible errors in the
implementation of the fitting package can also distort this distribution. Therefore, some
testing procedure is needed to avoid such errors.

In order to verify the fitting procedure, the authors introduce the technique of
Gaussian simulation. This verification technique eliminates the influence of the detector's
non-Gaussian response and the imperfect covariance matrix calibration on the
chi-square distribution. The idea of Gaussian simulation is to
redraw measurable parameters according to the multivariate normal distribution. A detailed
description of this procedure is given in the next paragraph.

Let us consider one event of some process from an experiment or Monte Carlo
simulation and assume that we have made a fit of this event under the
corresponding signal hypothesis.
The parameter values obtained after this fitting will be such that all
kinematic and vertex constraints are satisfied. Therefore, it is possible to use the measurable
parameters found as a result of this fitting as the corresponding mean values for random
generation of similar events according to the multivariate normal distribution. The covariance
matrix known from the initial event can be used as the covariance matrix corresponding to this
distribution. After generating a sufficient number of events, the fitting procedure can be applied
again to these events. As a result, one can obtain a chi-square distribution corresponding to
uncertainties that are of a Gaussian nature. Examples of Gaussian simulation are discussed in
section~\ref{sec:gaussian-simulation-examples}.

If the chi-square distribution obtained as a result of the Gaussian simulation turns out to be
distorted, then this is due either to the nonlinearity of the constraints or to errors in the
fitting package. The fact that in some cases the distortion of the distribution occurs due to the
nonlinearity of the constraints can also be verified. A related discussion is given in
section~\ref{sec:kskpi-gsim}.

\section{Vertices \label{sec:vertices}}
In this paper, vertices are considered as separate entities. The reason for
considering vertices as separate entities is that each vertex is always shared
by some set of particles. Each vertex has three coordinates. The kinematic and
vertex fitting package
includes vertex classes corresponding to coordinates in Cartesian and
cylindrical coordinate systems. The user can add vertices with custom
parametrization corresponding to a different coordinate system. To do this, one
needs to implement a new vertex class inherited from the base vertex class.
The base vertex class contains virtual abstract methods that return the
Cartesian coordinates of a vertex depending on its three parameters. This class
also contains virtual abstract methods for obtaining gradients and Hessians of
the Cartesian coordinates, depending on the vertex parametrization. In order to
create a custom vertex class, the user has to implement the above abstract
methods. In the case of the vertex corresponding to the Cartesian coordinate
system, the implementation of the methods listed above is trivial. For example,
functions that return the Cartesian coordinates of a vertex have the following
form:
\begin{equation}
  \label{eq:cartesian-vertex}
  \begin{split}
    x_{\text{v}}(\bm{a^{(v)}}) &= x^{(v)},\\
    y_{\text{v}}(\bm{a^{(v)}}) &= y^{(v)},\\
    z_{\text{v}}(\bm{a^{(v)}}) &= z^{(v)},\\
    \bm{a^{(v)}} &=
    \begin{bmatrix}
      x^{(v)}\\
      y^{(v)}\\
      z^{(v)}
    \end{bmatrix},
  \end{split}
\end{equation}
where $x_{\text{v}}$, $y_{\text{v}}$ and $z_{\text{v}}$ are functions that
return the Cartesian coordinates of a vertex depending on the vector
$\bm{a}^{(v)}$ of its parameters $x^{(v)}$, $y^{(v)}$ and $z^{(v)}$, which are
themselves the vertex Cartesian coordinates. In the case of a vertex
parametrized according to the cylindrical coordinate system, the functions
returning Cartesian coordinates are as follows:
\begin{equation}
  \label{eq:cylindrical-vertex}
  \begin{split}
    x_{\text{v}}(\bm{a^{(v)}}) &= \rho^{(v)}\cos{\phi^{(v)}},\\
    y_{\text{v}}(\bm{a^{(v)}}) &= \rho^{(v)}\sin{\phi^{(v)}},\\
    z_{\text{v}}(\bm{a^{(v)}}) &= z^{(v)},\\
    \bm{a^{(v)}} &=
    \begin{bmatrix}
      \rho^{(v)}\\
      \phi^{(v)}\\
      z^{(v)}
    \end{bmatrix},
  \end{split}
\end{equation}
where parameters $\rho^{(v)}$, $\phi^{(v)}$ and $z^{(v)}$ are the vertex
coordinates in the cylindrical coordinate system. Here and below, we do not
present gradients and Hessians for vertex coordinates, particle four-momenta and
trajectories, since the procedure for calculating them is well known. The reader
can find them in the fitting package. Links to the relevant repositories are
provided in section~\ref{sec:fitting-package}.

In the case of the CMD-3 experiment, vertex
parametrization~\eqref{eq:cartesian-vertex} is most often used. Such vertex
parametrizations as~\eqref{eq:cylindrical-vertex} can be used in specific cases.
For example, a vertex with parametrization~\eqref{eq:cylindrical-vertex} can be
used if it is required to set limits for the $\rho^{(v)}$ parameter. Such a need
may arise, for example, if one wants to reconstruct a vertex of a photon conversion
on the beam pipe.

All coordinates of the $e^+e^-$ interaction vertex are considered as free measurable
parameters. These parameters make an additional contribution to the chi-square given
by equation~\eqref{eq:chi-square-1}. The part of the covariance matrix $\hat{\tilde{C}}$
corresponding to the $e^+e^-$ interaction vertex is diagonal. Its diagonal elements are
the squares of the corresponding sizes of the interaction region. In the case of
the CMD-3 experiment a typical longitudinal~(along
the beam axis) size of the interaction region is about $2.5$-$3.0\text{
  cm}$. A typical transverse size of the interaction region is approximately
$60$-$100\;\upmu\text{m}$.

All coordinates of decay vertices are treated as free
non-measurable parameters and do not contribute to the chi-square.

\section{Particles \label{sec:particles}}
Like vertices, particles are considered as separate entities
in this work. Particle classes contain parametrizations of output four-momenta
with respect to the origin vertex and input four-momenta with respect to the
decay vertex. These four-momenta are used in the energy-momentum conservation
constraints, as well as in the constraints on the invariant mass of some
particle set. Energy-momentum constraints and mass constraints are described in
detail in sections~\ref{sec:energy-momentum-constraints} and
\ref{sec:mass-constraint}, respectively.

In the case of some particle classes, parametrization of particle trajectories
can also be implemented. These trajectories are required in
constraints implemented for vertex fitting (see section~\ref{sec:vertex-constraints}).

\subsection{Initial pseudo-particle \label{sec:initial-pseudo-particle}}
An initial pseudo particle is used in order to set the total four-momentum of
all particles present in a certain hypothesis. In the case of the
$\text{VEPP-}2000$ collider, the energies of the initial electrons and positrons
are the same with high accuracy, and the directions of their motion are
opposite. For this reason, the four-momentum parametrization for the initial
pseudo-particle in the case of the $\text{CMD-}3$ experiment can be written in
the following form:
\begin{equation}
  \label{eq:momentum-cm}
  \mathcal{P}^{(i)}_{\text{pseudo}} = (E_{\text{c.m.}}, \bm{0}),
\end{equation}
where $E_{\text{c.m.}}=2E_{\text{beam}}$ is the center-of-mass energy,
$E_{\text{beam}}$ is the beam energy, $i$ is the particle index. In the case of
the $\text{CMD-}3$, the energy is set constant at each event. In principle, the
user can release this parameter (make it free and measurable). In the case of
the $\text{VEPP-}2000$, the typical beam energy
spread is less than $1\text{ MeV}$. For this reason, using a non-constant center-of-mass energy
$E_{\text{c.m.}}$ in the case of the CMD-3 experiment will not have a
significant effect\footnote{This effect is within the detector resolution.} on
the fit result.

\subsection{Charged particles and photons}
The parametrizations of a charged particle and photon were placed in a separate
subsection, since the authors consider these parametrizations to be
detector-dependent. For example, in the case of a photon, the detector-dependent
part of the parametrization is contained at least in the description of the
photon conversion point inside a calorimeter. In the case of the $\text{CMD-}3$
experiment, a cylindrical coordinate system is used to set the photon conversion
point. The parametrization of a charged particle in the case of the
$\text{CMD-}3$ experiment is described in section~\ref{sec:charged-particle}.
This parametrization may also have some differences in the case of a different
experiment. For example, in the case of the charged particle parametrization
described in section~\ref{sec:charged-particle}, a constant
magnetic field\footnote{Furthermore, it is assumed that this field is directed
along the $Z\text{-axis}$.} at
any point of the drift chamber is used, i.e. the change in the magnetic field
near the detector end-caps is not taken into account. This is due to the fact
that the charged particle tracks themselves are reconstructed in the case of the
$\text{CMD-}3$ experiment under the assumption of a uniform magnetic field.
The inhomogeneity of the magnetic field in the CMD-3 experiment is usually taken
into account by introducing a corresponding systematic uncertainty. Since the
package of kinematic and vertex fitting discussed in this paper uses the
parameters of already reconstructed tracks, it makes no sense to take into
account the inhomogeneity of the magnetic field in this package.
Section~\ref{sec:charged-particle} discusses only the parametrization of a final
charged particle. The parametrization of an intermediate charged particle is
discussed in section~\ref{sec:intermediate-charged-particle}.

\subsubsection{Charged particle \label{sec:charged-particle}}
In the case of the experiment with the CMD-3 detector, the parametrizations of
four-momenta and trajectories of final charged particles depend on five
measurable parameters. These parameters are listed below:
\begin{itemize}
  \item $p^{(i)}_{\perp}$ is the radial component of momentum, i.e. the momentum
    component perpendicular to the magnetic field;
  \item $\rho^{(i)}_{\text{c}}$ is the distance between the beam axis and
    track\footnote{In this article,
      by track we mean the track of some charged particle, reconstructed using
      hits
      in the drift chamber.} of
    a charged particle;
  \item $z^{(i)}_{\text{c}}$ is $z$-coordinate of the track point closest to the
    beam axis;
  \item $\phi^{(i)}_{\text{c}}$ is the axial angle corresponding to the track point closest to
    the beam axis;
  \item $\theta^{(i)}_{\text{c}}$ is the polar angle of a charged particle momentum.
\end{itemize}
The parametrizations of a charged particle four-momentum and trajectory depend
also on one non-measurable parameter $ct^{(i)}_{\text{out}}$. This parameter
has a meaning of the distance along the charged particle trajectory from the
charged particle origin vertex to the track point closest to the beam axis. 

The parametrization of a charged particle four-momentum
$\mathcal{P}^{(i)}_{\text{c, out}}=\left(E^{(i)}_{\text{c}},\;\bm{p^{(i)}_{\text{c, out}}}\right)$
has the following form:
\begin{equation}
  \label{eq:charged-particle-momentum}
  \begin{split}
    E^{(i)}_{\text{c}} &= \sqrt{\left.p^{(i)}_{\perp}\right.^2(1 + \cot^2{\theta^{(i)}_{\text{c}}}) + \left.m^{(i)}_{\text{c}}\right.^2},\\
    \bm{p^{(i)}_{\text{c, out}}} &=
    \begin{bmatrix}
      p^{(i)}_{\perp}\cos{\left(\displaystyle\frac{\omega^{\mathrlap{(i)}}_{\mathrlap{\text{c}}}}{c}\hphantom{i}ct^{(i)}_{\text{out}} - \phi^{(i)}_{\text{c}}\right)}\\
      \addlinespace
      -p^{(i)}_{\perp}\sin{\left(\displaystyle\frac{\omega^{\mathrlap{(i)}}_{\mathrlap{\text{c}}}}{c}\hphantom{i}ct^{(i)}_{\text{out}} - \phi^{(i)}_{\text{c}}\right)}\\
      \addlinespace
      p^{(i)}_{\perp}\cot\theta^{(i)}_{\text{c}}
    \end{bmatrix}, \\
    \frac{\omega^{\mathrlap{(i)}}_{\mathrlap{\text{c}}}}{c}\hphantom{i} &= \frac{q^{(i)}_{\text{c}}B\kappa_{\text{c}}}{E^{(i)}_{\text{c}}},\\
    \kappa_{\text{c}} &\approx 2.9979\times10^{-3},
  \end{split}
\end{equation}
where the constant $q^{(i)}_{\text{c}}$ is the particle charge, measured in
elementary charges; the constant $m^{(i)}_{\text{c}}$ is the mass of a charged
particle, measured in $\text{GeV}/c^2$; the constant $B$ is the magnetic field,
measured in T. Fraction $\omega^{(i)}_{\text{c}}/c$ is the fraction of cyclotron
frequency to the speed of light; the constant  $\kappa_{\text{c}}\approx 2.9979\times10^{-3}$
is the proportionality
factor\footnote{The value of this factor is determined by the speed of light and units of measurement
of the quantities involved in equation~\eqref{eq:charged-particle-momentum}. Here and below, we assume
that energy is measured in $\text{GeV}$, magnetic field in $\text{T}$, time in seconds, and distance in
$\text{cm}$.}.

The parametrization of a charged particle trajectory
$\bm{r^{(i)}_{\text{c, out}}} = \begin{bmatrix} x^{(i)}_{\text{c, out}} & y^{(i)}_{\text{c, out}} & z^{(i)}_{\text{c, out}}\end{bmatrix}^{\intercal}$
is given by the following equation:
\begin{equation}
  \label{eq:charged-particle-trajectory}
  \begin{split}
    x^{(i)}_{\text{c, out}} &= x_{\text{beam}} + \left(R^{(i)}_{\text{c}} -
    \rho^{(i)}_{\text{c}}\right)\sin\phi^{(i)}_{\text{c}} +
    R^{(i)}_{\text{c}}\sin{\left(\frac{\omega^{\mathrlap{(i)}}_{\mathrlap{\text{c}}}}{c}\hphantom{i}ct^{(i)}_{\text{out}} -
      \phi^{(i)}_{\text{c}}\right)},\\
    y^{(i)}_{\text{c, out}} &= y_{\text{beam}} - \left(R^{(i)}_{\text{c}} -
    \rho^{(i)}_{\text{c}}\right)\cos\phi^{(i)}_{\text{c}} +
    R^{(i)}_{\text{c}}\cos{\left(\frac{\omega^{\mathrlap{(i)}}_{\mathrlap{\text{c}}}}{c}\hphantom{i}ct^{(i)}_{\text{out}} -
      \phi^{(i)}_{\text{c}}\right)},\\
    z^{(i)}_{\text{c, out}} &= z^{(i)}_{\text{c}} + \frac{p^{(i)}_{\perp}\cot\theta^{(i)}_{\text{c}}ct^{(i)}_{\text{out}}}{E^{(i)}_{\text{c}}},\\
    R^{(i)}_{\text{c}} &= \frac{p^{(i)}_{\perp}}{q^{(i)}_{\text{c}}B\kappa_{\rm c}},
  \end{split}
\end{equation}
where $x_{\text{beam}}$ are $y_{\text{beam}}$ the $x\text{-}$ and $y\text{-}$ coordinates of
the beam axis, respectively. In the case of the CMD-3 detector these coordinates
considered as constants, since they are known with a high accuracy.

\subsubsection{Photon \label{sec:photon}}
The parametrization of the photon four-momentum depends on seven parameters.
Four of the seven parameters belong to the photon. The remaining three
parameters are the coordinates $\bm{r_{\text{origin}}}$ of its origin
vertex. One of the four photon parameters is its energy $E^{(i)}_{\gamma}$. The
remaining three photon parameters are the coordinates $\rho^{(i)}_{\gamma}$,
$\phi^{(i)}_{\gamma}$ and $z^{(i)}_{\gamma}$ of the conversion point in a
cylindrical coordinate system. The photon four-momentum
$\mathcal{P}^{(i)}_{\gamma}=(E^{(i)}_{\gamma}, \bm{p^{(i)}_{\gamma}})$ is given
by the following equation\footnote{In this article, by norm we always
  mean the $\ell^2$-norm: $\|\bm{b}\|=\sqrt{\sum\limits^{\dim\bm{b}}_{i=1}b^2_i}$.}:
\begin{equation}
  \label{eq:photon-four-momentum}
  \begin{split}
    \bm{p^{(i)}_{\gamma}} &= E^{(i)}_{\gamma}\frac{\bm{r^{(i)}_{\text{conv}}} -
      \bm{r_{\text{origin}}}}{\lVert \bm{r^{(i)}_{\text{conv}}} - \bm{r_{\text{origin}}} \rVert},\\
    \bm{r^{(i)}_{\text{conv}}} &=
    \begin{bmatrix}
      \rho^{(i)}_{\gamma}\cos\phi^{(i)}_{\gamma}\\
      \rho^{(i)}_{\gamma}\sin\phi^{(i)}_{\gamma}\\
      z^{(i)}_{\gamma}
    \end{bmatrix}.
  \end{split}
\end{equation}
Since equation~\eqref{eq:photon-four-momentum} already requires a photon to fly
from the origin vertex $\bm{r_{\text{origin}}}$ to the conversion point
$\bm{r^{(i)}_{\text{conv}}}$, there is no need for a separate parametrization of
a photon trajectory.

\subsection{Intermediate particles \label{sec:intermediate-particles}}
Along all intermediate particles in the case of the $\text{CMD-}3$ experiment,
$K_S$ mesons require special attention due to their long lifetime.
Parametrization of neutral intermediate particles, such as $K_S$ mesons, is
given in section~\ref{sec:intermediate-neutral-particle}. Note that in the case
of such particles as $\pi^0$ or $\eta$ mesons, the parametrization
described in section~\ref{sec:intermediate-neutral-particle} should not be used.
In the terms of kinematic and vertex fitting, these particles decay at the
origin vertex due to their short lifetime. To take these particles into account,
it suffices to require a mass constraint on their decay products.

Due to the small radius ($30\text{ cm}$) of the drift chamber in the
$\text{CMD-}3$ experiment, it practically makes no sense to consider hypotheses
with intermediate charged particles. The probability of a charged $\pi$-meson or
$K$-meson decay inside the drift chamber is low. There is always an odd number
of charged particles among the decay products of a charged particle in order to
conserve charge. The tracks of an initial charged particle and the charged
products of its decay are well reconstructed if the decay vertex is near the
center of the drift chamber radius, since in this case the probability that all
tracks will have a sufficient number of hits is the highest. Thus, the number of
charged particle decay events in which kinematic and vertex fitting can be
successfully used is additionally limited by the track reconstruction efficiency.
However, intermediate charged particles have been implemented in the discussed
kinematic and vertex fitting package. In addition, it was verified that the use
of these intermediate particles makes it possible, for example, to find events
with $\pi^+\rightarrow\mu^+\nu_{\mu}$ decay\footnote{Since the detection of
intermediate charged particles in the case of $\text{CMD-}3$ is difficult
due to the small size of the drift chamber, the authors do not further consider
examples of hypotheses with such particles. The authors hope to study in more
detail the possibility of using intermediate charged
particles~\ref{sec:intermediate-charged-particle} in other experiments in the future.}.
The parametrization of intermediate charged particles is discussed in detail in
section~\ref{sec:intermediate-charged-particle}.

\subsubsection{Intermediate neutral particle \label{sec:intermediate-neutral-particle}}
The parametrization of the intermediate neutral particle depends on seven
parameters. Four parameters belong to this particle. The three remaining
parameters are the coordinates $\bm{r_{\text{origin}}}$ of its origin vertex.
The first three parameters of an intermediate neutral particle are the Cartesian
components of its momentum
\begin{equation}
  \label{eq:int-neutral-momentum}
\bm{p^{(i)}_{\text{int. n.}}}=\begin{bmatrix}p^{(i)}_{\text{x}} \\ p^{(i)}_{\text{y}} \\ p^{(i)}_{\text{z}}\end{bmatrix},
\end{equation}
the last parameter $\xi^{(i)}$ is proportional to the time interval between the particle origin and decay.
The four-momentum parametrization for an intermediate neutral particle has the following form:
\begin{equation}
    \mathcal{P}^{(i)}_{\text{int. n.}} = \left(\sqrt{\left.m^{(i)}_{\mathrlap{\text{int. n.}}}\right.^2\hphantom{i} +
      \left.\bm{p^{(i)}_{\mathrlap{\text{int. n.}}}}\right.^2\hphantom{i}},\;\bm{p^{(i)}_{\text{int. n.}}}\right),
\end{equation}
where the constant $m^{(i)}_{\text{int. n.}}$ is the particle mass.
Trajectory parametrization of an intermediate neutral particle can be written as follows:
\begin{equation}
  \label{eq:int-neutral-trajectory}
  \bm{r^{(i)}_{\text{int. n.}}} = \bm{r_{\text{origin}}} + \xi^{(i)}\bm{p^{(i)}_{\text{int. n.}}}.
\end{equation}
All parameters $p^{(i)}_{\text{x}}$, $p^{(i)}_{\text{y}}$, $p^{(i)}_{\text{z}}$
and $\xi^{(i)}$ of an intermediate neutral particle are considered to be
non-measurable.

\subsubsection{Intermediate charged particle \label{sec:intermediate-charged-particle}}
The parametrizations of four-momenta and the trajectories of intermediate
charged particles are very similar to the corresponding parametrizations in the
case of final charged particles~\ref{sec:charged-particle}. In the case of an
intermediate charged particle, the parametrizations of the output
four-momentum~\eqref{eq:charged-particle-momentum} and the output
trajectory~\eqref{eq:charged-particle-trajectory} with respect to the origin
vertex are exactly the same as for a final charged particle. The
parametrizations of the input four-momentum and the input trajectory with
respect to the decay vertex differ from the
parametrizations~\eqref{eq:charged-particle-momentum} and
\eqref{eq:charged-particle-trajectory} by replacing the parameter
$ct^{(i)}_{\text{out}}$ with the parameter $ct^{(i)}_{\text{in}}$.
As a result of this replacement,
the parametrization of the input four-momentum
$\mathcal{P}^{(i)}_{\text{c, in}}=\left(E^{(i)}_{\text{c}},\;\bm{p^{(i)}_{\text{c, in}}}\right)$
of an intermediate charged particle has the following form:
\begin{equation}
  \label{eq:input-charged-particle-momentum}
  \begin{split}
    E^{(i)}_{\text{c}} &= \sqrt{\left.p^{(i)}_{\perp}\right.^2(1 + \cot^2{\theta^{(i)}_{\text{c}}}) + \left.m^{(i)}_{\text{c}}\right.^2},\\
    \bm{p^{(i)}_{\text{c, in}}} &=
    \begin{bmatrix}
      p^{(i)}_{\perp}\cos{\left(\displaystyle\frac{\omega^{\mathrlap{(i)}}_{\mathrlap{\text{c}}}}{c}\hphantom{i}ct^{(i)}_{\text{in}} - \phi^{(i)}_{\text{c}}\right)}\\
      \addlinespace
      -p^{(i)}_{\perp}\sin{\left(\displaystyle\frac{\omega^{\mathrlap{(i)}}_{\mathrlap{\text{c}}}}{c}\hphantom{i}ct^{(i)}_{\text{in}} - \phi^{(i)}_{\text{c}}\right)}\\
      \addlinespace
      p^{(i)}_{\perp}\cot\theta^{(i)}_{\text{c}}
    \end{bmatrix},\\
    \frac{\omega^{\mathrlap{(i)}}_{\mathrlap{\text{c}}}}{c}\hphantom{i} &= \frac{q^{(i)}_{\text{c}}B\kappa_{\text{c}}}{E^{(i)}_{\text{c}}}.
  \end{split}
\end{equation}
The parametrization of input trajectory can be written as follows:
\begin{equation}
  \label{eq:input-charged-particle-trajectory}
  \begin{split}
    x^{(i)}_{\text{c, in}} &= x_{\text{beam}} + \left(R^{(i)}_{\text{c}} -
    \rho^{(i)}_{\text{c}}\right)\sin\phi^{(i)}_{\text{c}} +
    R^{(i)}_{\text{c}}\sin{\left(\frac{\omega^{\mathrlap{(i)}}_{\mathrlap{\text{c}}}}{c}\hphantom{i}ct^{(i)}_{\text{in}} -
      \phi^{(i)}_{\text{c}}\right)},\\
    y^{(i)}_{\text{c, in}} &= y_{\text{beam}} - \left(R^{(i)}_{\text{c}} -
    \rho^{(i)}_{\text{c}}\right)\cos\phi^{(i)}_{\text{c}} +
    R^{(i)}_{\text{c}}\cos{\left(\frac{\omega^{\mathrlap{(i)}}_{\mathrlap{\text{c}}}}{c}\hphantom{i}ct^{(i)}_{\text{in}} -
      \phi^{(i)}_{\text{c}}\right)},\\
    z^{(i)}_{\text{c, in}} &= z^{(i)}_{\text{c}} + \frac{p^{(i)}_{\perp}\cot\theta^{(i)}_{\text{c}}ct^{(i)}_{\text{in}}}{E^{(i)}_{\text{c}}},\\
    R^{(i)}_{\text{c}} &= \frac{p^{(i)}_{\perp}}{q^{(i)}_{\text{c}}B\kappa_{\rm c}}.
  \end{split}
\end{equation}
The parameter $ct^{(i)}_{\text{in}}$ in
equations~\eqref{eq:input-charged-particle-momentum} and
\eqref{eq:input-charged-particle-trajectory} has the meaning of the distance
along the charged particle trajectory from the track point closest to the beam
axis to the decay vertex of this particle. Like the parameter
$ct^{(i)}_{\text{out}}$, the parameter $ct^{(i)}_{\text{in}}$ is a
non-measurable parameter.

Practically, final~\ref{sec:charged-particle} and
intermediate~\ref{sec:intermediate-charged-particle} charged particles are
implemented as a single class in the kinematic and vertex fitting package. In
the case of a final charged particle the parameter $ct^{(i)}_{\text{in}}$ is
fixed, and the parameterizations of the input four-momentum and the input
trajectory given by equations~\eqref{eq:input-charged-particle-momentum}
and~\eqref{eq:input-charged-particle-trajectory} are not used.

\subsection{Lost particles \label{sec:lost-particles}}
An important requirement for the kinematic and vertex fitting package is the
ability to perform a fit with the hypothesis of lost particles\footnote{By lost
particles, we mean final particles that are not detected.}.
It should be noted that for a lost particle it makes sense to specify only a
four-momentum, while
specifying the trajectory of such a particle is meaningless. It should also be
noted that the four-momentum parameterization is different for massive and
massless particles. The four-momentum
parametrization of a massive lost particle
is discussed in section~\ref{sec:massive-lost-particle}, while the
parametrization in the case of a massless lost particle is given in
section~\ref{sec:massless-lost-particle}.

The parameters of the lost particles are non-measurable as they are not measured
directly with a detector. In the case of the discussed package, this is achieved
by specifying zero inverse covariance matrices corresponding to the parameters
of these particles. In some cases the
parameterizations from sections~\ref{sec:massive-lost-particle} and
\ref{sec:massless-lost-particle} can also be used for detected
particles. It is
assumed that the parametrization of the trajectories of such particles is of no
interest, and the corresponding inverse covariance matrices are non-zero.

\subsubsection{Massive particle \label{sec:massive-lost-particle}}
The parametrization of a massive lost particle has the following form:
\begin{equation}
  \label{eq:massive-particle-momentum}
    \mathcal{P}^{(i)}_{m\neq0} = (\sqrt{\left.m^{(i)}\right.^2 + \left.\bm{p^{(i)}}\right.^2},
    \bm{p^{(i)}}),\;
    \bm{p^{(i)}} = \begin{bmatrix} p^{(i)}_x \\ p^{(i)}_y \\ p^{(i)}_z \end{bmatrix},
\end{equation}
where the parameters $p^{(i)}_x$, $p^{(i)}_y$ and $p^{(i)}_z$ are the momentum
components of the particle, and the constant $m^{(i)}\neq0$ is its mass.

\subsubsection{Massless particle \label{sec:massless-lost-particle}}
The four-momentum parametrization of a lost massless particle cannot be chosen
in the form~\eqref{eq:massive-particle-momentum} with $m^{(i)}=0$. The
disadvantages of the parametrization~\eqref{eq:massive-particle-momentum} in the
case of a massless particle is that the denominators of the derivatives of this
parametrization become very small if $\bm{p^{(i)}}$ is close to zero.
For this reason, in the case of a lost massless particle, a slightly different
four-momentum parametrization is used. This parametrization has the following
form:
\begin{equation}
  \label{eq:massless-particle-momentum}
    \mathcal{P}^{(i)}_{m=0} = (E^{(i)}, E^{(i)}\bm{n^{(i)}}),\; 
    \bm{n^{(i)}} =
    \begin{bmatrix}
    \sin\theta^{(i)}\cos\phi^{(i)} \\
    \sin\theta^{(i)}\sin\phi^{(i)} \\
    \cos\theta^{(i)}
    \end{bmatrix},
\end{equation}
where the parameter $E^{(i)}$ is the particle energy, the parameters
$\theta^{(i)}$ and $\phi^{(i)}$ are the polar and axial angles, respectively.
These angles determine the direction of the particle momentum
$\bm{p^{(i)}} = E^{(i)}\bm{n^{(i)}}$.

\section{Constraints\label{sec:constraints}}
The discussed kinematic and vertex fitting package uses three different kinds of
constraints. These are energy-momentum conservation constraints, vertex
constraints and mass constraints.

The energy-momentum conservation constraints are imposed on a certain set of
particles, and are divided into four different constraints: one constraint is
needed for energy conservation, the other three are for momentum conservation. A
detailed discussion of the energy-momentum conservation constraints is given in
section~\ref{sec:energy-momentum-constraints}.

The mass constraint is imposed on a certain set of particles in order to require
that the invariant mass of these particles be equal to a certain value. The
detailed description of the mass constraint is given in
section~\ref{sec:mass-constraint}.

The vertex constraints are imposed on some individual particles at the vertices
of their origin and / or decay. These constraints require that the particle's
trajectory pass through the origin vertex and the decay vertex. A detailed
discussion of the vertex constraints is given in
section~\ref{sec:vertex-constraints}.

\subsection{Energy-momentum conservation constraints \label{sec:energy-momentum-constraints}}
The energy-momentum conservation constraints have the following form:
\begin{equation}
  \label{eq:energy-momentum-constraints}
  \sum\limits_{a\in\mathfrak{S}^{(j)}_{+}}\mathcal{P}^{(a)}_{\text{output}} -
  \sum\limits_{b\in\mathfrak{S}^{(j)}_{-}}\mathcal{P}^{(b)}_{\text{input}} = 0,
\end{equation}
where the sets $\mathfrak{S}^{(j)}_{\pm}$ are the sets of indices corresponding
to the particles involved in the considered constraints,
$\mathcal{P}^{(a)}_{\text{output}}$ is the output four-momentum of the
$a\text{-th}$ particle, $\mathcal{P}^{(b)}_{\text{input}}$ is the input
four-momentum of the $b\text{-th}$ particle. The set $\mathfrak{S}^{(j)}_{-}$
corresponds\footnote{Taking
into account section~\ref{sec:initial-pseudo-particle}, it can be concluded that
the cardinality of the set $\mathfrak{S}^{(j)}_{-}$ is equal to one~(as rule).}
to particles before interaction or decay, while the set $\mathfrak{S}^{(j)}_{+}$
corresponds to particles that are products of this interaction or decay.
The index $j$ in $\mathfrak{S}^{(j)}_{\pm}$ means the index of particle sets on
which the constraints~\eqref{eq:energy-momentum-constraints} are imposed. Thus,
it is implied that, in the same hypothesis, the constraints given by
equation~\eqref{eq:energy-momentum-constraints} can be imposed on several
different sets of particles. This circumstance is caused by the fact that the
energy-momentum conservation laws often need to be written at each vertex. Thus,
the sets $\mathfrak{S}^{(j)}_{\pm}$ usually correspond to particles having a
($j\text{-th}$) common vertex. However, in some cases described below, this
statement is not true. Note also that
equation~\eqref{eq:energy-momentum-constraints} is considered as four different
constraints: one constraint is for energy conservation, and remaining three
constraints are for conservation of three-dimensional momentum components. The
discussed kinematic and vertex fitting package allows using all these
constraints simultaneously, as well as only a part of them.

\begin{figure}[tbp]
  \centering
  \begin{tikzpicture}
    \SetTextStyle[TextRotation=0]
    \SetEdgeStyle[TextFont=\large]
    \SetVertexStyle[TextColor=white,TextFont=\large]
    \Vertices{trees/vertices_4p_element.csv}
    \Edges{trees/edges_4p_element.csv}
  \end{tikzpicture}
  \caption{An example of a sub-tree of some energy-momentum conservation
    constraint tree. The vertex $[P^{(j)}_{\alpha}]$ of the sub-tree corresponds
    to a constraint on the conservation of the $\alpha\text{-th}$ component of
    the four-momentum, $\alpha=x,y,z,t$. The edges of the sub-tree correspond to
    the particles on which the constraint $[P^{(j)}_{\alpha}]$ is imposed. The
    solid edge corresponds to a decaying particle, while the dashed edges
    correspond to its decay products. \label{fig:energy-momentum-constraints}}
\end{figure}
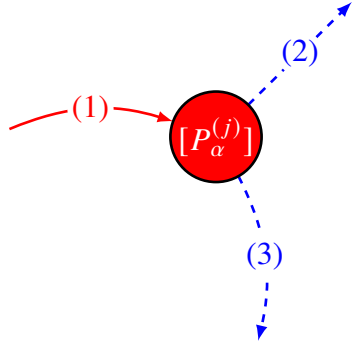
The energy-momentum constraints given by
equation~\eqref{eq:energy-momentum-constraints} are schematically shown in the
figure~\ref{fig:energy-momentum-constraints} when one particle decays into two
other particles. The vertex of the decay tree shown in this figure corresponds
to one of the components ($\alpha=x,y,z,t$) of the four-momentum conservation
constraints. The edges of this tree correspond to the particles on which
the constraints are imposed. The solid edge corresponds to the initial particle,
while the dashed edges correspond to its decay products.
Equation~\eqref{eq:energy-momentum-constraints} in the case of the example shown
in figure~\ref{fig:energy-momentum-constraints} takes the following form:
\begin{equation*}
  -\mathcal{P}^{(1)}_{\text{input}} + \mathcal{P}^{(2)}_{\text{output}} + \mathcal{P}^{(3)}_{\text{output}} = 0.
\end{equation*}

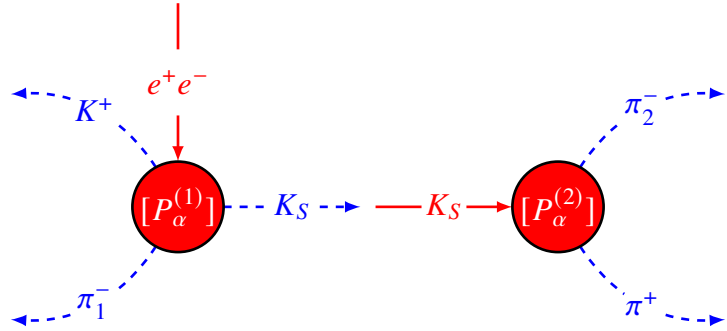
\begin{figure}[tbp]
  \centering
  \begin{tikzpicture}
    \SetTextStyle[TextRotation=0]
    \SetEdgeStyle[TextFont=\large]
    \SetVertexStyle[TextColor=white,TextFont=\large]
    \Vertices{trees/vertices_4p_kskpi.csv}
    \Edges{trees/edges_4p_kskpi.csv}
  \end{tikzpicture}
  \caption{The energy-momentum conservation constraint tree corresponding to the
    hypothesis $e^+e^-\rightarrow K_sK^+\pi^-,\;K_S\rightarrow\pi^+\pi^-$. Label
    $e^+e^-$ in the tree denotes the initial pseudo-particle with the input
    four-momentum
    $\mathcal{P}^{(e^+e^-)}_{\text{input}}=\left(2E_{\text{beam}},\;\bm{0}\right)$.
    The constraint $[P^{(1)}_{\alpha}]$ ensures the conservation of the
    $\alpha\text{-th}$ four-momentum component at the $e^+e^-$ interaction
    vertex, $\alpha=x,y,z,t$. Constraint $[P^{(2)}_{\alpha}]$ ensures the
    conservation of the $\alpha\text{-th}$ four-momentum component at the $K_S$
    meson decay vertex. Since the considered hypothesis contains two $\pi^-$
    mesons, they are denoted as $\pi^-_1$ and $\pi^-_2$. \label{fig:kskpi-tree}}
\end{figure}
Figure~\ref{fig:kskpi-tree} shows an example of the energy-momentum conservation
constraint tree in the case of the $e^+e^-\rightarrow K_SK^+\pi^-,\;K_s\rightarrow\pi^+\pi^-$ hypothesis. Label $e^+e^-$ denotes the
initial pseudo-particle~(see section~\ref{sec:initial-pseudo-particle}) with
the input four-momentum
$\mathcal{P}^{(e^+e^-)}_{\text{input}}=\left(2E_{\text{beam}},\;\bm{0}\right)$.
Constraint $[P^{(1)}_{\alpha}]$ ensures the conservation of the
$\alpha\text{-th}$ four-momentum component at the $e^+e^-$ interaction vertex,
while constraint $[P^{(2)}_{\alpha}]$ does this at the $K_S$ meson decay vertex.
According to the tree shown in figure~\ref{fig:kskpi-tree}, the intermediate
particle $K_S$ is included in the four-momentum conservation constraints. In
particular, this particle is included in the energy conservation constraint,
which leads to the fact that the invariant mass of the decay products of this
particle turns out to be fixed on the $K_S$ meson mass after the fit.

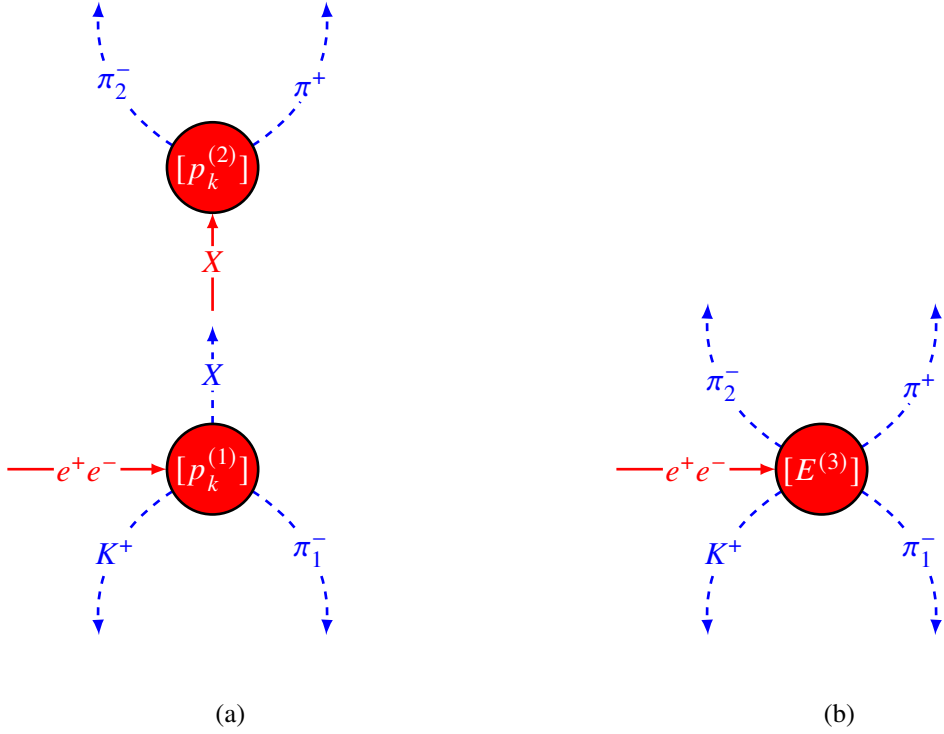
\begin{figure}[tbp]
  \centering
  \begin{subfigure}[t]{0.47\textwidth}
  \begin{tikzpicture}
    \SetTextStyle[TextRotation=0]
    \SetEdgeStyle[TextFont=\large]
    \SetVertexStyle[TextColor=white,TextFont=\large]
    \Vertices{trees/vertices_3p_xkpi.csv}
    \Edges{trees/edges_3p_xkpi.csv}
  \end{tikzpicture}
  \caption{\label{fig:xkpi-momentum-tree}}
  \end{subfigure}
  \hfill
  \begin{subfigure}[t]{0.47\textwidth}
    \begin{tikzpicture}
      \SetTextStyle[TextRotation=0]
      \SetEdgeStyle[TextFont=\large]
      \SetVertexStyle[TextColor=white,TextFont=\large]
      \Vertices{trees/vertices_en_xkpi.csv}
      \Edges{trees/edges_en_xkpi.csv}
    \end{tikzpicture}
    \caption{\label{fig:xkpi-energy-tree}}
  \end{subfigure}
  \caption{The momentum~(a) and energy~(b) conservation constraint trees
    corresponding to
    the hypothesis $e^+e^-\rightarrow X K^+\pi^-,\;X\rightarrow\pi^+\pi^-$,
    where $X$ is unknown intermediate massive particle. Vertices $[p^{(1)}_k]$
    and $[p^{(2)}_k]$ of the first tree represent the three-dimensional
    ($k=x,y,z$) momentum conservation constraints at the $e^+e^-$ interaction
    vertex and at the $X\rightarrow\pi^+\pi^-$ decay vertex, respectively. This tree contains all
    particles, including intermediate ones. Vertex $[E^{(3)}]$ of the second tree
    corresponds to the energy conservation constraint. This tree contains all
    particles except intermediate ones. \label{fig:xkpi-tree}}
\end{figure}
Sometimes it becomes necessary to use a hypothesis in which the invariant mass of
decay products is not fixed after the fit. This result can be achieved if the
momentum conservation constraints are applied to all particles, including
intermediate ones, while the energy conservation constraint is applied only to
the initial and final particles. Examples of energy-momentum conservation
constraint trees for such a hypothesis are shown in figure~\ref{fig:xkpi-tree}.
This hypothesis corresponds to the
$e^+e^-\rightarrow XK^+\pi^-,\;X\rightarrow\pi^+\pi^-$ process, where particle $X$ is
unknown\footnote{Particle $X$ is unknown in the sense that the value of its mass
does not affect the result of the fitting. This property of particle $X$ is determined
not by its parametrization, but by the specific configuration of energy-momentum
conservation constraints. As for the intermediate particle $K_S$,
the parameterizations described in section~\ref{sec:intermediate-neutral-particle}
are used for particle $X$. Thus, the particle $K_S$ can be used as a particle $X$.
The designation $X$ is introduced only to emphasize that the mass of such a particle
does not affect anything.}. Two
trees are shown in the figure~\ref{fig:xkpi-tree}. The tree shown in
Figure~\ref{fig:xkpi-momentum-tree} describes how three  momentum
conservation constraints are applied, while the tree in
figure~\ref{fig:xkpi-energy-tree} describes how the energy conservation
constraint is applied. A example of using this hypothesis can be found in
section~\ref{sec:xkpi}.

\subsection{Mass constraint \label{sec:mass-constraint}}
The mass constraint is used in order to require that the invariant mass of some
set of particles be equal to a certain value. This constraint can be written as
follows:
\begin{equation}
  \label{eq:mass-constraint}
  \left(\sum\limits_{i\in\mathfrak{F}^{(j)}_{+}}\mathcal{P}^{(i)}_{\text{output}}\right)^2 - m^2_{\text{target}} = 0,
\end{equation}
where $\mathfrak{F}^{(j)}_{+}$ is some set of particle indices,
$\mathcal{P}^{(i)}_{\text{output}}$ is the output four-momentum of the
$i\text{-th}$ particle, and $m_{\text{target}}$ is the target value of the
invariant mass.

On the one hand, the introduction of mass constraint~\eqref{eq:mass-constraint}
is redundant, since the energy-momentum conservation
constraints~\ref{sec:energy-momentum-constraints} together with the use of
intermediate particles~\ref{sec:intermediate-particles} make it possible to
achieve the same result. Indeed, suppose that we are dealing with a hypothesis
in which the $\pi^0$ meson decays into two photons. Since the $\pi^0$ meson
lifetime is short, the $\pi^0$ meson decay vertex is close to its origin vertex.
The resolution of the ($\text{CMD-}3$) detector does not allow distinguishing
one vertex from another. Thus, in terms of the kinematic and vertex fitting, the
$\pi^0$ meson decays at the origin vertex. In this case, the use of vertex
constraints~\ref{sec:vertex-constraints} for the intermediate $\pi^0$ meson does
not make sense. The parameter $\xi^{(i)}$ of the intermediate $\pi^0$ meson will
not be used in this case and can be fixed (see
equation~\eqref{eq:int-neutral-trajectory}). However, eight energy-momentum
conservation constraints can be imposed on the intermediate $\pi^0$ meson in
this case. Four constraints correspond to the laws of energy-momentum
conservation at the origin vertex, and four others correspond to such laws at the
decay vertex.

On the other hand, the use of four-momentum parametrizations from
section~\eqref{sec:intermediate-particles} in the case of short-lived
intermediate particles is not justified because it leads to a needless increase
in the number of minimization parameters. In addition to the parameters of the
intermediate particle, eight more Lagrange multipliers will be involved in the
minimization procedure in this case. If, however, a short-lived intermediate
particle is not introduced directly and the mass
constraint~\eqref{eq:mass-constraint} is used, then only one additional Lagrange
multiplier corresponding to this constraint will be involved in the minimization
procedure. Thus, the conclusion is that in the case of short-lived intermediate
particles, it is better to use the mass constraint~\eqref{eq:mass-constraint} if
necessary.

\subsection{Vertex constraints \label{sec:vertex-constraints}}
Vertex constraints are used to require particle trajectories to pass through
origin and decay vertices. The vertex constraints that require a particle to fly
out of its origin vertex have the following form:
\begin{equation}
  \bm{r^{(i)}_{\text{output}}} - \bm{r_{\text{origin}}} = \bm{0},
\end{equation}
where $\bm{r^{(i)}_{\text{output}}}$ is the parametrization of the output
trajectory of the $i\text{-th}$ particle, i.e. the trajectory corresponding to
the escape of $i\text{-th}$ particle from its origin vertex
$\bm{r_{\text{origin}}}$. The vertex constraints, which require a particle to
fly to its decay vertex, can be written as follows:
\begin{equation}
  \bm{r^{(i)}_{\text{input}}} - \bm{r_{\text{decay}}} = \bm{0},
\end{equation}
where $\bm{r^{(i)}_{\text{input}}}$ is the parametrization of the input
trajectory of the $i\text{-th}$ particle, i.e. the trajectory along which the
particle flies into its decay vertex $\bm{r_{\text{decay}}}$.

Among all particle kinds discussed in section~\ref{sec:particles}, there are
particle kinds for which trajectories are not defined. Such particle kinds are
the initial pseudo particle from section~\ref{sec:initial-pseudo-particle}, the
lost particles from section~\ref{sec:lost-particles}, and the photon from
section~\ref{sec:photon}. However, it should be noted that the photon flies from
its origin vertex to its conversion point according to the parametrization of
its momentum~\eqref{eq:photon-four-momentum}. For the final charged
particle~\ref{sec:charged-particle}, only the output
trajectory~\eqref{eq:charged-particle-trajectory} is given, while for the
intermediate charged particle~\ref{sec:intermediate-charged-particle}, both the
output~\eqref{eq:charged-particle-trajectory} and
input~\eqref{eq:input-charged-particle-trajectory} trajectories are given. In
the case of the discussed package of kinematic and vertex fitting, only the
input trajectory~\eqref{eq:int-neutral-trajectory} is specified for the
neutral intermediate particle~\ref{sec:intermediate-neutral-particle}.

\section{Examples \label{eq:mc-examples}}
This section provides examples of applying the kinematic and vertex fitting
package in various hypotheses to the
simulated\footnote{By simulated events in this section, we mean the events of
Monte Carlo simulation, which takes into account the response of the
$\text{CMD-}3$ detector.} events of various $e^+e^-$
annihilation processes. In all examples considered in this section,
with the exception of the example described in subsection~\ref{sec:xkpi-noisir},
simulations are done taking into account initial state radiation~(ISR).
Since the kinematic hypotheses considered in the article do not contain ISR
photons, they are not strict signal hypotheses if they are applied to events
with ISR photons. As a result, some of the distributions given in this section
are distorted. However, the distortion of chi-square distributions is not fully
explained by neglecting ISR photons in kinematic hypotheses. Verification of
this statement is given in subsection~\ref{sec:xkpi-noisir}.

Among the examples below, hypotheses with intermediate long-lived neutral
particles ($K_S$ mesons), hypotheses containing charged particles only,
hypotheses containing photons only, hypotheses containing both charged particles
and photons, as well as hypotheses with lost particles are
discussed in detail. The following subsections present mainly the results of
applying the fitting algorithm under signal hypotheses. However, in some cases,
examples of the fitting under background hypotheses are also given.

In some cases, the kinematic and vertex fitting procedure includes several fits
per event. This is due to the particle combinatorics. For example, it may turn
out that there are several ways to match tracks from the drift chamber with charged
particles of some hypothesis. Further, the set of all fits in an event is
called the fitting procedure. If at least one of the fits in an event converges to
a local minimum, then the corresponding fitting procedure is said to have converged
to a local minimum. If none of the fits converge, then the fitting procedure is said
to have failed. If among all convergent fits there are only those that have
converged to a local maximum, then the fitting procedure is said to have converged to
a local maximum. The figures given in this section correspond only to those events in
which the kinematic and vertex fitting procedure converged to a local minimum.

\subsection{Hypotheses $e^+e^-\rightarrow{X}K^{\pm}\pi^{\mp}$, $X\rightarrow\pi^+\pi^-$ \label{sec:xkpi}}
\subsubsection{Description of the $e^+e^-\rightarrow{X}K^{\pm}\pi^{\mp}$, $X\rightarrow\pi^+\pi^-$ hypotheses \label{sec:xkpi-desc}}
Section~\ref{sec:xkpi} presents the results of applying
the package of kinematic and
vertex fitting under two charged conjugate hypotheses
$e^+e^-\rightarrow XK^+\pi^-$ and $e^+e^-\rightarrow X K^-\pi^+$,
where $X$ is an unknown
intermediate neutral particle decaying into two charged pions:
$X\rightarrow\pi^+\pi^-$. These hypotheses were developed in order to select the
events of the $e^+e^-\rightarrow K_S K^{\pm}\pi^{\mp},\;K_S\rightarrow\pi^+\pi^-$
process. The particle $X$ is called unknown because the energy conservation
constraints are not applied at the $e^+e^-$ interaction vertex and the decay
vertex. The energy conservation constraint is applied only to final and
initial particles
in these hypotheses. Therefore, the mass of the particle $X$ is not
contained in the Lagrange function~\eqref{eq:lagrange-function} and, as a consequence,
does not participate in the
constrained minimization of the chi-square
function~\eqref{eq:chi-square-1}. For this reason, the invariant mass of the
$X\rightarrow\pi^+\pi^-$ decay products, calculated using parameters obtained from the fitting
is not fixed (e.g. on the $K_S$ mass). For more details, see
sections~\ref{sec:intermediate-neutral-particle},
\ref{sec:energy-momentum-constraints}
and figure~\ref{fig:xkpi-tree}.

The hypotheses under discussion have the following
structure.
\begin{itemize}
  \item Each hypothesis has two vertices. The first vertex is the $e^+e^-$
    interaction vertex, the second vertex is the decay vertex of the particle
    $X$. All coordinates of the $X\rightarrow\pi^+\pi^-$ decay vertex are free non-measurable parameters.
    All coordinates of the $e^+e^-$ interaction
    vertex are considered as free measurable parameters and contribute to the
    chi-square.
  \item The hypothesis $e^+e^-\rightarrow X K^{+}\pi^{-},\;X\rightarrow\pi^+\pi^-$
    requires the presence of four final particles
    and one intermediate particle $X$. The
    intermediate particle $X$ originates at the $e^+e^-$ interaction vertex and
    decays at its decay vertex into two final particles $\pi^+$ and $\pi^-$. Another
    final $\pi^-$ comes from the $e^+e^-$ interaction vertex. The final $K^+$
    also originates at this vertex. The charge conjugate hypothesis
    $e^+e^-\rightarrow X K^{-}\pi^{+}$ differs from the
    $e^+e^-\rightarrow X K^{+}\pi^{-}$ hypothesis in that $K^-$ and $\pi^+$ propagate
    from the $e^+e^-$ interaction region instead of $K^+$ and $\pi^-$. Both
    hypotheses also contain the initial pseudo-particle that provides the
    $e^+e^-$ four-momentum~(see section~\ref{sec:initial-pseudo-particle}).
  \item At each vertex, three-dimensional momentum conservation constraints are
    imposed. In total, each of two hypotheses has $6$ constraints on the
    conservation of the three-dimensional momentum components, i.e. three
    constraints per each vertex.
  \item Each hypothesis has only one energy conservation constraint. Only the
    final particles and the initial pseudo-particle are involved in this
    constraint.
  \item Each hypothesis requires three vertex constraints for each final
    particle (one constraint for each trajectory component). Three
    vertex constraints are also required for the intermediate particle $X$. See,
    sections~\ref{sec:charged-particle}, \ref{sec:intermediate-neutral-particle}
    and~\ref{sec:vertex-constraints} for more details. In total, each of two
    hypotheses contains $15$ vertex constraints.
\end{itemize}
In total, each of two discussed hypotheses contains $22$ constraints. Each of the
hypotheses contains $34$ free parameters: $23$ measurable and $11$ non-measurable
parameters.

The similar hypothesis with $K_S$ instead of $X$ is discussed in section~\ref{sec:kskpi}.

\subsubsection{Fitting procedure details \label{sec:xkpi-fitting-details}}
\begin{table}[tbp]
  \centering
  \caption{Mappings between charged particles and their tracks in
    the case of the $e^+e^-\rightarrow X K^{\pm}\pi^{\mp}$ hypotheses.
    $1$ --- the $e^+e^-$ interaction vertex,
    $2$ --- the $X\rightarrow\pi^+\pi^-$ decay vertex,
    $t^+_1$ and $t^+_2$ are tracks of positively charged
    particles,
    $t^-_1$ and $t^-_2$ are tracks of negatively charged
    particles.\label{tab:xkpi-track-mapping}}
  \begin{tabular}[t]{V{4}cV{3}c|c|c|cV{3}c|c|c|cV{4}}
    \hlineB{4}
    \multicolumn{1}{V{4}lV{3}}{Hypothesis} &
    \multicolumn{4}{cV{3}}{$e^+e^- \rightarrow X K^+\pi^-$} &
    \multicolumn{4}{cV{4}}{$e^+e^-\rightarrow X K^-\pi^+$} \\
    \hline
    \multicolumn{1}{V{4}lV{3}}{Origin vertex} &
    \multicolumn{2}{c|}{$1$} &
    \multicolumn{2}{cV{3}}{$2$} &
    \multicolumn{2}{c|}{$1$} &
    \multicolumn{2}{cV{4}}{$2$} \\
    \hline
    \multicolumn{1}{V{4}lV{3}}{Particle} &
    $K^+$ & $\pi^-$ &
    $\pi^+$ & $\pi^-$ &
    $K^-$ & $\pi^+$ &
    $\pi^+$ & $\pi^-$ \\
    \hlineB{3}

    \multirow{4}{*}{\begin{minipage}{2cm} Track combinations\end{minipage}} &
    $t^+_1$ & $t^-_1$ &
    $t^+_2$ & $t^-_2$ &

    $t^-_1$ & $t^+_1$ &
    $t^+_2$ & $t^-_2$\\

    &
    $t^+_1$ & $t^-_2$ &
    $t^+_2$ & $t^-_1$ &

    $t^-_1$ & $t^+_2$ &
    $t^+_1$ & $t^-_2$\\

    &
    $t^+_2$ & $t^-_1$ &
    $t^+_1$ & $t^-_2$ &

    $t^-_2$ & $t^+_1$ &
    $t^+_2$ & $t^-_1$\\

    &
    $t^+_2$ & $t^-_2$ &
    $t^+_1$ & $t^-_1$ &

    $t^-_2$ & $t^+_2$ &
    $t^+_1$ & $t^-_1$\\
    \hlineB{4}
  \end{tabular}
\end{table}
Further, the considered hypotheses are applied to the simulated events of the
signal process $e^+e^-\rightarrow K_S K^{\pm}\pi^{\mp}$ and the background
process $e^+e^-\rightarrow\pi^+\pi^-\pi^+\pi^-$. Only four-track events are used
in order to demonstrate how the fitting works under the discussed hypotheses.
Moreover, only such four tracks are considered, which correspond to the total electric
charge equal to zero. At the same time, for demonstration purposes,
the pre-separation of tracks into
kaon and pion ones is not performed. The presence of four charged particles
implies the presence of combinatorics associated with the fact that it is not
known in advance which track from a certain event corresponds to one or another
charged particle. Therefore, it is necessary to perform a separate fit for each
mapping of particles to their tracks. All possible variants of such mappings are
given in table~\ref{tab:xkpi-track-mapping}. The table shows for each hypothesis
there are four possible mappings of charged particles into tracks. Since it is not
known in advance which of the signal processes $e^+e^-\rightarrow K_S K^{+}\pi^{-}$
or $e^+e^-\rightarrow K_S K^{-}\pi^{+}$ took place in a particular event, in each
event the fit is performed in both hypotheses
$e^+e^-\rightarrow X K^{\pm}\pi^{\mp}$. Thus, in each event, the fit is performed
eight times.

In order to demonstrate how the fitting works, for each hypothesis, among the
four mappings of particles to their tracks, the one in which the fit gives the
smallest chi-square is selected. Fits in the charge conjugate hypotheses
$e^+e^-\rightarrow X K^{+}\pi^{-}$ and $e^+e^-\rightarrow X K^{-}\pi^{+}$ give
the same chi-square distribution when applied to
signal\footnote{The signal events for the $e^+e^-\rightarrow X K^{+}\pi^{-}$
hypothesis are the events of the $e^+e^-\rightarrow K_S K^{+}\pi^{-}$ process,
while the signal events for the $e^+e^-\rightarrow X K^{-}\pi^{+}$ hypothesis are
the events of the $e^+e^-\rightarrow K_S K^{-}\pi^{+}$ process.} events.
Therefore, among these hypotheses, the
hypothesis corresponding to the minimum chi-square is
selected\footnote{Thus, among the
results of eight fits in each event, the result that corresponds to the minimum
chi-square is accepted.}. It should be noted that only those fits that have
converged to a local minimum are accepted for consideration. If in the event
none of the fits converged to a local minimum, then such an event is ignored.

\subsubsection{Fitting the
  $e^+e^-\rightarrow K_S K^{\pm}\pi^{\mp}$
  events \label{sec:kskpi-xkpi}}
\begin{figure}[tbp]
  \centering
  \begin{subfigure}[t]{0.47\textwidth}
    \centering
    \includegraphics[width=\textwidth]{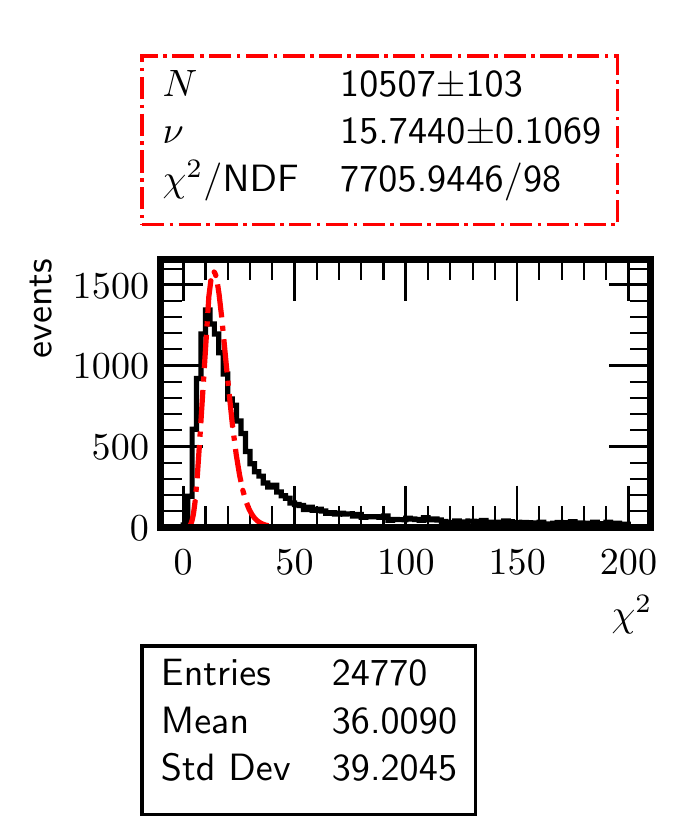}
    \caption{\label{fig:kf-chi2-xkpi}}
  \end{subfigure}
  \hspace{1em}
  \begin{subfigure}[t]{0.47\textwidth}
    \centering
    \includegraphics[width=\textwidth]{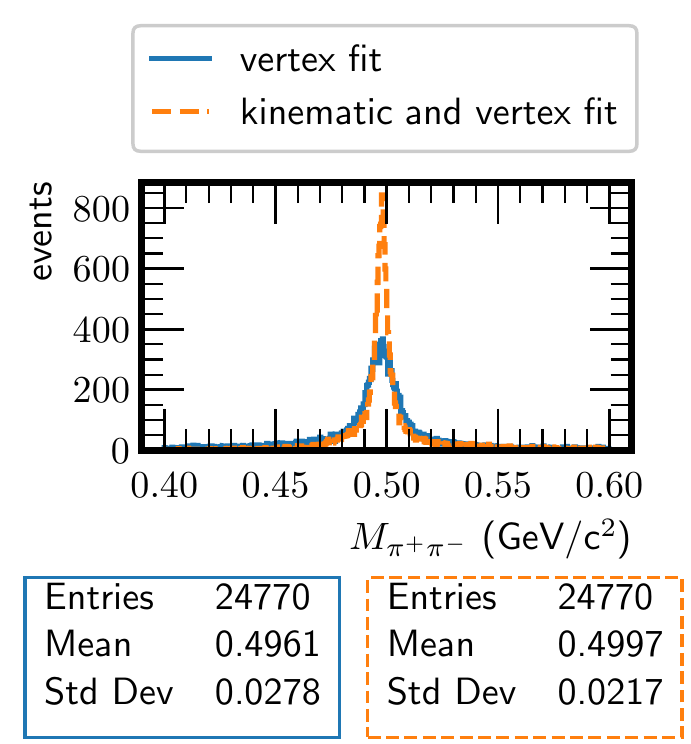}
    \caption{\label{fig:inv-mass-xkpi}}
  \end{subfigure}
 \caption{Chi-square distribution (a) and invariant mass distributions
   of the $X\rightarrow\pi^+\pi^-$ decay products (b). All distributions
   were obtained for the  events of the $e^+e^-\rightarrow K_S K^{\pm}\pi^{\mp}$
   simulation. The chi-square distribution is shown as black solid
   line and  corresponds to the fitting procedure
   described in section~\ref{sec:xkpi-fitting-details}. The red
   dash-dotted line shown in figure~(a) is the result of fitting
   the chi-square distribution with the chi-squared probability density
   function given by equation~\eqref{eq:chi-square-pdf}. The fitting
   function depends on two parameters. One of these parameters is the
   normalization factor~$N$, the other is the number of degrees of
   freedom~$\nu$. Both parameters were free during fitting.
   The dashed line in
   figure~(b) corresponds to the invariant mass distribution obtained using
   the parameters found with the fitting procedure described in
   section~\ref{sec:xkpi-fitting-details}. The solid line
   corresponds to the invariant mass distribution obtained using the vertex fit
   only. This vertex fit is applied to the pair of pions from the
   $X\rightarrow\pi^+\pi^-$ decay. The tracks corresponding to these pions are
   considered to be known from the fitting procedure described in
   section~\ref{sec:xkpi-fitting-details}. \label{fig:chi2-and-mass-xkpi}}
\end{figure}
This subsection presents the results of the fitting the
simulated $e^+e^-\rightarrow K_S K^{\pm}\pi^{\mp}$ events under the
$e^+e^-\rightarrow X K^{\pm}\pi^{\mp},\;X\rightarrow\pi^+\pi^-$ hypotheses.
The $e^+e^-\rightarrow K_S K^{\pm}\pi^{\mp}$ events were simulated at the
center-of-mass energy of $1792.9\text{ MeV}$ and the magnetic field of
$1\text{ T}$. The fitting procedure converged to a local minimum in
$24770$ events out of $25224$. In $147$ events, the fitting procedure
converged to a local maximum. In the rest $307$ events, this procedure
did not converge.

Figure~\ref{fig:kf-chi2-xkpi} shows the chi-square distribution~(black solid
line) for events
where the fitting procedure described in section~\ref{sec:xkpi-fitting-details}
has converged to a
local minimum. If the influence of the constraint nonlinearity is small,
then according to sections~\ref{sec:case-of-nonzero-l} and~\ref{sec:xkpi-desc},
we expect the number of degrees of freedom to be $m-l=22-11=11$. Chi-square
distribution that corresponds to the chi-squared probability density function given
by equation~\eqref{eq:chi-square-pdf} must have a mean value equal to the number
of degrees of freedom. However, the mean value of the chi-square distribution
shown in figure~\ref{fig:kf-chi2-xkpi} is significantly larger than the expected
number of degrees of freedom. The reason is that the chi-square distribution
shown in figure~\ref{fig:kf-chi2-xkpi} is distorted due to the non-Gaussian
response of the detector and the imperfect calibration of the covariance matrix
$\hat{\tilde{C}}$. Partially, the distortion of this distribution is caused by
the nonlinearity of the constraints, as well as neglecting ISR photons in
the kinematic hypotheses.

Although the mean value of the chi-square
distribution shown in figure~\ref{fig:kf-chi2-xkpi} is significantly larger than
expected one, the peak position of this distribution is quite close to the expected
number of degrees of freedom. This is easy to see by comparing the chi-square
distribution~(black solid line) with the red dash-dotted line. The
dash-dotted line is obtained by fitting the chi-square distribution with the
chi-square probability density function given by
equation~\eqref{eq:chi-square-pdf}. It can be seen from the figure that the
fitting curve does not describe the chi-square distribution well. The fitting
curve reaches its maximum at point $\chi^2=\nu\approx{15.7}$. The maximum of the
chi-square distribution~(black solid line) is located slightly to the left of
this point.

Effects related with the non-Gaussian
response of the detector, the imperfect calibration of the covariance matrix
$\hat{\tilde{C}}$, and constraint nonlinearity are discussed in section~\ref{sec:gaussian-simulation-examples}.
This discussion is based on the examples of the Gaussian simulation. The
technique of the Gaussian simulation is introduced in
section~\ref{sec:gaussian-simulation}.
This technique eliminates the influence of the detector response and the
calibration of the covariance matrix on the shape of the chi-square
distribution. The effects associated with the absence of ISR photons
in kinematic hypotheses are also excluded in the Gaussian simulation.
Therefore, in cases where the nonlinearity of the constraints can
be neglected, the chi-square distribution obtained using the Gaussian simulation
is well described by the chi-squared probability density function. In particular,
section~\ref{sec:kskpi-gsim} shows
that in the case of a Gaussian response, the number of degrees of freedom of the
chi-square distribution corresponding to this example is indeed $11$.

Figure~\ref{fig:inv-mass-xkpi} shows the invariant mass distributions of the
two pions from the $X\rightarrow\pi^+\pi^-$ decay. The dashed line corresponds
to the invariant mass distribution obtained using the particle parameters found
with the kinematic and vertex fitting procedure described in
section~\ref{sec:xkpi-fitting-details}. The solid line
corresponds to the invariant mass distribution obtained using the vertex fit
only. This vertex fit is applied to the pair of the pions from the
$X\rightarrow\pi^+\pi^-$ decay. The tracks corresponding to these pions are considered
to be known from the kinematic and vertex fitting procedure. It can be
seen from the figure that kinematic fitting significantly improves the
resolution of the invariant mass. The mean values of the invariant mass for the
both distributions are close to the $K_S$ meson invariant mass.

\begin{figure}[tbp]
  \centering
  \begin{subfigure}[t]{0.47\textwidth}
    \centering
    \includegraphics[width=\textwidth]{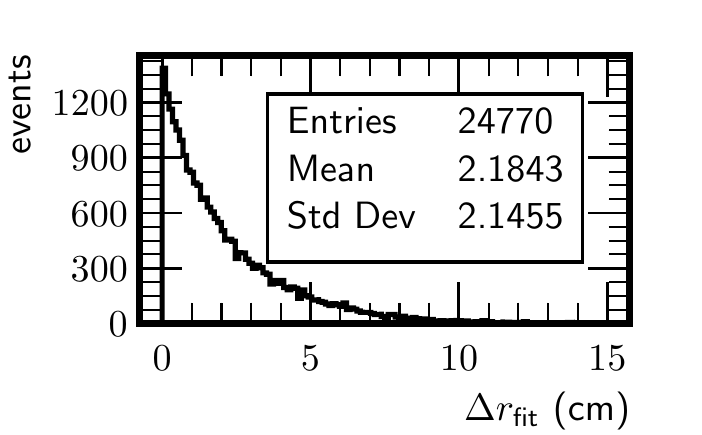}
    \caption{\label{fig:dr-fit-xkpi}}
  \end{subfigure}
  \hspace{1em}
  \begin{subfigure}[t]{0.47\textwidth}
    \centering
    \includegraphics[width=\textwidth]{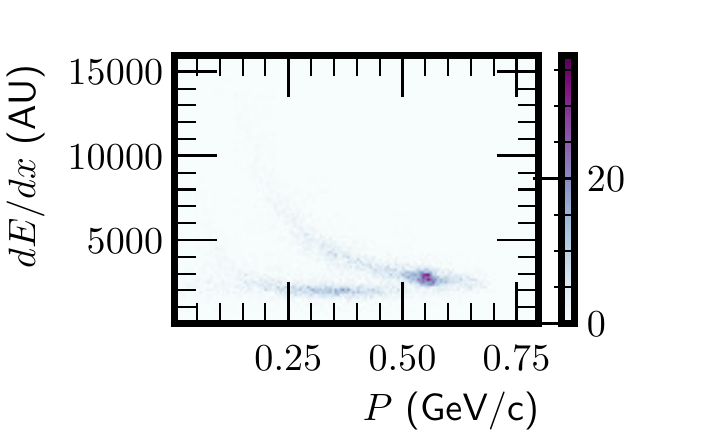}
    \caption{\label{fig:dedx-kaon-v1-xkpi}}
  \end{subfigure}
  \begin{subfigure}[t]{0.47\textwidth}
    \centering
    \includegraphics[width=\textwidth]{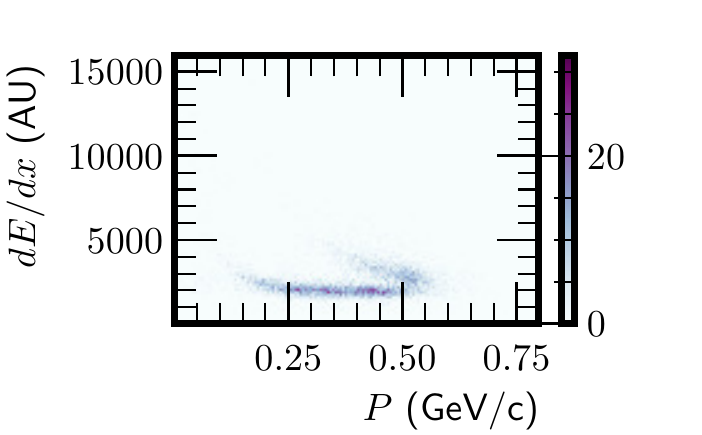}
    \caption{\label{fig:dedx-pion-v1-xkpi}}
  \end{subfigure}
  \hspace{1em}
  \begin{subfigure}[t]{0.47\textwidth}
    \centering
    \includegraphics[width=\textwidth]{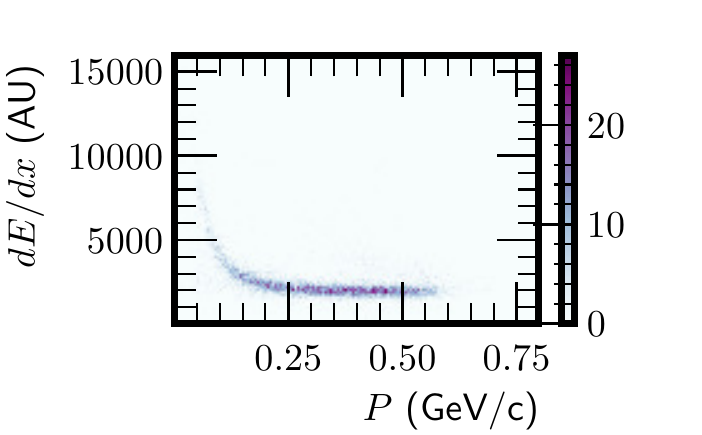}
    \caption{\label{fig:dedx-pion-v2-xkpi}}
  \end{subfigure}
  \caption{Some distributions obtained as a result of applying the fitting
    procedure described in section~\ref{sec:xkpi-fitting-details} to the simulated
    $e^+e^-\rightarrow K_SK^{\pm}\pi^{\mp}$ events. Distribution of the distance
    between the $e^+e^-$ interaction vertex and the $X\rightarrow\pi^+\pi^-$ decay
    vertex are shown in figure~(a). Other figures show dependencies of $dE / dx$ on
    the momentum of particles identified by the fitting procedure as (b)~kaons,
    (c)~pions from the $e^+e^-$ interaction vertex, (d)~pions from the
    $X\rightarrow\pi^+\pi^-$ decay vertex. The units of $dE / dx$ in figures (b), (c)
    and (d) are arbitrary.}
\end{figure}
Figure~\ref{fig:dr-fit-xkpi} shows the distribution of the distance between the
$e^+e^-$ interaction vertex and the $X\rightarrow\pi^+\pi^-$ decay vertex. The
coordinates of both vertices are found using the kinematic and vertex
fitting procedure. The shape of the considered distribution, as expected,
is close to exponential. However, this distribution is not exactly exponential,
since $K_s$-mesons have a non-trivial energy spectrum.

The kinematic fitting technique makes it possible to separate charged kaons from
pions, since the masses of these particles differ significantly. However, this
technique does not allow achieving the desired level of the $K\text{-}\pi$
separation\footnote{At least when using the considered hypotheses.}.
For example, in the case of fitting the
$e^+e^-\rightarrow K_SK^{\pm}\pi^{\mp}$ events under the $e^+e^-\rightarrow X K^{\pm}\pi^{\mp}$
hypotheses, kaons and pions can be misassigned at the $e^+e^-$ interaction vertex.
These misassignments can be seen in figures~\ref{fig:dedx-kaon-v1-xkpi}
and~\ref{fig:dedx-pion-v1-xkpi}. The figures show the dependencies of $dE/dx$ on
the particle momentum for those particles that the fitting procedure identified
as kaons and pions. Figure~\ref{fig:dedx-kaon-v1-xkpi} corresponds to particles
that have been identified as kaons, while figure~\ref{fig:dedx-pion-v1-xkpi}
corresponds to pions. In each of the figures, one can see typical dependencies
corresponding to both particle types (true kaons and pions).
Figure~\ref{fig:dedx-pion-v2-xkpi} shows the dependence of $dE/dx$ on the
particle momentum for those particles that were identified as pions from the
$X\rightarrow\pi^+\pi^-$ decay. In this figure, one can see only a typical
dependence corresponding to true pions. Thus, we can conclude that in this case
there is a significant misassignment of kaons and pions at the $e^+e^-$
interaction vertex, while pions from the $X\rightarrow\pi^+\pi^-$ decay are not
misassigned as kaons.

Since it is known from the simulation which of processes $e^+e^-\rightarrow K_SK^+\pi^-$
or $e^+e^-\rightarrow K_SK^-\pi^+$ took place in each event, it is possible to indicate
the proportion of events in which the fitting procedure incorrectly found the signal
hypothesis. This proportion is approximately equal to~$34.6\%$.

\begin{figure}[tbp]
  \centering
  \begin{subfigure}[t]{0.47\textwidth}
    \centering
    \includegraphics[width=\textwidth]{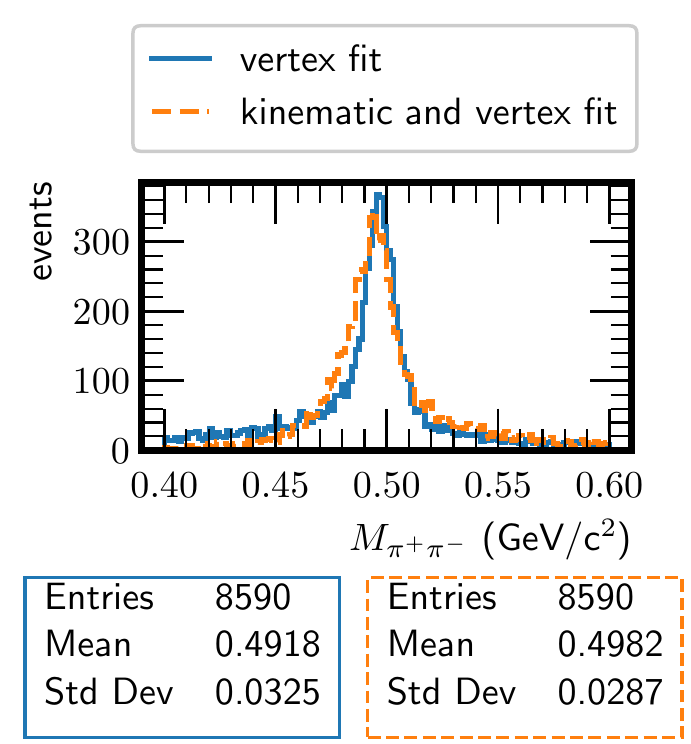}
    \caption{\label{fig:mks_wrong_xkpi}}
  \end{subfigure}
  \hspace{1em}
  \begin{subfigure}[t]{0.47\textwidth}
    \centering
    \includegraphics[width=\textwidth]{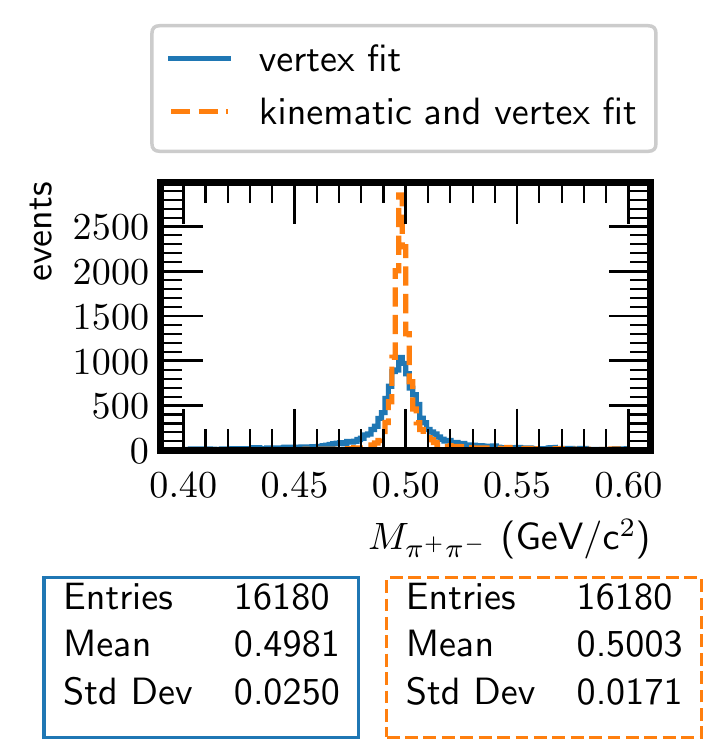}
    \caption{\label{fig:mks_good_xkpi}}
  \end{subfigure}
  \caption{Invariant mass distributions of the $X\rightarrow\pi^+\pi^-$ decay
    products. The distributions shown in the figure
    were obtained for the simulated events of the
    $e^+e^-\rightarrow K_S K^{\pm}\pi^{\mp}$ process. The meaning of the solid
    and dashed lines is the same as in figure~\ref{fig:inv-mass-xkpi}.
    Figure~(a) corresponds to events in which the fitting
    procedure converged to an incorrect signal
    hypothesis, while figure~(b) corresponds to events in which the procedure
    converged to the correct signal hypothesis.}
\end{figure}
It should be noted that the misassignment of kaons and pions has a dramatic
effect on the results of the fitting procedure described in
section~\ref{sec:xkpi-fitting-details}.
This misassignment leads to a significant distortion of the fitting
parameters. In the considered case, not only the parameters of particles flying from the
$e^+e^-$ interaction vertex are distorted, but also the parameters of pions
from the $X\rightarrow\pi^+\pi^-$ decay vertex. This happens because the
parameters of different particles are related through the constraints.
Figures~\ref{fig:mks_wrong_xkpi} and~\ref{fig:mks_good_xkpi} show the
distributions of the invariant masses of pions from the
$X\rightarrow\pi^+\pi^-$ decay. Both figures were obtained using the
$e^+e^-\rightarrow K_SK^{\pm}\pi^{\mp}$ simulation events. However,
figure~\ref{fig:mks_wrong_xkpi}
was obtained for events in which the fitting procedure chose the wrong signal
hypothesis, and figure~\ref{fig:mks_good_xkpi} corresponds to those events
in which the signal hypothesis was chosen correctly. It can be seen from
these figures that the resolution of the invariant mass is significantly
worse if the signal hypothesis was chosen incorrectly.

Thus, one can conclude that an additional $K\text{-}\pi$ separation procedure is
required to avoid the misassignment of kaons and pions. This procedure may be
based on the use of $dE / dx$, for example. The $K\text{-}\pi$ separation
procedure must be used before the kinematic and vertex fitting. After performing
such a procedure, it will be known which tracks correspond to kaons with the
highest probability. Thus, in each event, one know the mapping of the kaon to
its track. The sign of the kaon charge determines the signal hypothesis that
will be used for the fitting in a certain event. As a result, the number
of fits in a
single event can be significantly reduced. Instead of mappings from
table~\ref{tab:xkpi-track-mapping}, it is enough to
consider only permutations of tracks for the pions of the same charge, i.e.
there will be only two fits per event. Among these two fits, as in the case of
fitting procedure described in section~\ref{sec:xkpi-fitting-details}, one can
choose the fit with the smallest chi-square.

\begin{figure}[tbp]
  \centering
  \begin{subfigure}[t]{0.47\textwidth}
    \centering
    \includegraphics[width=\textwidth]{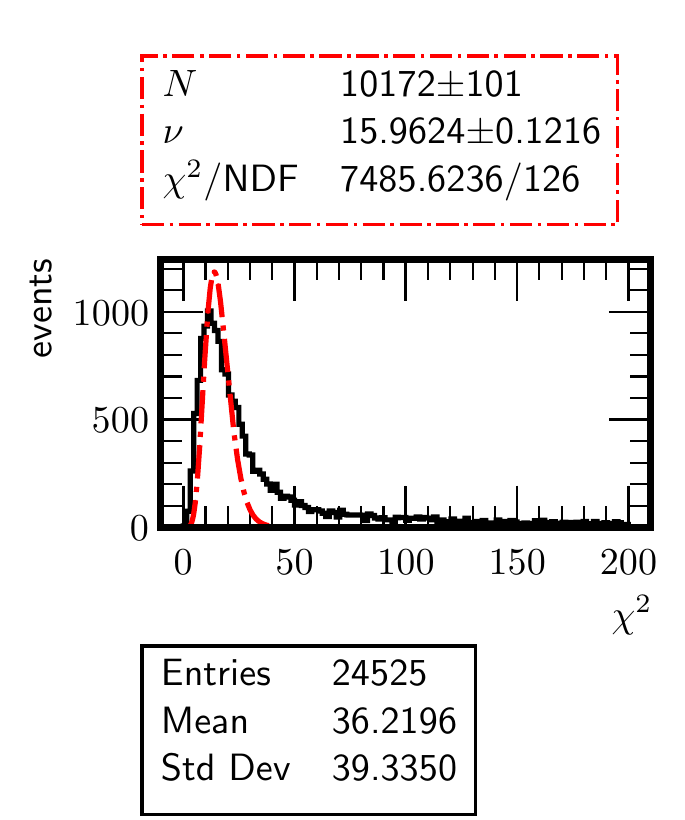}
    \caption{\label{fig:kf_chi2_simhypo_xkpi}}
  \end{subfigure}
  \begin{subfigure}[t]{0.47\textwidth}
    \centering
    \includegraphics[width=\textwidth]{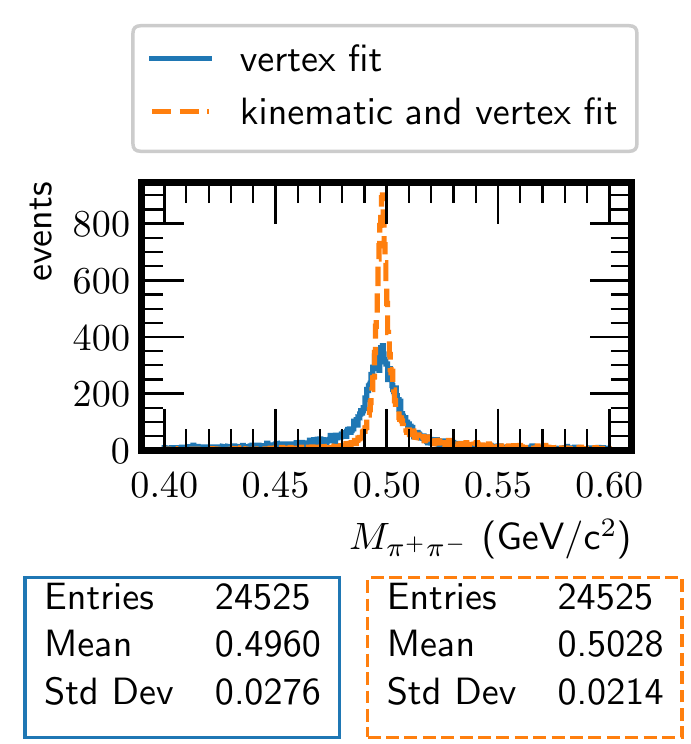}
    \caption{\label{fig:inv-mass-simhypo-xkpi}}
  \end{subfigure}
  \caption{Chi-square distribution (a) and invariant mass distributions
    of the $X\rightarrow\pi^+\pi^-$ decay products (b).
    This figure is similar to figure~\ref{fig:chi2-and-mass-xkpi}, but
    the fitting procedure is performed under the correct signal
    hypothesis only.
    In each event of the $e^+e^-\rightarrow K_SK^{\pm}\pi^{\mp}$ simulation,
    it is known which of the two signal hypotheses ($e^+e^-\rightarrow{X}K^+\pi^-$
    or $e^+e^-\rightarrow{X}K^-\pi^+$) is the correct one.\label{fig:sim-hypo-xkpi}}
\end{figure}
In this work, we do not use any additional $K\text{-}\pi$ separation procedure,
but for demonstration purposes we can use the fact that in each event of the simulation
we know which of the final states $K_S K^+\pi^-$ or $K_S K^-\pi^+$ occurred. Thus,
it is possible to run fitting procedure for each event of the
$e^+e^-\rightarrow K_S K^{\pm}\pi^{\mp}$ simulation under the right signal
hypothesis. For a given signal hypothesis, it is necessary to consider only four
mappings from table~\ref{tab:xkpi-track-mapping}, i.e. do four fits in each event.
Using the correct signal hypothesis in each event eliminates the misassignment of
kaons and pions at the $e^+e^-$ interaction vertex. Taking into account that there
is also no misassignment of kaons and pions from the $X\rightarrow\pi^+\pi^-$
decay (see figure~\ref{fig:dedx-pion-v2-xkpi}), it
can be obtained that with this approach there is no misassignment of kaons and
pions. The fitting procedure in this case converges to a local minimum in
$24525$ events out of $25224$, i.e. in most events where the wrong hypothesis was
previously chosen, there was a fit under the correct hypothesis with a slightly
larger chi-square.

Some of the results of using the fitting procedure described in the previous paragraph
are shown in figure~\ref{fig:sim-hypo-xkpi}. Figure~\ref{fig:kf_chi2_simhypo_xkpi} shows
the chi-square distribution. Figure~\ref{fig:inv-mass-simhypo-xkpi} shows the distributions
of the invariant mass of pions from the $X\rightarrow\pi^+\pi^-$ decay. Unlike
figure~\ref{fig:inv-mass-xkpi}, figure~\ref{fig:inv-mass-simhypo-xkpi} was
obtained by fitting each simulation event using the correct signal hypothesis
only. It can be seen from figure~\ref{fig:inv-mass-simhypo-xkpi} that the
distribution of the invariant mass obtained using the pion parameters after
fitting has somewhat changed compared to the similar distribution shown in
figure~\ref{fig:inv-mass-xkpi} (see dashed line). The left slope of the peak
became steeper, and the peak itself is slightly higher, which indicates a slight
improvement in mass resolution. However, the distribution in
figure~\ref{fig:inv-mass-simhypo-xkpi} has non-Gaussian tails, as in the case of
a similar distribution in figure~\ref{fig:inv-mass-xkpi}, which leads to the fact
that the standard deviations for the histograms in both figures are approximately
the same.

\begin{figure}[tbp]
  \centering
  \begin{subfigure}[t]{0.47\textwidth}
    \centering
    \includegraphics[width=\textwidth]{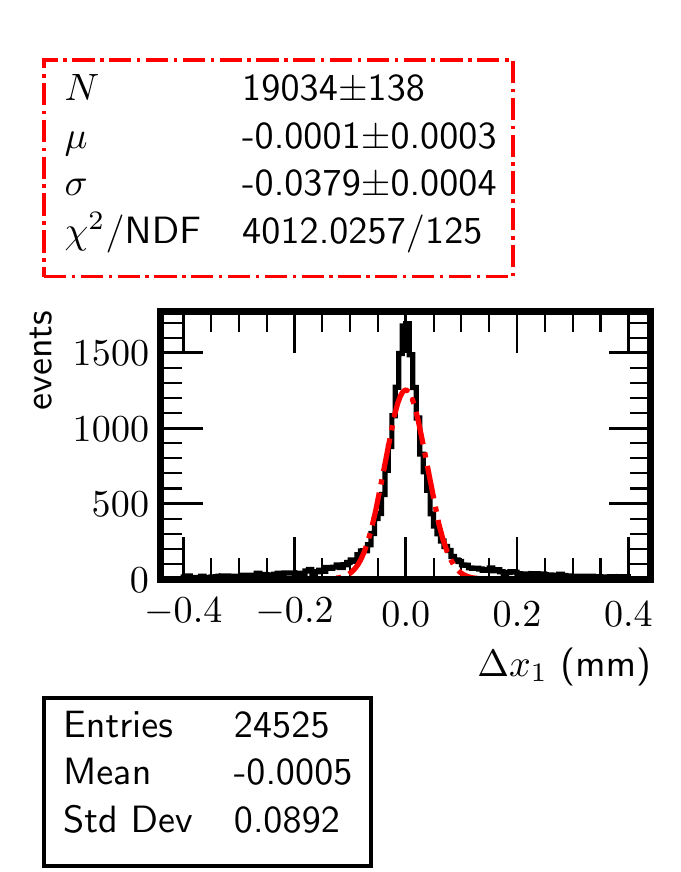}
    \caption{\label{fig:kf_vtx_dx1_simhypo_xkpi}}
  \end{subfigure}
  \hspace{1em}
  \begin{subfigure}[t]{0.47\textwidth}
    \centering
    \includegraphics[width=\textwidth]{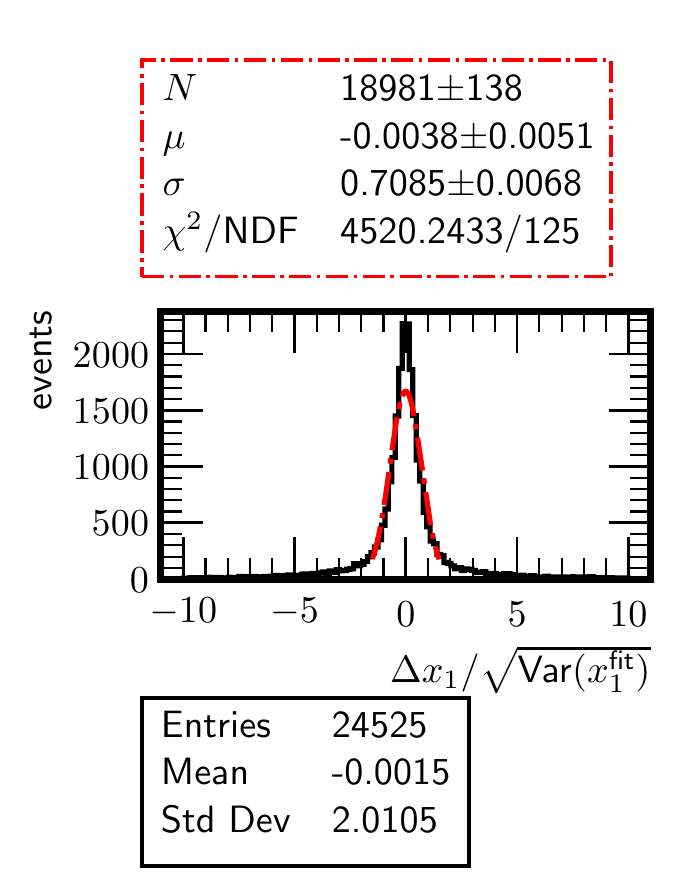}
    \caption{\label{fig:pull_x1_simhypo_xkpi}}
  \end{subfigure}
  \caption{Comparison of the $x$-coordinate of the $e^+e^-$ interaction vertex
    found using the fitting procedure with the same coordinate known
    from the simulation. This figure
    corresponds to the events of the $e^+e^-\rightarrow{K_S}K^{\pm}\pi^{\mp}$,
    $K_S\rightarrow\pi^+\pi^-$ simulation. The kinematic and vertex fitting procedure is applied to these events under
    the correct signal hypothesis~($e^+e^-\rightarrow{X}K^{+}\pi^{-}$ or
    $e^+e^-\rightarrow{X}K^{-}\pi^{+}$), known from the simulation. The
    difference between the $x$-coordinate of the $e^+e^-$ interaction vertex
    found using the fitting procedure and the same coordinate known from
    the simulation is shown in figure~(a) with a black solid line. The pull distribution for this
    coordinate is shown in figure~(b) with a black solid line. The dash-dotted
    lines in both
    figures~(a) and~(b) indicate the results of fitting the listed distributions
    using the
    probability density function of the normal distribution multiplied by the
    normalization factor~$N$. This factor, mean value~$\mu$ and
    standard deviation~$\sigma$ are free fitting parameters.\label{fig:vtx_x1_simhypo_xkpi}}
\end{figure}

\begin{figure}[tbp]
  \centering
  \begin{subfigure}[t]{0.47\textwidth}
    \centering
    \includegraphics[width=\textwidth]{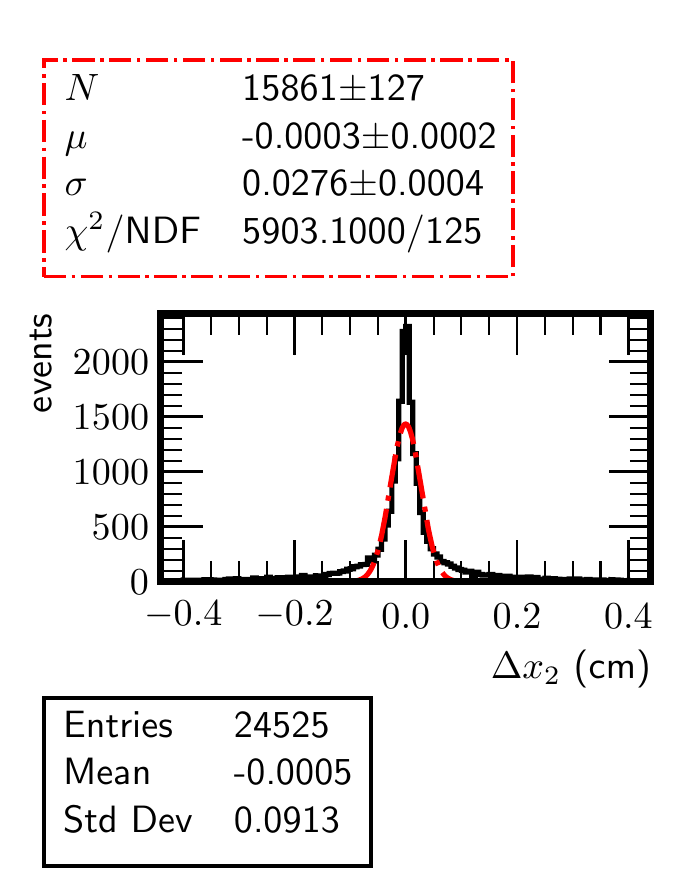}
    \caption{\label{fig:kf_vtx_dx2_simhypo_xkpi}}
  \end{subfigure}
  \hspace{1em}
  \begin{subfigure}[t]{0.47\textwidth}
    \centering
    \includegraphics[width=\textwidth]{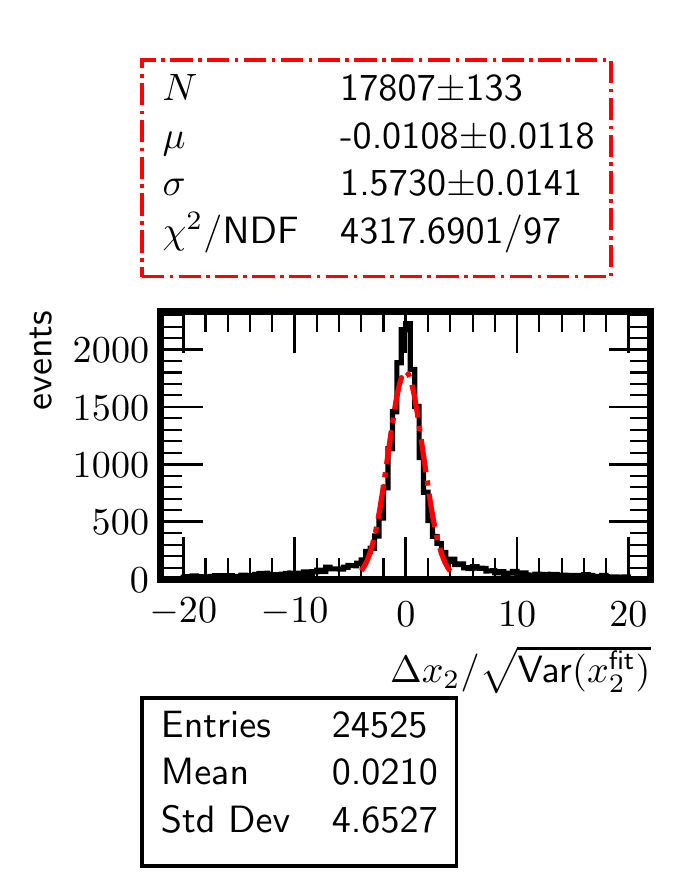}
    \caption{\label{fig:pull_x2_simhypo_xkpi}}
  \end{subfigure}
  \caption{Comparison of the $x$-coordinate of the $X\rightarrow\pi^+\pi^-$ decay vertex
    found using the fitting procedure with the same coordinate known
    from the simulation. This figure
    corresponds to the events of the $e^+e^-\rightarrow{K_S}K^{\pm}\pi^{\mp}$,
    $K_S\rightarrow\pi^+\pi^-$ simulation. The kinematic and vertex fitting procedure is applied to these events under
    the correct signal hypothesis~($e^+e^-\rightarrow{X}K^{+}\pi^{-}$ or
    $e^+e^-\rightarrow{X}K^{-}\pi^{+}$), known from the simulation. The
    difference between the $x$-coordinate of the $X\rightarrow\pi^+\pi^-$ decay vertex
    found using the fitting procedure and the same coordinate known from
    the simulation is shown in figure~(a) with a black solid line. The pull distribution for this
    coordinate is shown in figure~(b) with a black solid line. The dash-dotted
    lines in both
    figures~(a) and~(b) indicate the results of fitting the listed distributions
    using the
    probability density function of the normal distribution multiplied by the
    normalization factor~$N$. This factor, mean value~$\mu$ and
    standard deviation~$\sigma$ are free fitting parameters.\label{fig:vtx_x2_simhypo_xkpi}}
\end{figure}
It can be concluded what is the accuracy of vertex reconstruction using the
kinematic and vertex fitting procedure by comparing the vertex coordinates known
from the simulation with similar coordinates found using the fitting procedure.
Figure~\ref{fig:kf_vtx_dx1_simhypo_xkpi} shows the distribution of the
difference between the $x$-coordinate of the $e^+e^-$ interaction vertex
found using the kinematic and vertex fitting procedure with the same coordinate
known from the simulation. To rule out misassignment of kaons and pions in the
$X\rightarrow\pi^+\pi^-$ decay vertex, the fitting procedure was carried out under the correct signal
hypothesis.

Figure~\ref{fig:pull_x1_simhypo_xkpi} shows the pull distribution
for the $x$-coordinate of the $e^+e^-$ interaction vertex. Further, by pull
distribution we mean the distribution of the quantity
$\Delta\zeta/\sqrt{\text{Var}(\zeta^{\text{fit}})}$, where
$\zeta=x\text{, }y\text{ or }z$ is the vertex coordinate, $\Delta\zeta$ is the
difference between the $\zeta$-coordinate of the vertex found using the
kinematic and vertex fitting procedure with the same coordinate
known from the simulation, $\text{Var}(\zeta^{\text{fit}})$ is the variance of
the $\zeta$-coordinate found using the fitting package. This variance is
the corresponding diagonal element of the covariance matrix, defined as
$2\hat{\mathcal{H}}^{-1}$, where $\hat{\mathcal{H}}$ is the Hessian of
the Lagrange function calculated at the last iteration of Newton's method~(see
equation~\eqref{eq:newton-method}).

In the case of the pull distribution shown
in figure~\ref{fig:pull_x1_simhypo_xkpi}, it can be seen that the corresponding
standard deviation is about twice greater than one. As in the case of the
chi-square distribution, this fact is mainly explained by the non-Gaussian
response of the detector and the imperfect calibration of the matrix $\hat{\tilde{C}}$.
The discussed standard deviation should be close to one if the detector
response is Gaussian and the covariance matrix is well calibrated. The
verification of the latter statement is given in section~\ref{sec:kskpi-gsim}
using Gaussian simulation of the $e^+e^-\rightarrow{K_S}K^{\pm}\pi^{\mp}$ process. As
a result of this verification, the standard deviation of the
quantity~$\Delta\zeta/\sqrt{\text{Var}(\zeta^{\text{fit}})}$ is close to one, which
indicates that the fitting procedure correctly reconstructs the
vertices.

\begin{figure}[tbp]
  \begin{subfigure}[t]{0.47\textwidth}
    \centering
    \includegraphics[width=\textwidth]{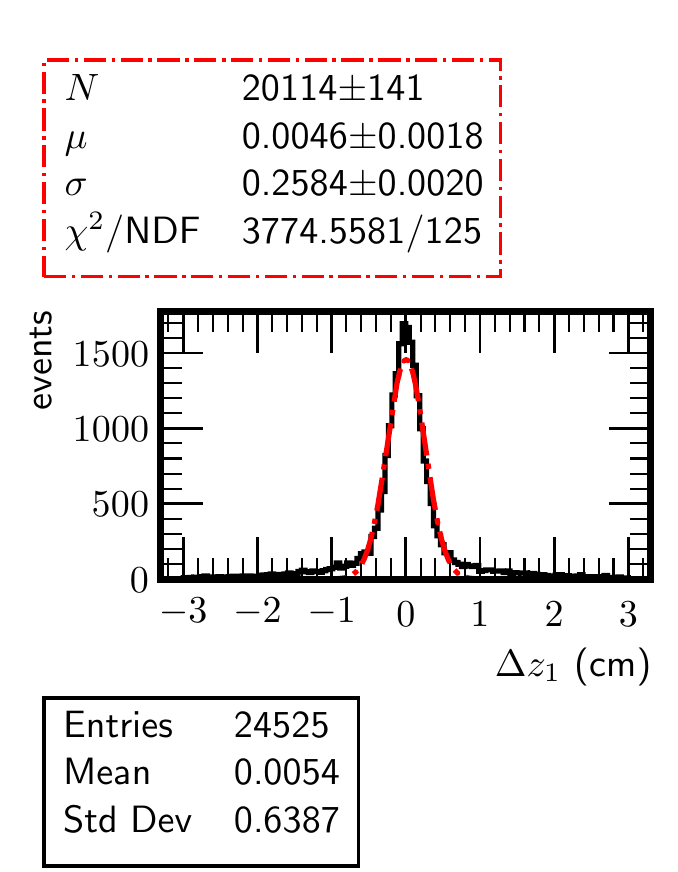}
    \caption{\label{fig:kf_vtx_dz1_simhypo_xkpi}}
  \end{subfigure}
  \hspace{1em}
  \begin{subfigure}[t]{0.47\textwidth}
    \centering
    \includegraphics[width=\textwidth]{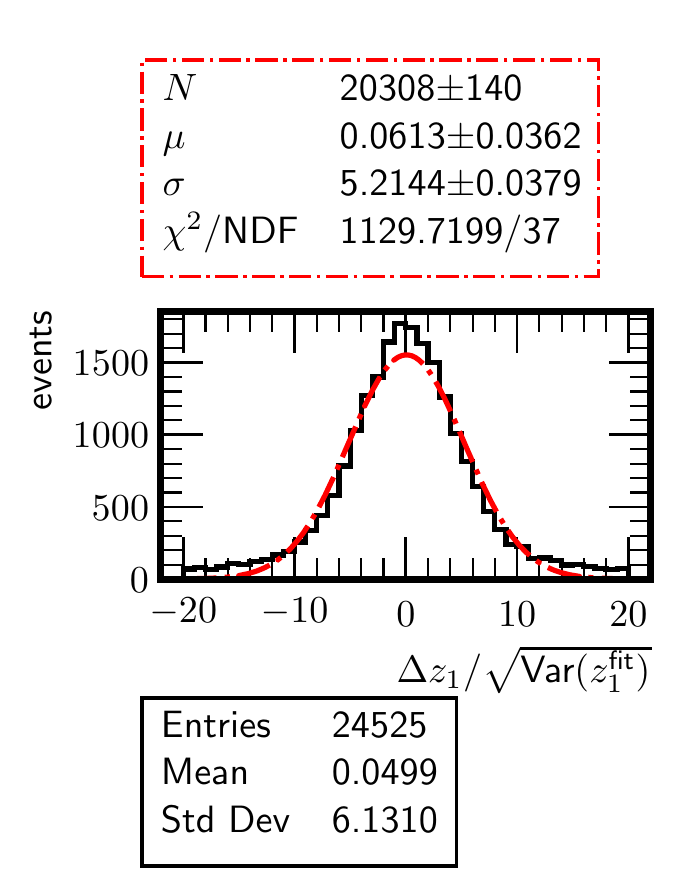}
    \caption{\label{fig:pull_z1_simhypo_xkpi}}
  \end{subfigure}
  \caption{Comparison of the $z$-coordinate of the $e^+e^-$ interaction vertex
    found using the fitting procedure with the same coordinate known
    from the simulation. This figure
    corresponds to the events of the $e^+e^-\rightarrow{K_S}K^{\pm}\pi^{\mp}$,
    $K_S\rightarrow\pi^+\pi^-$ simulation. The kinematic and vertex fitting procedure is applied to these events under
    the correct signal hypothesis~($e^+e^-\rightarrow{X}K^{+}\pi^{-}$ or
    $e^+e^-\rightarrow{X}K^{-}\pi^{+}$), known from the simulation. The
    difference between the $z$-coordinate of the $e^+e^-$ interaction vertex
    found using the fitting procedure and the same coordinate known from
    the simulation is shown in figure~(a) with a black solid line. The pull distribution for this
    coordinate is shown in figure~(b) with a black solid line. The dash-dotted
    lines in both
    figures~(a) and~(b) indicate the results of fitting the listed distributions
    using the
    probability density function of the normal distribution multiplied by the
    normalization factor~$N$. This factor, mean value~$\mu$ and
    standard deviation~$\sigma$ are free fitting parameters.\label{fig:vtx_z1_simhypo_xkpi}}
\end{figure}

\begin{figure}[tbp]
  \begin{subfigure}[t]{0.47\textwidth}
    \centering
    \includegraphics[width=\textwidth]{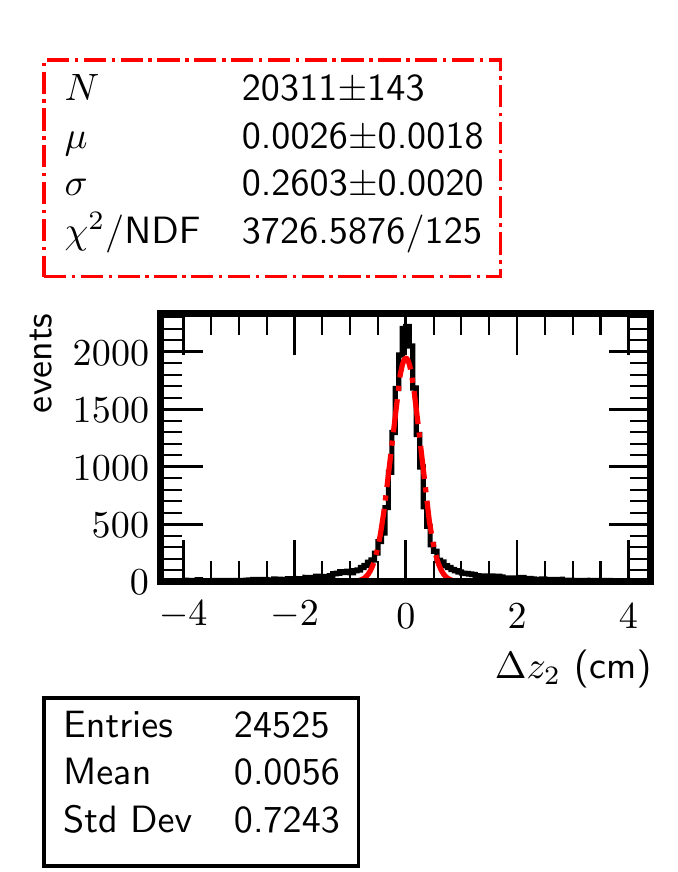}
    \caption{\label{fig:kf_vtx_dz2_simhypo_xkpi}}
  \end{subfigure}
  \hspace{1em}
  \begin{subfigure}[t]{0.47\textwidth}
    \centering
    \includegraphics[width=\textwidth]{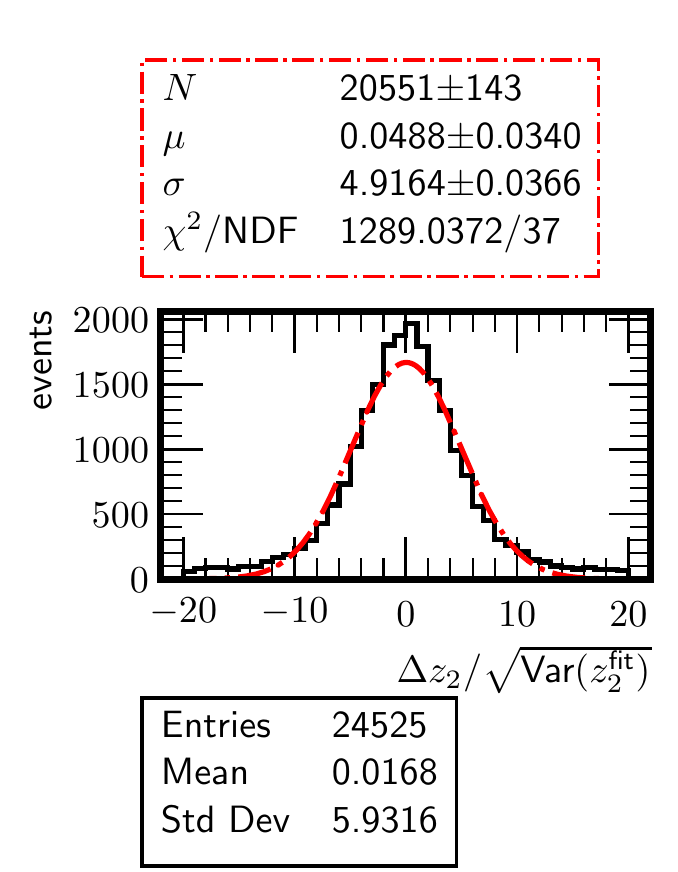}
    \caption{\label{fig:pull_z2_simhypo_xkpi}}
  \end{subfigure}
  \caption{Comparison of the $z$-coordinate of the $X\rightarrow\pi^+\pi^-$ decay vertex
    found using the fitting procedure with the same coordinate known
    from the simulation. This figure
    corresponds to the events of the $e^+e^-\rightarrow{K_S}K^{\pm}\pi^{\mp}$,
    $K_S\rightarrow\pi^+\pi^-$ simulation. The kinematic and vertex fitting procedure is applied to these events under
    the correct signal hypothesis~($e^+e^-\rightarrow{X}K^{+}\pi^{-}$ or
    $e^+e^-\rightarrow{X}K^{-}\pi^{+}$), known from the simulation. The
    difference between the $z$-coordinate of the $X\rightarrow\pi^+\pi^-$ decay vertex
    found using the fitting procedure and the same coordinate known from
    the simulation is shown in figure~(a) with a black solid line. The pull distribution for this
    coordinate is shown in figure~(b) with a black solid line. The dash-dotted
    lines in both
    figures~(a) and~(b) indicate the results of fitting the listed distributions
    using the
    probability density function of the normal distribution multiplied by the
    normalization factor~$N$. This factor, mean value~$\mu$ and
    standard deviation~$\sigma$ are free fitting parameters.\label{fig:vtx_z2_simhypo_xkpi}}
\end{figure}
Figure~\ref{fig:kf_vtx_dx2_simhypo_xkpi} shows the distribution of the
difference between the $x$-coordinate of the $X\rightarrow\pi^+\pi^-$ decay vertex
found using the kinematic and vertex fitting procedure with the same coordinate
known from the simulation. Figure~\ref{fig:pull_x2_simhypo_xkpi} shows the pull
distribution corresponding to this coordinate.
Figure~\ref{fig:vtx_z1_simhypo_xkpi} is similar to
figure~\ref{fig:vtx_x1_simhypo_xkpi}, but corresponds to the $z\text{-}$
coordinate of the $e^+e^-$ interaction vertex. Figure~\ref{fig:vtx_z2_simhypo_xkpi} corresponds to the
$z\text{-}$ coordinate of the $X\rightarrow\pi^+\pi^-$ decay vertex.
For the same reason as in the case of figure~\ref{fig:pull_x1_simhypo_xkpi},
the standard deviations
corresponding to the pull distributions shown in
figures~\ref{fig:pull_x2_simhypo_xkpi},~\ref{fig:pull_z1_simhypo_xkpi}
and~\ref{fig:pull_z2_simhypo_xkpi} are greater than one.

\subsubsection{Fitting the events of the %
  $e^+e^-\rightarrow{K_S}K^{\pm}\pi^{\mp}$ simulation with %
  ISR disabled \label{sec:xkpi-noisir}}
\begin{figure}[tbp]
  \centering
  \begin{subfigure}[t]{0.47\textwidth}
    \centering
    \includegraphics[width=\textwidth]{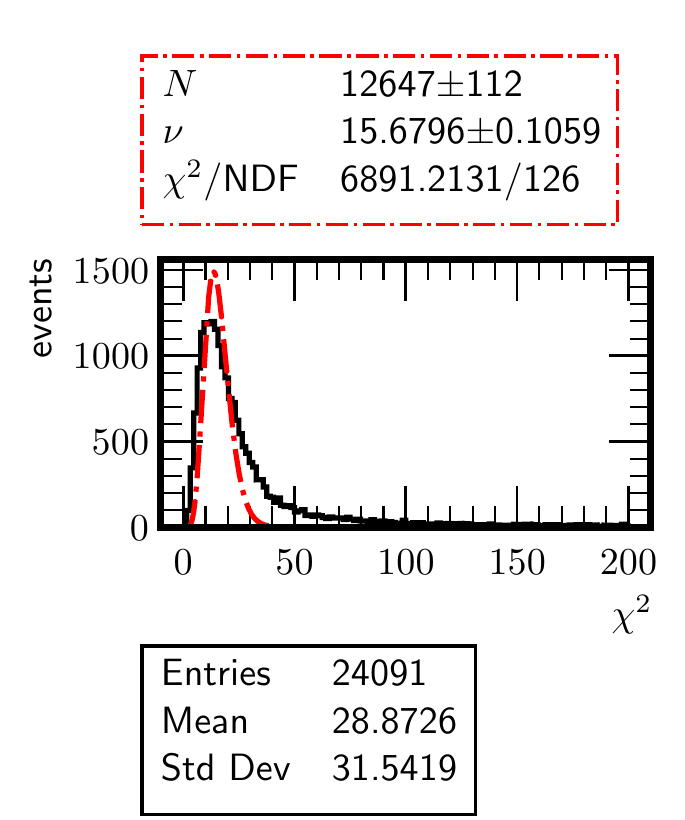}
    \caption{\label{fig:kf_chi2_simhypo_xkpi-noisr}}
  \end{subfigure}
  \hspace{1em}
  \begin{subfigure}[t]{0.47\textwidth}
    \centering
    \includegraphics[width=\textwidth]{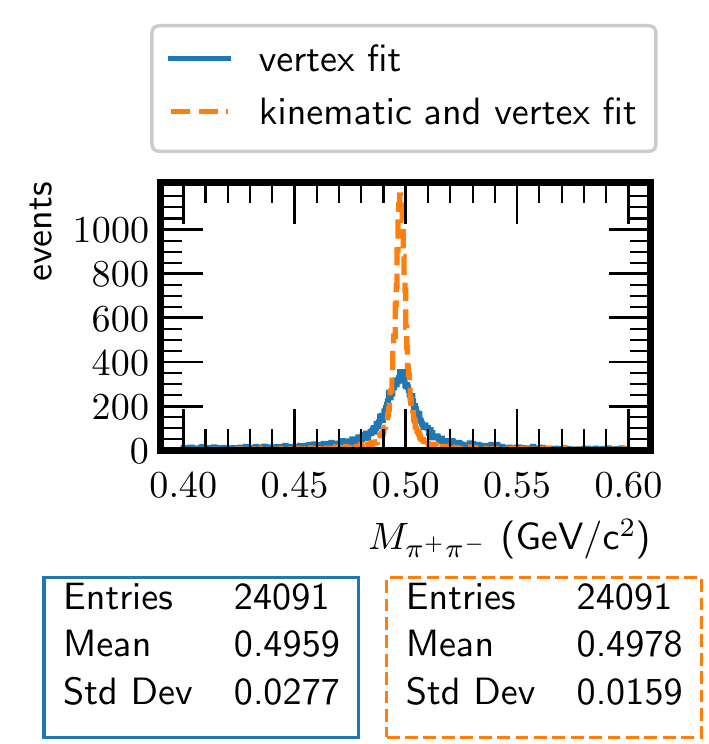}
    \caption{\label{fig:kf_mks_simhypo_xkpi-noisr}}
  \end{subfigure}
  \caption{Chi-square distribution (a) and invariant mass distributions
    of the $X\rightarrow\pi^+\pi^-$ decay products (b). The distribution is
    obtained for the events of the $e^+e^-\rightarrow{K_S}K^{\pm}\pi^{\mp}$,~$K_S\rightarrow\pi^+\pi^-$
    simulation fitted under the $e^+e^-\rightarrow{X}K^{\pm}\pi^{\mp}$,~$X\rightarrow\pi^+\pi^-$ hypotheses.  This figure is similar to
    figure~\ref{fig:sim-hypo-xkpi},
    but ISR is disabled in the events of the $e^+e^-\rightarrow{K_S}K^{\pm}\pi^{\mp}$ simulation.\label{fig:sim-hypo-xkpi-noisr}}
\end{figure}
In the previous subsection, it is shown that the chi-square distribution obtained
as a result of fitting the events of the
$e^+e^-\rightarrow{K_S}K^{\pm}\pi^{\mp}$,~$K_S\rightarrow\pi^+\pi^-$ simulation under
the $e^+e^-\rightarrow{X}K^{\pm}\pi^{\mp}$,~$X\rightarrow\pi^+\pi^-$ hypotheses is not
described by the chi-square
probability density function given by equation~\eqref{eq:chi-square-pdf}.
In this regard, it is interesting to know what contribution to this distortion
is caused by the absence of ISR photons in the kinematic hypotheses.
In order to answer this question, we consider the events of the
$e^+e^-\rightarrow{K_S}K^{\pm}\pi^{\mp}$,~$K_S\rightarrow\pi^+\pi^-$ simulation, in
which initial state radiation is turned off. The center-of-mass energy and
the magnetic field in this simulation are the same as in the previous subsection.
The events of this simulation are fitted under the
$e^+e^-\rightarrow{X}K^{\pm}\pi^{\mp}$,~$X\rightarrow\pi^+\pi^-$ hypotheses~(see
sections~\ref{sec:xkpi-desc} and~\ref{sec:xkpi-fitting-details}). To avoid the misassignment of kaons and
pions~(see the previous section), each event is fitted under the correct signal
hypothesis~($e^+e^-\rightarrow{X}K^{+}\pi^{-}$ or $e^+e^-\rightarrow{X}K^{-}\pi^{+}$)
known from the simulation. In the
case of the considered example, the fitting procedure converged to a local minimum in $24091$ events
out of $24839$. This procedure converged to a local maximum in $279$ events. In the remaining $469$ events,
the procedure did not converge. The results of the fitting are shown in figure~\ref{fig:sim-hypo-xkpi-noisr}.
The distributions shown in this figure should be compared with similar distributions shown in
figure~\ref{fig:sim-hypo-xkpi}~(in the case of figure~\ref{fig:sim-hypo-xkpi}
ISR is enabled in the events of the $e^+e^-\rightarrow{K_S}K^{\pm}\pi^{\mp}$ simulation).

Figure~\ref{fig:kf_chi2_simhypo_xkpi-noisr} shows the chi-square distribution corresponding to the
events of the $e^+e^-\rightarrow{K_S}K^{\pm}\pi^{\mp}$ simulation with ISR disabled.
Although neither the simulated events nor the kinematic hypotheses contain ISR
photons, this distribution is not consistent with the chi-squared probability
density function.
As a result, it can be concluded that the distortion of the chi-square
distribution is not fully explained by the effects associated with initial state
radiation. The same conclusion can be drawn for other kinematic hypotheses
discussed below.

Figure~\ref{fig:kf_mks_simhypo_xkpi-noisr} shows the distributions of the
invariant mass of the $X\rightarrow\pi^+\pi^-$ decay products. The solid line
corresponds to the distribution obtained as a result of the vertex fitting,
while the dashed line corresponds to the distribution obtained as a
result of kinematic and vertex fitting under the
$e^+e^-\rightarrow{X}K^{\pm}\pi^{\mp}$,~$X\rightarrow\pi^+\pi^-$
hypotheses. Both distributions were obtained for the events of the
$e^+e^-\rightarrow{K_S}K^{\pm}\pi^{\mp}$ simulation with ISR disabled. The
distribution indicated by the dashed line
in figure~\ref{fig:kf_mks_simhypo_xkpi-noisr} can be compared with the similar
distribution shown in
figure~\ref{fig:inv-mass-simhypo-xkpi}. The distribution shown in
figure~\ref{fig:inv-mass-simhypo-xkpi}
is distorted with respect to the distribution shown in
figure~\ref{fig:kf_mks_simhypo_xkpi-noisr} because the simulated
events in the case of the distribution shown in
figure~\ref{fig:inv-mass-simhypo-xkpi}
contain ISR photons, which are not taken into account in the
$e^+e^-\rightarrow{X}K^{\pm}\pi^{\mp}$,~$X\rightarrow\pi^+\pi^-$ hypotheses. To avoid
such distortion of invariant mass distribution, one can use
kinematic hypotheses containing ISR photons. However the use
of such hypotheses usually worsens the resolution of invariant
masses, since ISR photons in the case of CMD-3 are
considered as lost particles.

\subsubsection{Fitting the $e^+e^-\rightarrow\pi^+\pi^-\pi^+\pi^-$ events \label{sec:4pi-xkpi}}
\begin{figure}[tbp]
  \centering
  \begin{subfigure}[t]{0.47\textwidth}
    \centering
    \includegraphics[width=\textwidth]{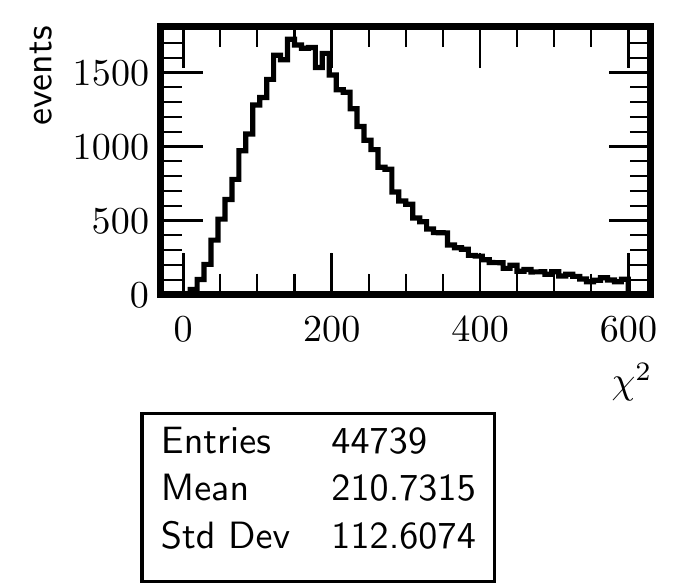}
    \caption{\label{fig:kf-chi2-4pi-xkpi}}
  \end{subfigure}
  \hspace{1em}
  \begin{subfigure}[t]{0.47\textwidth}
    \centering
    \includegraphics[width=\textwidth]{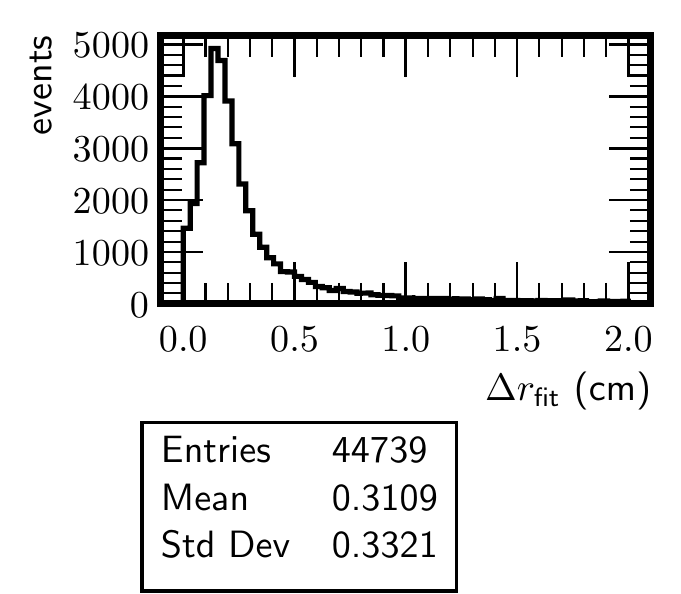}
    \caption{\label{fig:dr-fit-4pi-xkpi}}
  \end{subfigure}
  \caption{Chi-square distribution (a) and distribution of the distance between the
    $e^+e^-$ interaction vertex and the $X\rightarrow\pi^+\pi^-$ decay vertex (b). Both
    distributions are obtained by applying the kinematic and vertex fitting procedure
    to the events of the $e^+e^-\rightarrow\pi^+\pi^-\pi^+\pi^-$. The corresponding
    fitting procedure is described in section~\ref{sec:xkpi-fitting-details}.
    \label{fig:fit-4pi-xkpi}}
\end{figure}
This subsection provides an example of applying the fitting
procedure described in section~\ref{sec:xkpi-fitting-details} under the
$e^+e^-\rightarrow X K^{\pm}\pi^{\mp},\;X\rightarrow\pi^+\pi^-$ hypotheses to the simulated
$e^+e^-\rightarrow\pi^+\pi^-\pi^+\pi^-$ events. The center-of-mass energy and the
magnetic field for these events are the same as
in the previous section. The process
$e^+e^-\rightarrow\pi^+\pi^-\pi^+\pi^-$ is one of the main background processes in the
study of the $e^+e^-\rightarrow K_SK^{\pm}\pi^{\mp}$ process with the $\text{CMD-}3$ detector.
It should be noted that the $e^+e^-\rightarrow\pi^+\pi^-\pi^+\pi^-$ process does not
contain charged kaons in the final state and, therefore, is also a background process
with respect to the $e^+e^-\rightarrow X K^{\pm}\pi^{\mp}$ hypotheses. The fitting procedure
converged to a local minimum in $44739$ events out of $45199$. Some results of the fitting
are shown in figure~\ref{fig:fit-4pi-xkpi}.

Figure~\ref{fig:kf-chi2-4pi-xkpi}
shows the chi-square distribution obtained by applying the fitting procedure to the
events of the $e^+e^-\rightarrow\pi^+\pi^-\pi^+\pi^-$ simulation. Since the
$e^+e^-\rightarrow\pi^+\pi^-\pi^+\pi^-$ is the background process with respect to the
considered hypotheses, the chi-square distribution turned out to be wide~(compare with
figure~\ref{fig:kf-chi2-xkpi}).

Figure~\ref{fig:dr-fit-4pi-xkpi} shows the distribution of the distance
$\Delta r_{\text{fit}}$ between
the $e^+e^-$ interaction vertex and the $X\rightarrow\pi^+\pi^-$ decay vertex.
The distribution was obtained for the $e^+e^-\rightarrow\pi^+\pi^-\pi^+\pi^-$
events using the fitting procedure described in section~\ref{sec:xkpi-fitting-details}.
The distance
is calculated using the vertex coordinates found as a result of the fitting.
Since in the $e^+e^-\rightarrow\pi^+\pi^-\pi^+\pi^-$ process all pions
originate from the $e^+e^-$ interaction vertex, the large $\Delta r_{\text{fit}}$
distances are not expected. Indeed, it can be seen from the figure that most of
the events belong to region $\Delta r_{\text{fit}} \lesssim 5\text{ mm}$.

Finally, we can conclude that the chi-square and
$\Delta r_{\text{fit}}$ distributions can be successfully used
to select the events of the $e^+e^-\rightarrow K_SK^{\pm}\pi^{\mp}$
process and suppress the corresponding background processes, such as
the $e^+e^-\rightarrow\pi^+\pi^-\pi^+\pi^-$ process.

\subsection{Hypotheses $e^+e^-\rightarrow K_SK^{\pm}\pi^{\mp},\;K_S\rightarrow\pi^+\pi^-$
  \label{sec:kskpi}}
\subsubsection{Description of the
  $e^+e^-\rightarrow K_SK^{\pm}\pi^{\mp}$ hypotheses}
In this section we present the results of fitting under
$e^+e^-\rightarrow K_S K^{\pm}\pi^{\mp},\;K_S\rightarrow\pi^+\pi^-$ hypotheses.
The difference between section~\ref{sec:xkpi} and this section is that
in this section the intermediate neutral particle ($K_S$) is known.
By the fact that the intermediate particle is known, we mean that this
particle directly participates in the energy-momentum conservation constraints,
i.e. its mass influences the fitting results. The energy-momentum
conservation constraint tree corresponding to the
$e^+e^-\rightarrow K_SK^+\pi^-,\;K_S\rightarrow\pi^+\pi^-$
hypothesis has the form shown in figure~\ref{fig:kskpi-tree}. For the
charge-conjugate hypothesis, there is a similar tree, but with oppositely
charged particles. Thus, energy conservation constraints are imposed in
each of the two vertices and include all particles related with one or
another vertex. This is the only difference between the hypotheses
$e^+e^-\rightarrow K_SK^{\pm}\pi^{\mp}$ and
$e^+e^-\rightarrow X K^{\pm}\pi^{\mp}$ (see subsection~\ref{sec:xkpi-desc}).

The fitting procedure with the $e^+e^-\rightarrow K_SK^{\pm}\pi^{\mp}$
hypotheses is completely similar to the fitting
procedure with
the $e^+e^-\rightarrow X K^{\pm}\pi^{\mp}$
hypotheses described in section~\ref{sec:xkpi-fitting-details}. This procedure does not
contain any additional $K\text{-}\pi$
pre-separation algorithm and is used for demonstration
purposes only.

\subsubsection{Fitting the
  $e^+e^-\rightarrow K_SK^{\pm}\pi^{\mp}$
  events \label{sec:kskpi-kskpi}}
\begin{figure}[tbp]
  \centering
  \begin{subfigure}[t]{0.47\textwidth}
    \centering
    \includegraphics[width=\textwidth]{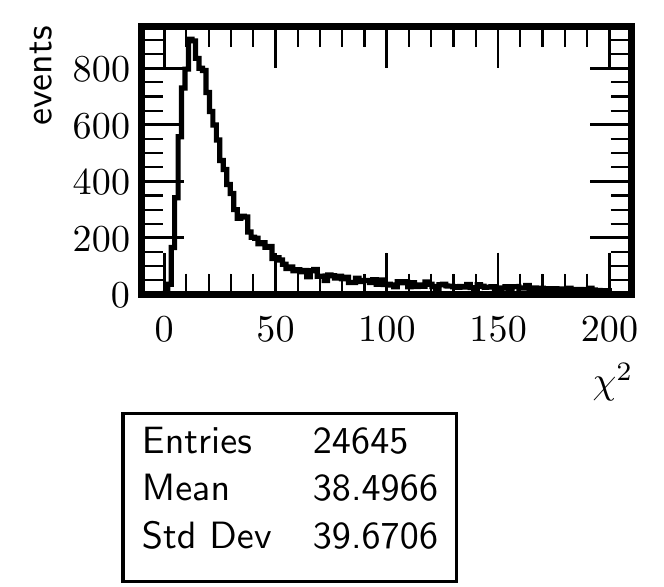}
    \caption{\label{fig:kf-chi2-kskpi}}
  \end{subfigure}
  \hspace{1em}
  \begin{subfigure}[t]{0.47\textwidth}
    \centering
    \includegraphics[width=\textwidth]{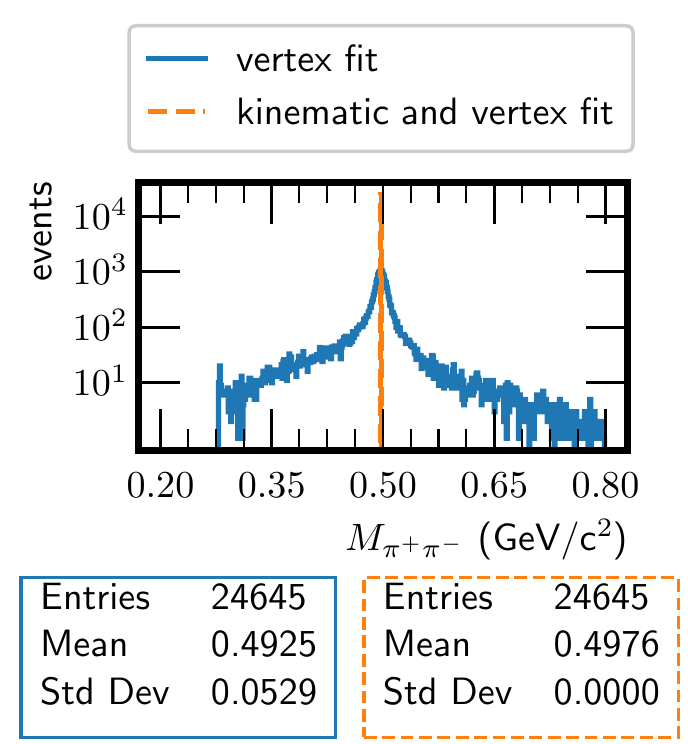}
    \caption{\label{fig:inv-mass-kskpi}}
  \end{subfigure}
  \caption{Chi-square distribution (a) and invariant mass distributions
    of the $X\rightarrow\pi^+\pi^-$ decay products (b). This figure is
    similar to figure~\ref{fig:chi2-and-mass-xkpi}, but the kinematic and
    vertex fitting is done under the $e^+e^-\rightarrow{K_S}K^{\pm}\pi^{\mp}$,
    $K_S\rightarrow\pi^+\pi^-$ hypotheses.\label{fig:fit-kskpi}}
\end{figure}
This subsection discusses the results of applying the fitting procedure under the
$e^+e^-\rightarrow K_SK^{\pm}\pi^{\mp}$ hypotheses to the events of the
$e^+e^-\rightarrow K_SK^{\pm}\pi^{\mp}$ process. The fitting procedure converged to a
local minimum in $24645$ events out of $25224$. In $217$ events, this procedure
converged to a local maximum. In the rest $362$ events, the fitting procedure did
not converge. Some of the fitting results are shown in figure~\ref{fig:fit-kskpi}.

Figure~\ref{fig:kf-chi2-kskpi} shows the chi-square distribution for
the events of the $e^+e^-\rightarrow K_SK^{\pm}\pi^{\mp}$ simulation. This
distribution corresponds to the $e^+e^-\rightarrow K_SK^{\pm}\pi^{\mp}$
hypotheses. The expected number of degrees of freedom is $l - m = 23 - 11 = 12$.
However, as in the case of figure~\ref{fig:kf-chi2-xkpi}, the distribution
in figure~\ref{fig:kf-chi2-kskpi} does not correspond to the expected number of
degrees of freedom due to the reasons listed in section~\ref{sec:kskpi-xkpi}.

Figure~\ref{fig:inv-mass-kskpi} shows the two-pion invariant mass distribution
for pions from the $K_S\rightarrow\pi^+\pi^-$ decay. As in the case of
figure~\ref{fig:inv-mass-xkpi}, the dashed line indicates the invariant mass
distribution obtained using the pion parameters found with the kinematic and
vertex fitting procedure. Since the intermediate particle ($K_S$) in the
considered hypotheses has a certain mass and participates in the energy-momentum
conservation constraints, this distribution turned out to be fixed on the
value of the $K_S$-meson mass. The solid line denotes the invariant mass
distribution obtained using the $K_S\rightarrow\pi^+\pi^-$ decay vertex fit.
The pion tracks from the $K_S$-meson decay are assumed to be known from the
kinematic and vertex fitting procedure.

\subsection{Hypothesis $e^+e^-\rightarrow\gamma\gamma\gamma$ \label{sec:3gamma}}
\subsubsection{Description of the $e^+e^-\rightarrow\gamma\gamma\gamma$ hypothesis}
In section~\ref{sec:3gamma}, we present the results of
applying the kinematic and vertex fitting
package under the $e^+e^-\rightarrow\gamma\gamma\gamma$ hypothesis to the events of
the $e^+e^-\rightarrow\pi^0\gamma,\;\pi^0\rightarrow\gamma\gamma$ simulation.
The $e^+e^-\rightarrow\gamma\gamma\gamma$ hypothesis has the following form.
\begin{itemize}
 \item The hypothesis contains a single vertex. This vertex is the
   $e^+e^-$ interaction vertex. All coordinates of this vertex
   are considered as free measurable parameters and contribute to the
   chi-square.
 \item The hypothesis contains three final particles. These particles are
   photons described in section~\ref{sec:photon}. The hypothesis contains
   also the initial pseudo-particle (see section~\ref{sec:initial-pseudo-particle}),
   which is used in order to provide the total
   four-momentum of the initial particles ($e^+e^-$).
  \item All of the above particles are involved in four energy-momentum conservation
    constraints.
  \item The vertex constraints described in section~\ref{sec:vertex-constraints} are
    not imposed on the photons, because the photons are already emitted from the
    vertex of their origin due to the parametrization of their
    momentum~\eqref{eq:photon-four-momentum}.
\end{itemize}
In total, the $e^+e^-\rightarrow\gamma\gamma\gamma$ hypothesis
contains $4$ constraints and $15$ free parameters. All of
these parameters are measurable.

\subsubsection{Fitting the $e^+e^-\rightarrow\pi^0\gamma$ events\label{sec:fitting-pi0gamma}}
\begin{figure}[tbp]
  \centering
  \begin{subfigure}[t]{0.47\textwidth}
    \centering
    \includegraphics[width=\textwidth]{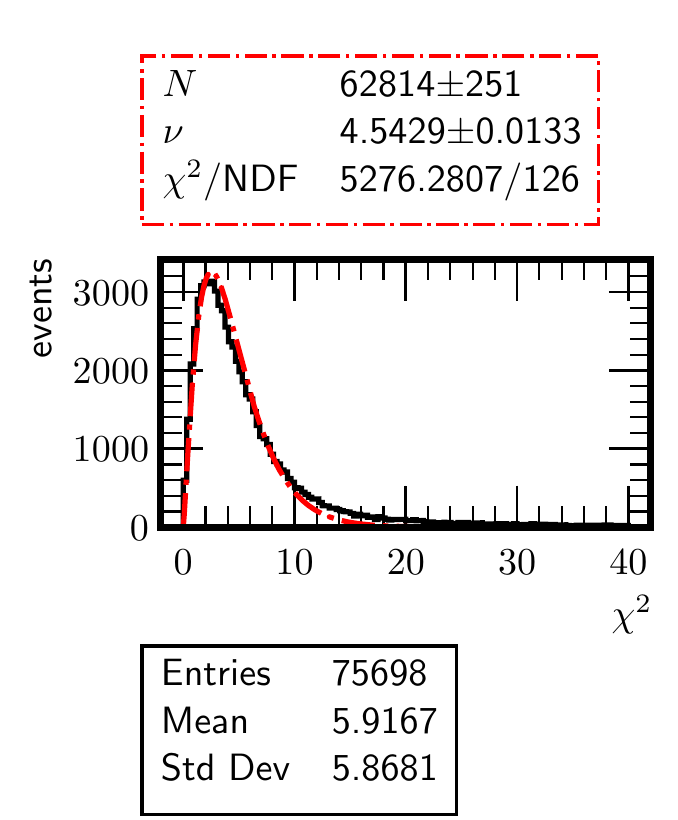}
    \caption{\label{fig:kf-chi2-3gamma}}
  \end{subfigure}
  \hspace{1em}
  \begin{subfigure}[t]{0.47\textwidth}
    \centering
    \includegraphics[width=1\textwidth]{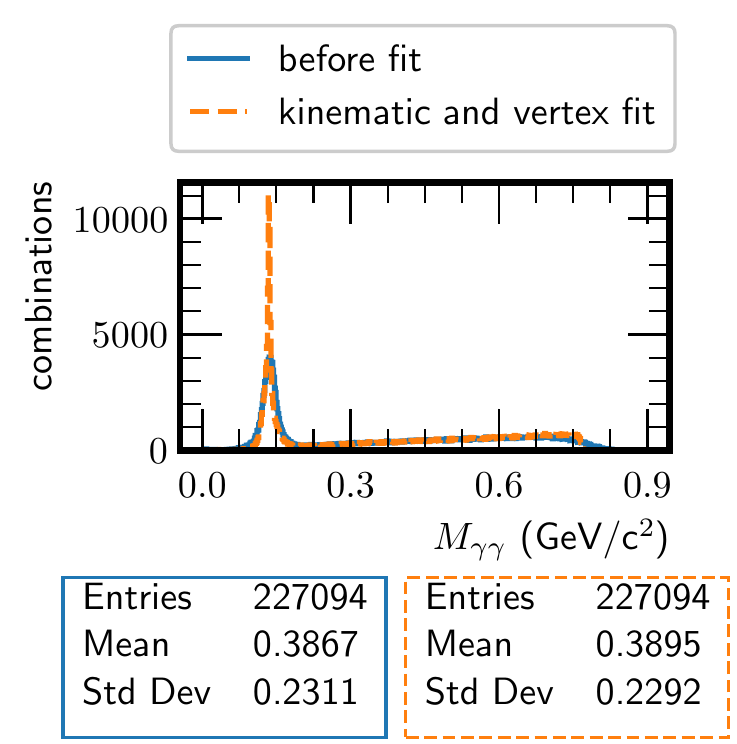}
    \caption{\label{fig:kf-mgg-3gamma}}
  \end{subfigure}
  \caption{Chi-square distribution (a) and two-photon invariant mass distributions (b).
    All distributions are obtained as a result of fitting the events of the $e^+e^-\rightarrow\pi^0\gamma$,
    $\pi^0\rightarrow\gamma\gamma$ simulation under the $e^+e^-\rightarrow\gamma\gamma\gamma$  hypothesis.
    In figure~(b), the dashed line corresponds to the invariant mass distribution obtained with kinematic
    and vertex fitting procedure. The distribution indicated by the solid curve in the same figure is
    obtained using initial photon parameters and assuming that photons propagate from the
    center of the detector.}
\end{figure}
In this subsection, we discuss the results of fitting the
simulated
$e^+e^-\rightarrow\pi^0\gamma,\;\pi^0\rightarrow\gamma\gamma$
events under the $e^+e^-\rightarrow\gamma\gamma\gamma$
hypothesis. The events of this simulation correspond to the center-of-mass
energy equal to $782.7$~MeV. For the demonstration purposes, only events with three
photon clusters in the calorimeter are considered. Thus, since all three photons
are equivalent, there is a single mapping of photons to their clusters, i.e.
only one fit is performed in each event. The fitting procedure converged to a
local minimum in all events.

Figure~\ref{fig:kf-chi2-3gamma} shows a chi-square distribution corresponding to the
example discussed in this section. The mean value of the distribution shown in the
figure is approximately $5.9$. This value is greater than the expected number of degrees
of freedom~($m - l = 4 - 0 = 4$). The fact that the chi-square distribution is
distorted is due to the reasons listed in section~\ref{sec:kskpi-xkpi}.

Figure~\ref{fig:kf-mgg-3gamma} shows histograms with the two-photon
invariant mass distributions. The histogram indicated by the solid line is
obtained using the measured photon parameters, and the histogram indicated by
the dashed line is obtained using the photon parameters found using the fitting
procedure. In the case of the histogram marked with the solid line, it is assumed
that photons propagate from the center of the detector. Two-photon invariant mass
distributions are shown for the reason that we would like to see a
pion peak on them. Since in the $e^+e^-\rightarrow\gamma\gamma\gamma$ hypothesis,
it is unknown which photons are the pion decay products, these histograms contain
the invariant masses of all different photon pairs,
i.e. each histogram has three entries per event. Thus, in these histograms, in
addition to the pion peak, there is a contribution from the combinatorial
background. It is seen from the figure that the fitting procedure significantly
improves the two-photon invariant mass resolution.

\begin{figure}[tbp]
  \centering
  \begin{subfigure}[t]{0.47\textwidth}
    \centering
    \includegraphics[width=\textwidth]{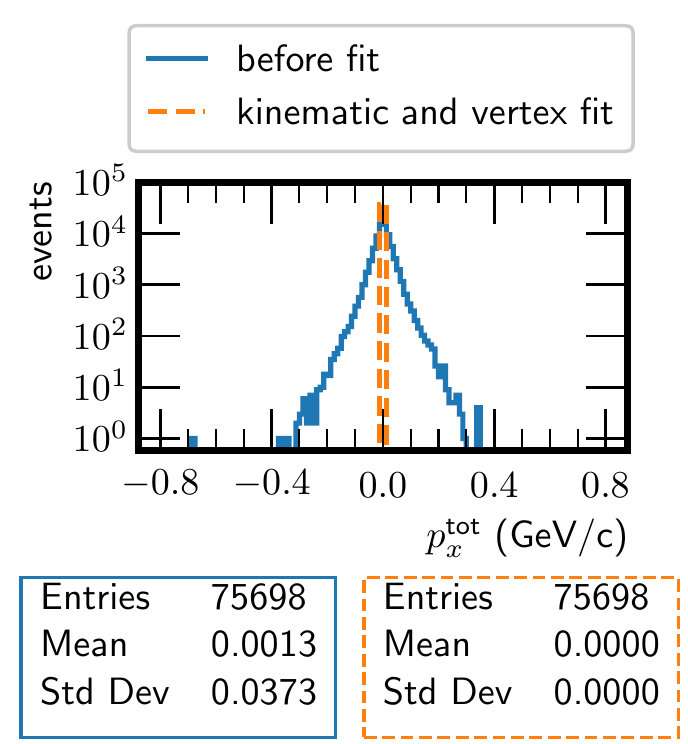}
    \caption{\label{fig:kf-px-tot-3gamma}}
  \end{subfigure}
  \hspace{1em}
  \begin{subfigure}[t]{0.47\textwidth}
    \centering
    \includegraphics[width=\textwidth]{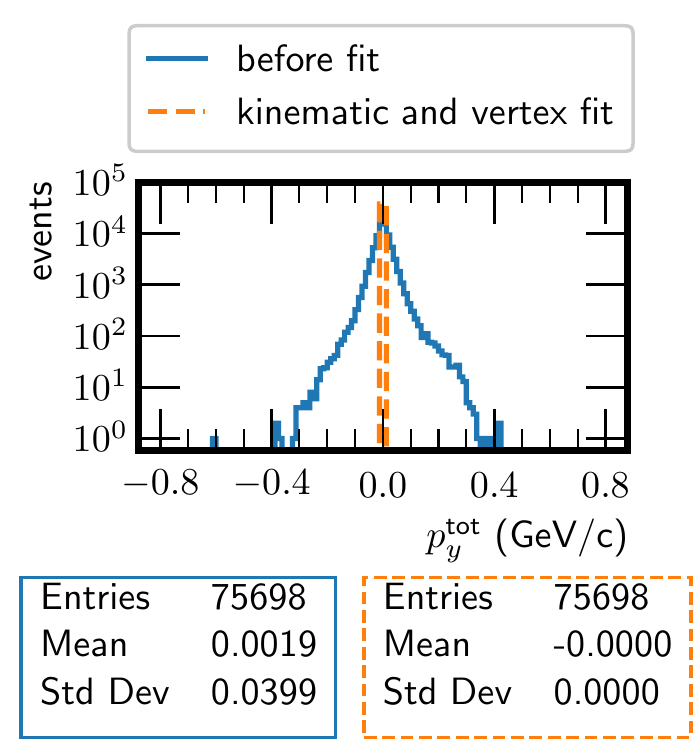}
    \caption{\label{fig:kf-py-tot-3gamma}}
  \end{subfigure}
  \begin{subfigure}[t]{0.47\textwidth}
    \centering
    \includegraphics[width=\textwidth]{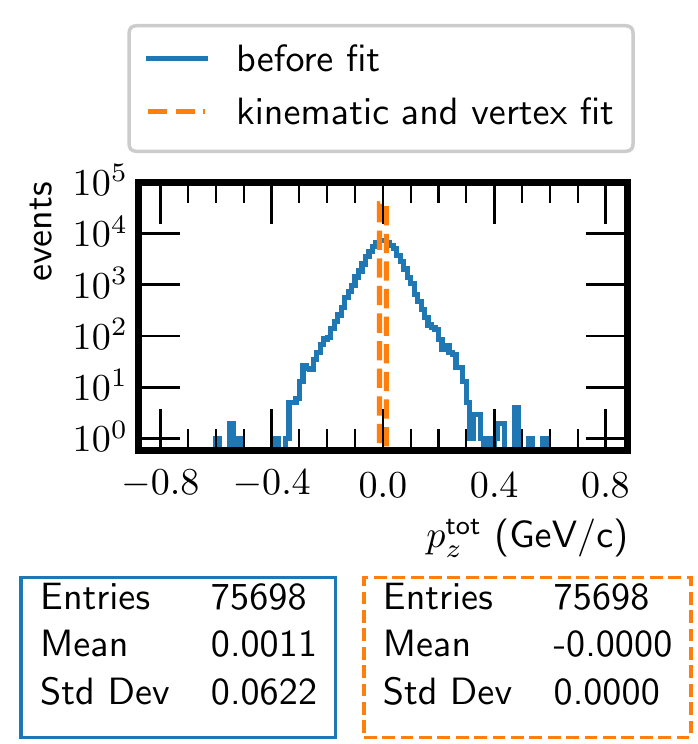}
    \caption{\label{fig:kf-pz-tot-3gamma}}
  \end{subfigure}
  \hspace{1em}
  \begin{subfigure}[t]{0.47\textwidth}
    \centering
    \includegraphics[width=\textwidth]{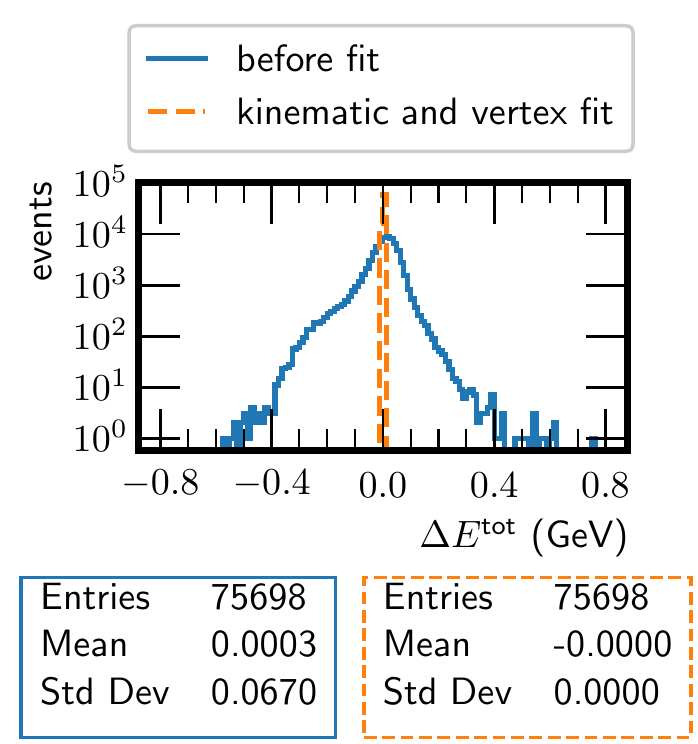}
    \caption{\label{fig:kf-e-tot-3gamma}}
  \end{subfigure}
  \caption{Distributions demonstrating the fulfillment of the energy-momentum
    conservation constraints the case of the
    $e^+e^-\rightarrow\gamma\gamma\gamma$ hypothesis. The distributions are
    obtained for the events of the $e^+e^-\rightarrow\pi^0\gamma$,
    $\pi^0\rightarrow\gamma\gamma$ simulation. Figures~(a), (b), and (c) show
    the $x\text{-}$, $y{\text{-}}$, and $z\text{-}$ components of the total
    momentum, respectively. Figure~(d) shows the difference between total
    energy and center-of-mass energy. The solid lines show the distributions
    obtained using the initial parameters of the photons under the assumption
    that the photons propagate from the center of the detector. The dashed
    lines indicate the distributions obtained using the parameters found by
    fitting.\label{fig:kf-tot-4-momentum-3gamma}}
\end{figure}
The hypothesis $e^+e^-\rightarrow\gamma\gamma\gamma$ is quite simple. In this
hypothesis, the energy-momentum conservation constraints are only imposed,
and there are no explicit vertex constraints\footnote{There are no vertex
constraints described in section~\ref{sec:vertex-constraints} and corresponding
to their own Lagrange multipliers. The vertex constraints hold implicitly due
to the parametrization of the photon
momentum~(see equation~\eqref{eq:photon-four-momentum}).}.
Let us show how the energy-momentum conservation constraints hold.
Figure~\ref{fig:kf-tot-4-momentum-3gamma} shows the distributions of the
left-hand sides of the energy-momentum conservation
constraints~\eqref{eq:energy-momentum-constraints}. The solid lines indicate the
distributions obtained using the measured particle parameters. In the case of
distributions indicated by solid lines, it is assumed that photons propagate
from the center of the detector. The dashed lines
denote the distributions obtained using the particle parameters found with the
fitting procedure. The dashed distributions are fixed at zero, i.e.
energy-momentum conservation constraints are satisfied with high accuracy.
Similar distributions for the energy-momentum conservation constraints, as well
as for vertex constraints, can also be shown in the case of more complex
hypotheses. However, in this paper, we restrict ourselves to the distributions
shown in figure~\ref{fig:kf-tot-4-momentum-3gamma}.

\subsection{Hypothesis $e^+e^-\rightarrow\pi^+\pi^-\gamma\gamma$}
\subsubsection{Description of the $e^+e^-\rightarrow\pi^+\pi^-\gamma\gamma$
  hypothesis \label{sec:2pi2gamma}} Section~\ref{sec:2pi2gamma} discusses the
results of applying kinematic and vertex fitting under the
$e^+e^-\rightarrow\pi^+\pi^-\gamma\gamma$ hypothesis. This hypothesis has the
following form.
\begin{itemize}
  \item The hypothesis has a single vertex, which is the $e^+e^-$ interaction
    vertex. All coordinates of this vertex are considered as free
    measurable parameters and contribute to the chi-square.
  \item The hypothesis requires the presence of four final particles and one
    initial pseudo-particle~\ref{sec:initial-pseudo-particle}. The list of final
    particles consists of two photons~(see section~\ref{sec:photon}) and two
    oppositely charged pions~(see section~\ref{sec:charged-particle}).
  \item There are four constraints on the energy-momentum conservation in this
    hypothesis. All particles are involved into these constraints.
  \item Three vertex constraints are imposed on each charged particle. No vertex
    constraints are imposed on the photons, because the photons already fly out
    of the vertex due to the parametrization of their momenta.
\end{itemize}

In total, the $e^+e^-\rightarrow\pi^+\pi^-\gamma\gamma$ hypothesis contains $10$
constraints. This hypothesis contains $23$ free parameters. Among these parameters,
$21$ parameters are measurable and $2$ parameters are non-measurable.

\subsubsection{Fitting the $e^+e^-\rightarrow\eta\pi^+\pi^-$, $\eta\rightarrow\gamma\gamma$
  events \label{sec:etapipi-2pi2gamma}}
\begin{figure}[tbp]
  \centering
  \begin{subfigure}[t]{0.47\textwidth}
    \centering
    \includegraphics[width=\textwidth]{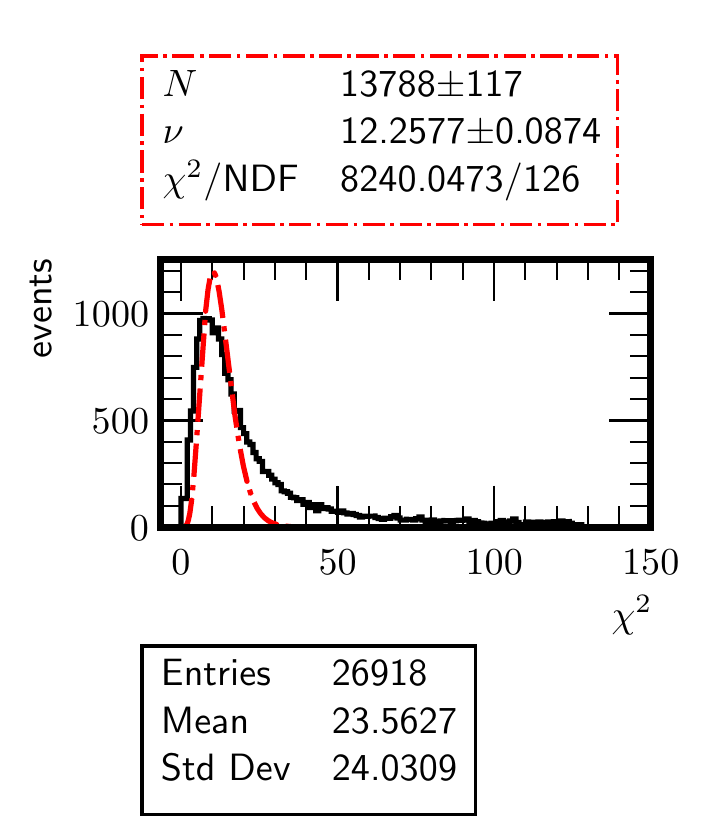}
    \caption{\label{fig:kf-chi2-2pi2gamma}}
  \end{subfigure}
  \hspace{1em}
  \begin{subfigure}[t]{0.47\textwidth}
    \centering
    \includegraphics[width=\textwidth]{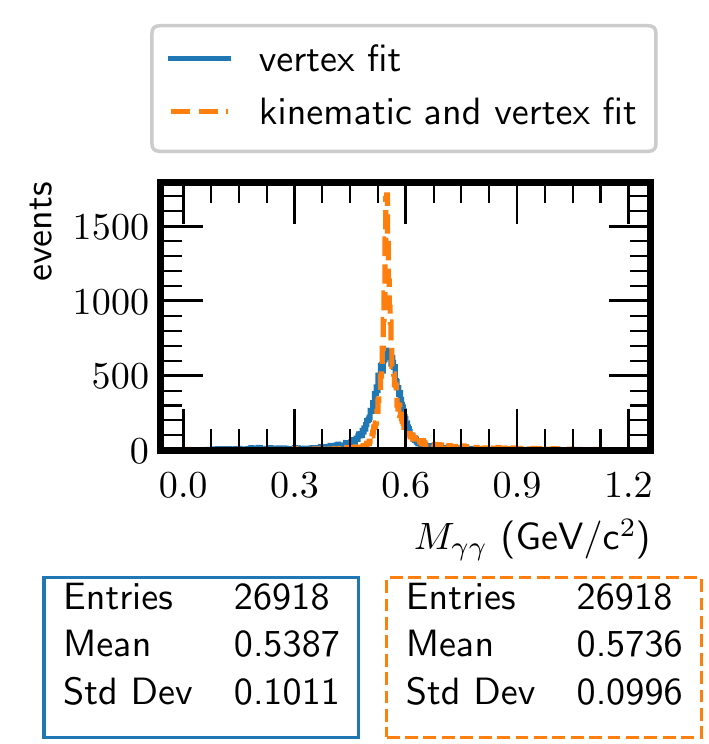}
    \caption{\label{fig:mgg-2pi2gamma}}
  \end{subfigure}
  \caption{Chi-square distribution (a) and two-photon invariant mass distributions (b).
    The distributions are obtained by fitting the events of the
    $e^+e^-\rightarrow\eta\pi^+\pi^-$, $\eta\rightarrow\gamma\gamma$ simulation under
    the $e^+e^-\rightarrow\pi^+\pi^-\gamma\gamma$ hypothesis. The dashed line in
    figure~(b) corresponds to the invariant mass distribution obtained using the
    parameters found with the kinematic and vertex fitting procedure. The solid line
    corresponds to the invariant mass distribution obtained using the vertex fit only.
    This fit is applied to the two tracks of charged pions in the drift chamber. In the
    case of the latter distribution, the photons are assumed to propagate from the
    vertex found by vertex fitting.\label{fig:fit-2pi2gamma}}
  \end{figure}

In this subsection we present the results of applying the fitting package to
the events of the
$e^+e^-\rightarrow\eta\pi^+\pi^-$, $\eta\rightarrow\gamma\gamma$ simulation. These
events correspond to the center-of-mass energy of $1.84$~GeV. For
demonstration purposes, only those events are passed to the input of the fitting
procedure, in which there are at least two photon clusters in the calorimeter.
If there are only two photon clusters in an event, then a single fit is
performed in this event. If the number of photon clusters is more than two, then
separate fits are performed for each different mapping of photons to their
clusters. Permutations of two photon clusters in each pair are not considered,
because they do not affect the fitting result. In the case of the considered
example, the fitting procedure converged in $26918$ events out of $29199$.
In $2199$ events the fitting procedure did not converge, while in $82$ events
this procedure converged to a local maximum. Some fitting results are shown
in figure~\ref{fig:fit-2pi2gamma}.

Figure~\ref{fig:kf-chi2-2pi2gamma} shows the chi-square distribution corresponding to the
example discussed in this section. The mean value of the distribution is approximately
equal to~$23$, while the excepted number of degrees of freedom is~$m - l = 10 - 2 = 8$.
As in the previous examples, the mean value
of this distribution does not match the expected number of degrees of
freedom. The reasons for the distribution being distorted are the same as those
listed in section~\ref{sec:kskpi-xkpi}.

The fitting procedure significantly improves the resolution of the
two-photon invariant mass. The distributions of the two-photon invariant mass are shown in
figure~\ref{fig:mgg-2pi2gamma}. The distribution indicated by the dashed line was obtained
using the particle parameters found with the kinematic and vertex fitting procedure. The
distribution shown by the solid line was obtained using the measured photon parameters.
In the case of the latter distribution, the vertex is found from the vertex fit of the pair
of tracks in the drift chamber.

\subsection{Hypothesis $e^+e^-\rightarrow\eta\pi^+\pi^-,\;\eta\rightarrow\pi^+\pi^-\pi^0_{\text{lost}}$ \label{sec:etapipi-etato3pi}}
\subsubsection{Description of the $e^+e^-\rightarrow\eta\pi^+\pi^-,\;\eta\rightarrow\pi^+\pi^-\pi^0_{\text{lost}}$ hypothesis}
In section~\ref{sec:etapipi-etato3pi}, we discuss the
$e^+e^-\rightarrow\eta\pi^+\pi^-,\;\eta\rightarrow\pi^+\pi^-\pi^0_{\text{lost}}$
hypothesis. In this hypothesis the $\eta$-meson is not
explicitly contained in the form of an intermediate particle. This is due to its
short life time. The presence of a short-lived intermediate particle in this
case is emulated by the mass constraint~\eqref{eq:mass-constraint}. The
considered hypothesis has the following structure.

\begin{itemize}
\item The hypothesis has a single vertex. This vertex is the $e^+e^-$ interaction vertex.
  All coordinates of this vertex are considered as free
  measurable parameters and contribute to the chi-square.
\item The hypothesis contain five final particles. The four final particles are
  charged pions~(see section~\ref{sec:charged-particle}), and the fifth particle is
  a lost $\pi^0$. This $\pi^0$ is represented as the lost massive
  particle described in section~\ref{sec:massive-lost-particle}. In order to provide the
  four-momentum of initial particles, the hypothesis also contains the initial
  pseudo-particle~(see section~\ref{sec:initial-pseudo-particle}).
\item All of the particles listed above are involved in four energy-momentum
  conservation constraints.
\item Each charged pion has three vertex constraints. In total, the hypothesis
  contains $12$ vertex constraints.
\item Three pions ($\pi^+\pi^-\pi^0$) are involved in the mass constraint.
\end{itemize}
In total, the $e^+e^-\rightarrow\eta\pi^+\pi^-$,
$\eta\rightarrow\pi^+\pi^-\pi^0_{\text{lost}}$ hypothesis contains $17$ constraints.
This hypothesis contains $30$ free parameters. Among these parameters, $23$ parameters are measurable and
$7$ parameters are non-measurable.

\subsubsection{Fitting procedure details \label{sec:etapipi-etato3pi-fitting-procedure}}
\begin{table}[tbp]
  \centering
  \caption{Mappings between charged particles and their tracks in the case of
    the $e^+e^-\rightarrow\eta\pi^+\pi^-,\;\eta\rightarrow\pi^+\pi^-\pi^0_{\text{lost}}$ hypothesis.
    $t^+_1$ and $t^+_2$ are tracks of positively charged particles, $t^-_1$ and $t^-_2$ are tracks
    of negatively charged particles.
    \label{tab:etapipi-etato3pi}}
  \begin{tabular}[t]{V{4}cV{3}c|cV{2}c|c|cV{4}}
    \hlineB{4}
    \multicolumn{1}{V{4}lV{3}}{Intermediate particle} &
    \multicolumn{2}{cV{2}}{} &
    \multicolumn{3}{cV{4}}{$\eta$}\\
    \hlineB{2}
    \multicolumn{1}{V{4}lV{3}}{Final particle} &
    $\pi^+$ & $\pi^-$ &
    $\pi^+$ & $\pi^-$ &
    $\pi^0_{\text{lost}}$ \\
    \hlineB{3}
    \multirow{4}{*}{\begin{minipage}{2cm} Track combinations\end{minipage}} &
    $t^+_1$ & $t^-_1$ &
    $t^+_2$ & $t^-_2$ &\\

    & $t^+_1$ & $t^-_2$ &
    $t^+_2$ & $t^-_1$ &\\

    & $t^+_2$ & $t^-_1$ &
    $t^+_1$ & $t^-_2$ &\\

    & $t^+_2$ & $t^-_2$ &
    $t^+_1$ & $t^-_1$ &\\
    \hlineB{4}
  \end{tabular}
\end{table}
For demonstration purposes, only events with four tracks are passed to
the input of the fitting procedure. Moreover, these tracks must correspond
to a neutral combination of charged particles.
Due to the presence of the mass constraint, charged pions are not equivalent in the
$e^+e^-\rightarrow\eta\pi^+\pi^-,\;\eta\rightarrow\pi^+\pi^-\pi^0_{\text{lost}}$
hypothesis. This is due to the fact that only two charged pions are involved in
this constraint. Therefore, the fitting procedure under the considered
hypothesis must take into account different mappings of charged pions into their
tracks. These mappings are listed in table~\ref{tab:etapipi-etato3pi}. Among all
possible combinations from this table, the combination corresponding to the minimum
chi-square is selected. Note that in order for the point of initial optimization parameters
to lie closer to the conditional minimum point, the initial momentum of the lost
$\pi^0$ is set equal to the initial missing momentum of four charged pions.

\subsubsection{Fitting the $e^+e^-\rightarrow\eta\pi^+\pi^-,\;\eta\rightarrow\pi^+\pi^-\pi^0$ events}
\begin{figure}[tbp]
  \centering
  \begin{subfigure}[t]{0.47\textwidth}
    \centering
    \includegraphics[width=\textwidth]{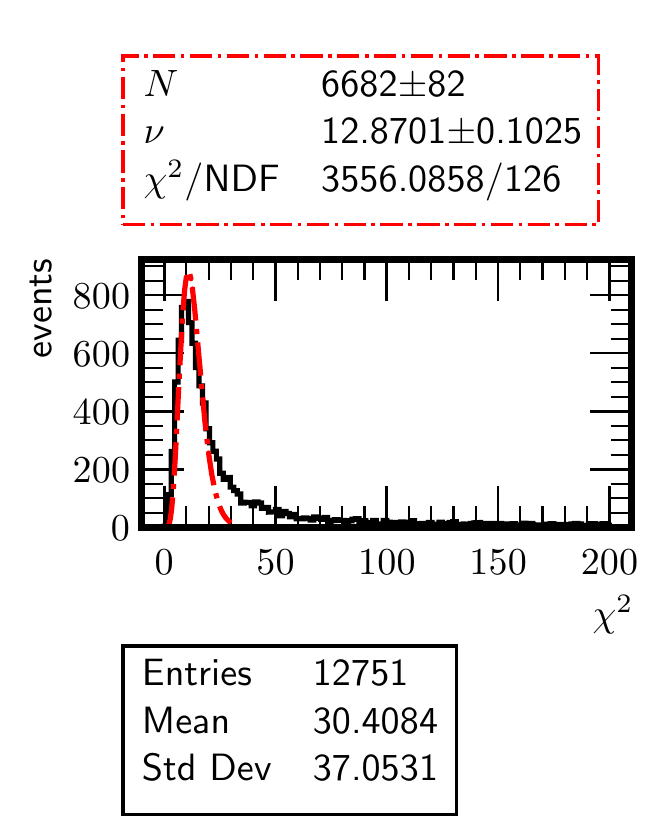}
    \caption{\label{fig:kf-chi2-4pipi0lost-eta-mass}}
  \end{subfigure}
  \hspace{1em}
  \begin{subfigure}[t]{0.47\textwidth}
    \centering
    \includegraphics[width=\textwidth]{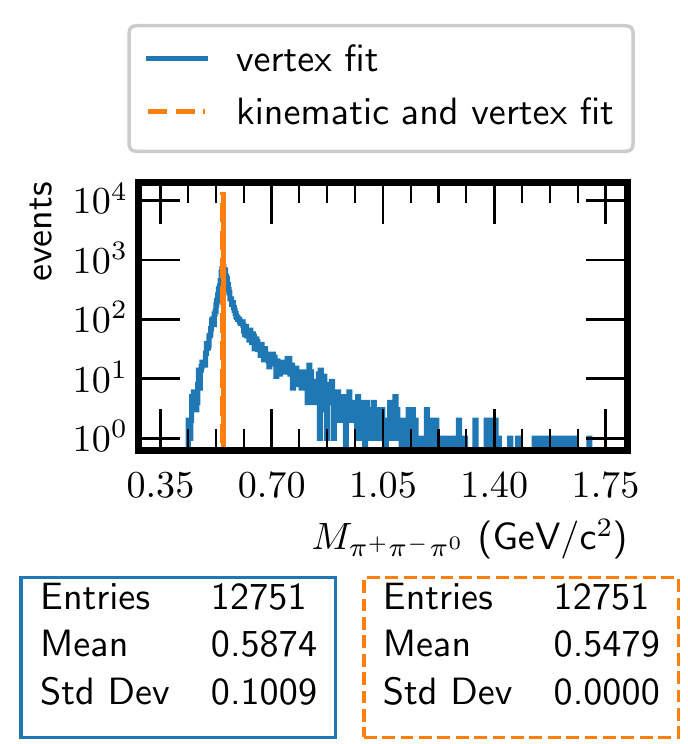}
    \caption{\label{fig:m3pi-3pipi0lost-eta-mass}}
  \end{subfigure}
  \caption{Chi-square distribution (a) and three-pion invariant mass distributions (b).
    The distributions are obtained by fitting the events of the
    $e^+e^-\rightarrow\eta\pi^+\pi^-$, $\eta\rightarrow\pi^+\pi^-\pi^0$ simulation under
    the $e^+e^-\rightarrow\eta\pi^+\pi^-$, $\eta\rightarrow\pi^+\pi^-\pi^0_{\text{lost}}$
    hypothesis. The dashed line in figure~(b) corresponds to the invariant mass
    distribution obtained using the parameters found with the kinematic and vertex
    fitting procedure. The solid line corresponds to the invariant mass distribution
    obtained using the vertex fit only. This fit is applied to the four tracks of
    charged pions in the drift chamber. In the case of the latter distribution, the
    missing momentum of the four charged pions is chosen as the momentum of the
    neutral pion.}
\end{figure}
The fitting procedure described above is applied to the events of the
$e^+e^-\rightarrow\eta\pi^+\pi^-,\;\eta\rightarrow\pi^+\pi^-\pi^0,\;\pi^0\rightarrow\gamma\gamma$
simulation. These events correspond to the center-of-mass energy of $1.88$~GeV.
For demonstration purposes, we completely ignore the information about the presence
of photons, i.e. even if photons from the $\pi^0\rightarrow\gamma\gamma$ decay were detected,
we still consider the neutral pion to be lost. The fitting
procedure converged to a local minimum in $12751$ events out of $12796$. This
procedure converged to a local maximum in $8$ events and did not converge in
$37$ events.

The corresponding chi-square distribution is shown in
figure~\ref{fig:kf-chi2-4pipi0lost-eta-mass}. The mean value of this
distribution is approximately equal to $37$, while the expected number
of degrees of freedom is $m - l = 17 - 7 = 10$. That is, the distribution
shown in the figure does not fit the chi-squared probability density function
with $10$ degrees of freedom. The fact that the chi-square distribution is
distorted is due to the reasons listed in section~\ref{sec:kskpi-xkpi}.

Figure~\ref{fig:m3pi-3pipi0lost-eta-mass} shows the distributions of the
three-pion invariant mass. These three pions are the
$\eta\rightarrow\pi^+\pi^-\pi^0$ decay products. The distribution indicated by
the dashed line was obtained using the particle parameters found with the
kinematic and vertex fitting procedure. This distribution turned out to be fixed
on the $\eta$-meson mass, since the corresponding mass constraint is used in the
hypothesis. The distribution indicated by the solid line was obtained using the
parameters of the charged pions found from the corresponding vertex fit. In the case of
the latter distribution, it is assumed that the momentum of the lost neutral pion
was set equal to the missing momentum of the four charged pions.

\section{Examples of Gaussian simulation\label{sec:gaussian-simulation-examples}}
In this section, we discuss kinematic and vertex fitting of events obtained using
Gaussian simulation described in section~\ref{sec:gaussian-simulation}.
Subsection~\ref{sec:etapipi-gsim} provides examples of Gaussian simulations of the
$e^+e^-\rightarrow\eta\pi^+\pi^-,\;\eta\rightarrow\gamma\gamma$ process. The events of
these simulations are fitted under the $e^+e^-\rightarrow\pi^+\pi^-\gamma\gamma$
hypothesis. Subsection~\ref{sec:kskpi-gsim} provides examples of Gaussian simulations
of the $e^+e^-\rightarrow K_S K^{\pm}\pi^{\mp},\;K_S\rightarrow\pi^+\pi^-$ process. The events
of these simulations are fitted under the
$e^+e^-\rightarrow X K^{\pm}\pi^{\mp},\;X\rightarrow\pi^+\pi^-$ hypotheses. The examples
discussed in this section are of interest for testing the fitting package, as well as for
observing how constraint nonlinearity can distort a chi-square distribution.

\subsection{Hypothesis $e^+e^-\rightarrow\pi^+\pi^-\gamma\gamma$ \label{sec:etapipi-gsim}}
\begin{figure}[tbp]
  \centering
  \includegraphics[width=0.8\textwidth]{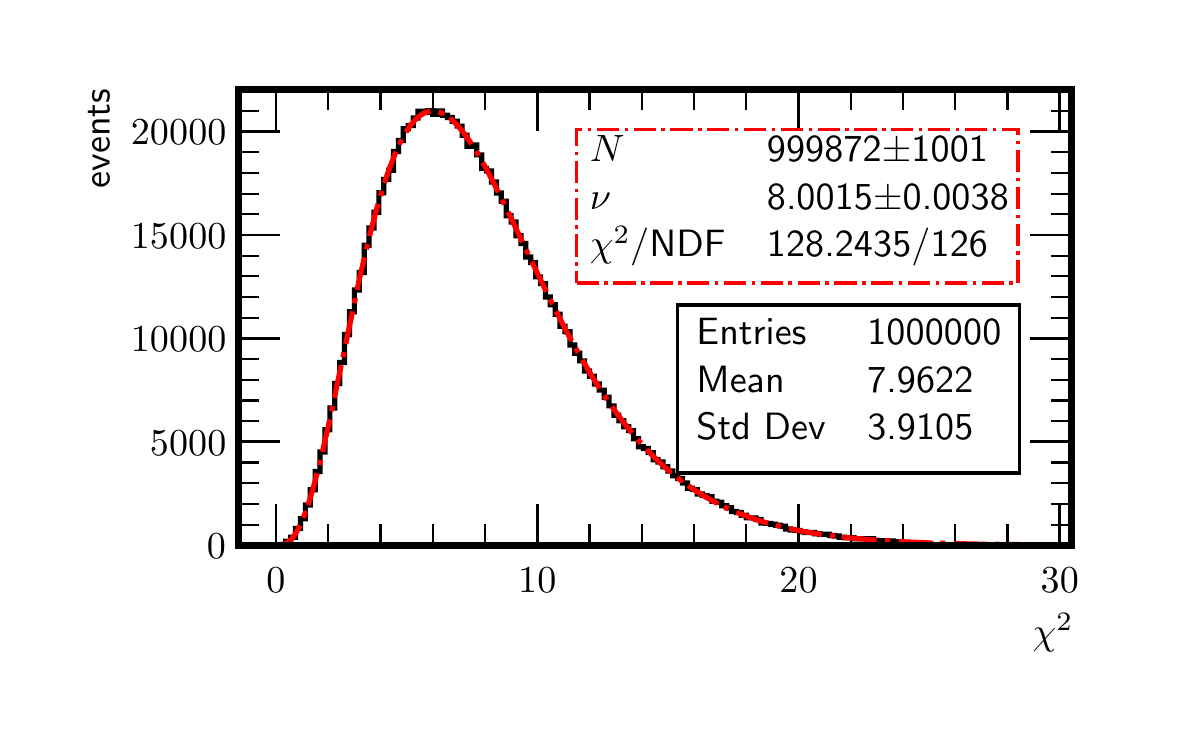}
  \caption{Chi-square distribution in the case of the
    $e^+e^-\rightarrow\eta\pi^+\pi^-,\;\eta\rightarrow\gamma\gamma$ Gaussian simulation.
    The events of this simulation were fitted under the $e^+e^-\rightarrow\pi^+\pi^-\gamma\gamma$
    hypothesis~(see section~\ref{sec:2pi2gamma}). All the constraints
    corresponding to this hypothesis were enabled. The solid line is the $\chi^2$ histogram, and
    the dash-dotted line is the fitting function $b{N}f_{\chi^2}(t;\nu)$ proportional to the chi-square
    PDF with $\nu$ degrees of freedom. Constant $b$ is the bin width. Number of events
    $N$ and the number degrees of freedom $\nu$ are the free fitting parameters.
    \label{fig:chi2-gsim-etapipi}}
\end{figure}
This subsection presents the results of fitting the events of the Gaussian simulation of
the $e^+e^-\rightarrow\eta\pi^+\pi^-,\;\eta\rightarrow\gamma\gamma$ process. As the initial
event, a certain event of the $e^+e^-\rightarrow\eta\pi^+\pi^-,\;\eta\rightarrow\gamma\gamma$
Monte Carlo simulation is chosen. It should be noted that the same Monte Carlo simulation
is used as in section~\ref{sec:etapipi-2pi2gamma}. The kinematic and vertex fitting procedure
under the $e^+e^-\rightarrow\pi^+\pi^-\gamma\gamma$ hypothesis~(see section~\ref{sec:2pi2gamma}) was applied
once to the initial event. This procedure made it possible to obtain measurable particle parameters
satisfying the constraints. Then the initial event was redefined: the measured parameters
of the particles were replaced by the corresponding parameters obtained with the fitting
procedure (see section~\ref{sec:gaussian-simulation}). Then these
parameters were drawn multiple times according to the corresponding multivariate normal
distribution. As a result of such draws, the events of the Gaussian simulation were obtained.
These events were fitted under the $e^+e^-\rightarrow\pi^+\pi^-\gamma\gamma$ hypothesis.
Figure~\ref{fig:chi2-gsim-etapipi} shows the chi-square distribution corresponding to the
events of this Gaussian simulation. This distribution corresponds to the case when all constraints are
enabled in the $e^+e^-\rightarrow\pi^+\pi^-\gamma\gamma$ hypothesis. A list of these constraints
is given in section~\ref{sec:2pi2gamma}. The hypothesis contains $m = 10$ constraints
and $l = 2$ free non-measurable parameters. If all constraints were linear, one would expect
the chi-square distribution to be described by the chi-squared probability density
function~\eqref{eq:chi-square-pdf} with $m -l = 8$ degrees of
freedom (see section~\ref{sec:case-of-nonzero-l}). However, all the constraints of the considered
hypothesis are nonlinear. Despite this, the chi-square distribution is well described by the chi-squared
probability density function. This statement was verified by fitting the distribution to the probability
density function~\eqref{eq:chi-square-pdf} multiplied by the normalization factor $bN$, where $b$ is
the bin width of the chi-square histogram and $N$ is the number of events in the Gaussian simulation. In
the fitting function, the number of degrees of freedom and $N$ are considered as free parameters. The chi-square
histogram in figure~\ref{fig:chi2-gsim-etapipi} is indicated by a solid line, while the fitting function
is indicated by a dash-dotted line. The fitting parameters are shown in the
figure inside the dash-dotted box.
The fitting
function indeed describes the histogram well. The number of degrees found with the fit is $\nu=8.0015\pm0.0038$,
i.e. it is equal to $8$ with high accuracy.

\begin{figure}[tbp]
  \centering
  \begin{subfigure}[t]{0.47\textwidth}
    \centering
    \includegraphics[width=\textwidth]{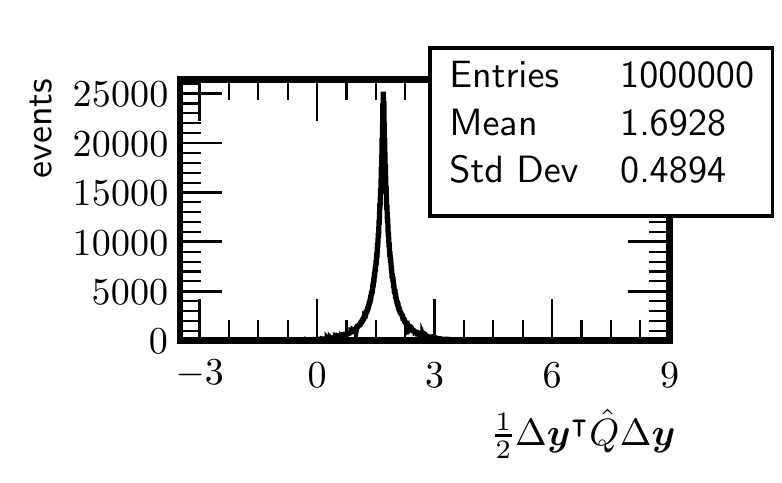}
    \caption{\label{fig:q-form-gsim-etapipi}}
  \end{subfigure}
  \hspace{1em}
  \begin{subfigure}[t]{0.47\textwidth}
    \centering
    \includegraphics[width=\textwidth]{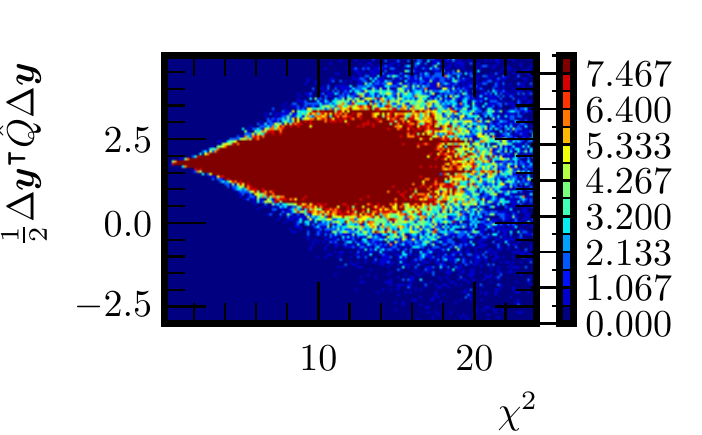}
    \caption{\label{fig:q-form-vs-chi2-gsim-etapipi}}
  \end{subfigure}
  \caption{Distribution~(a) of the quadratic form $\frac{1}{2}\Delta\bm{y}^{\intercal}\hat{Q}\Delta\bm{y}$ and
    two-dimensional distribution~(b): $\frac{1}{2}\Delta\bm{y}^{\intercal}\hat{Q}\Delta\bm{y}$ versus $\chi^2$.
    These distributions correspond to the events
    of the $e^+e^-\rightarrow\eta\pi^+\pi^-,\;\eta\rightarrow\gamma\gamma$ Gaussian simulation.
    These events were fitted under $e^+e^-\rightarrow\pi^+\pi^-\gamma\gamma$
    hypothesis. All constraints of this hypothesis were enabled.}
\end{figure}
To estimate the magnitude of the nonlinearity of the constraints, one can obtain
the distribution of the quadratic form
$\frac{1}{2}\Delta\bm{y}^{\intercal}\hat{Q}\Delta\bm{y}$, where $\hat{Q}$ is
the $\hat{Q}_{s - 1}$ from equations~\eqref{eq:hessian} and~\eqref{eq:q-s-1}
calculated after the minimization is completed. As noted above, the matrix
$\hat{Q}$ contains the constraint Hessians. In the case of linear constraints,
this matrix is zero. The distribution of the quadratic form
$\frac{1}{2}\Delta\bm{y}^{\intercal}\hat{Q}\Delta\bm{y}$ is shown in
figure~\ref{fig:q-form-gsim-etapipi}. This distribution was obtained under the conditions
described in the previous paragraph. It can be seen from the figure that the
mean value of the quadratic form
$\frac{1}{2}\Delta\bm{y}^{\intercal}\hat{Q}\Delta\bm{y}$ is approximately equal
to $1.7$, which is significantly less than the mean value~($\nu=8$) of the chi-square
distribution. The standard deviation of the considered distribution  is approximately
equal to $0.49$. From equation~\eqref{eq:hessian} it is clear that in order to estimate the magnitude
of the nonlinearity of the constraints, the quadratic form
$\frac{1}{2}\Delta\bm{y}^{\intercal}\hat{Q}\Delta\bm{y}$ should be compared with
the chi-square $\chi^2=\Delta\bm{x}^{\intercal}\hat{\tilde{C}}^{-1}\Delta\bm{x}$.
The corresponding two-dimensional histogram is shown in
figure~\ref{fig:q-form-vs-chi2-gsim-etapipi}. It can be seen from the figure that the
quadratic form $\frac{1}{2}\Delta\bm{y}^{\intercal}\hat{Q}\Delta\bm{y}$ is
usually several times less than the chi-square. It is shown below that in those
cases where the nonlinearity of the constraints leads to a significant distortion
of the chi-square distribution, the
$\frac{1}{2}\Delta\bm{y}^{\intercal}\hat{Q}\Delta\bm{y}$ distribution is
significantly wider.

\begin{figure}[tbp]
  \centering
  \begin{subfigure}[t]{0.47\textwidth}
    \centering
    \includegraphics[width=\textwidth]{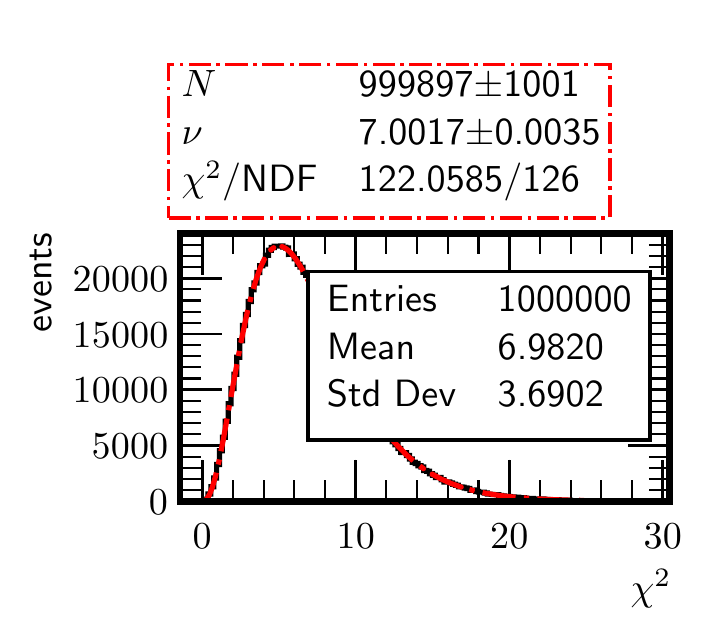}
    \caption{\label{fig:chi2-vertex-and-momentum-gsim-etapipi}}
  \end{subfigure}
  \hspace{1em}
  \begin{subfigure}[t]{0.47\textwidth}
    \centering
    \includegraphics[width=\textwidth]{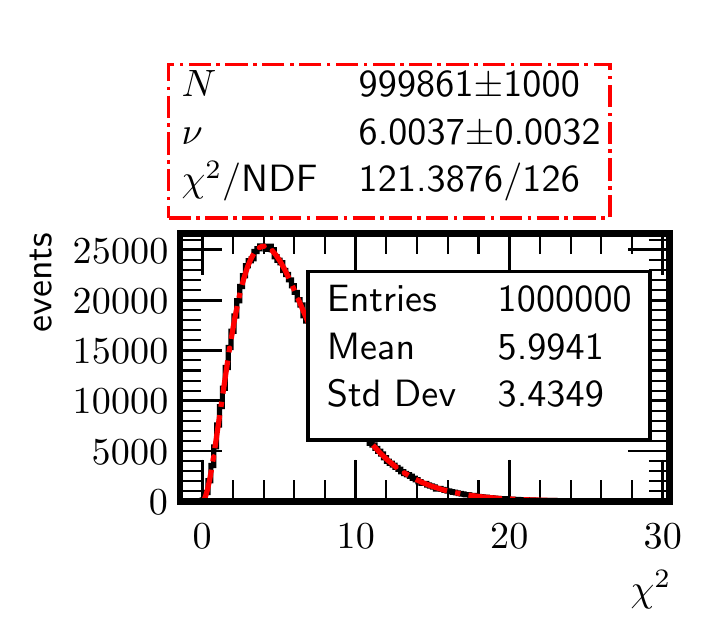}
    \caption{\label{fig:chi2-vertex-and-xy-momentum-gsim-etapipi}}
  \end{subfigure}
  \begin{subfigure}[t]{0.47\textwidth}
    \centering
    \includegraphics[width=\textwidth]{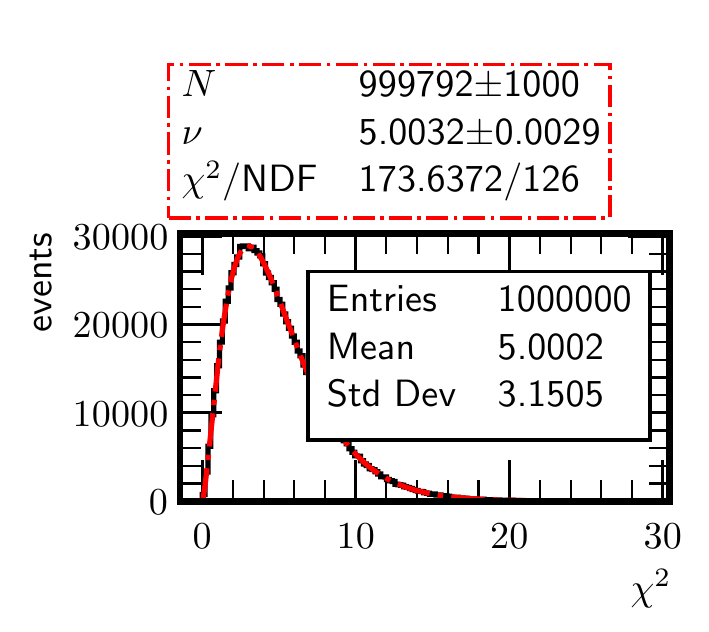}
    \caption{\label{fig:chi2-vertex-and-z-momentum-gsim-etapipi}}
  \end{subfigure}
  \hspace{1em}
  \begin{subfigure}[t]{0.47\textwidth}
    \centering
    \includegraphics[width=\textwidth]{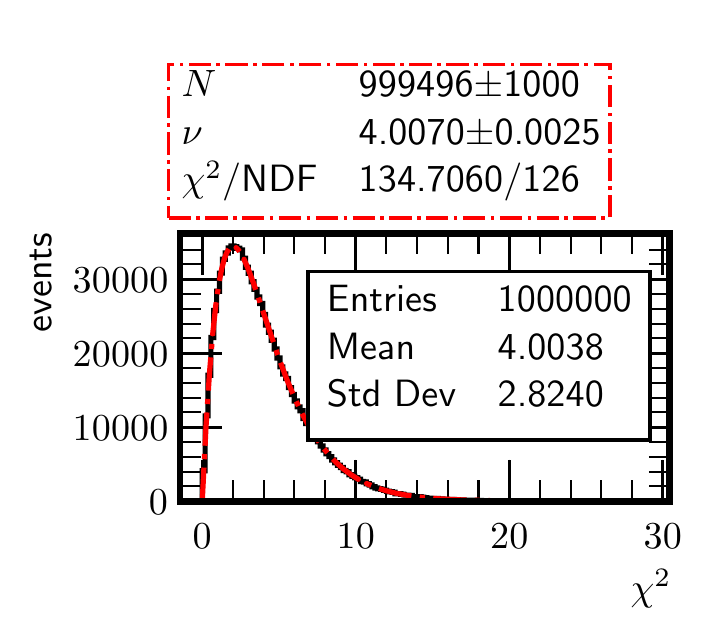}
    \caption{\label{fig:chi2-vertex-gsim-etapipi}}
  \end{subfigure}
  \caption{Chi-square distributions. These distributions correspond to the events of the
    $e^+e^-\rightarrow\eta\pi^+\pi^-,\;\eta\rightarrow\gamma\gamma$ Gaussian simulation. The events of
    this simulation were fitted under the $e^+e^-\rightarrow\pi^+\pi^-\gamma\gamma$
    hypothesis. During the fitting, some constraints were disabled.
    Figure~(a) corresponds to the case when the energy conservation constraint is disabled.
    Figure~(b) corresponds to the case when the energy and $z$-momentum conservation constraints are disabled.
    Figure~(c) corresponds to the case when the  energy, $x\text{-}$ and $y$-momentum conservation constraints are disabled.
    Finally, figure~(d) corresponds to the case when all four-momentum conservation constraints are disabled.
    The solid line indicates the $\chi^2$ histogram, while the dash-dotted line indicates the fitting function,
    similar to that used in figure~\ref{fig:chi2-gsim-etapipi}. The parameters of the fitting functions,
    as well as the fit quality ratios $\chi^2/\text{NDF}$, are shown in the dash-dotted boxes.}
\end{figure}
Chi-square distributions similar to those shown in figure~\ref{fig:chi2-gsim-etapipi} can be obtained when some
of the constraints are disabled while fitting the events of the Gaussian simulation.
So, for example, figure~\ref{fig:chi2-vertex-and-momentum-gsim-etapipi} shows
the distribution in the case when the energy conservation constraint is disabled.
Figure~\ref{fig:chi2-vertex-and-xy-momentum-gsim-etapipi} shows the chi-square distribution in the case when the
energy conservation and $z$-momentum conservation constraints are disabled.
Figure~\ref{fig:chi2-vertex-and-z-momentum-gsim-etapipi} shows the chi-square
distribution in the case when the energy, $x$-momentum, and $y$-momentum conservation constraints are disabled.
Finally, figure~\ref{fig:chi2-vertex-gsim-etapipi} shows the distribution in the case when all four-momentum
conservation constraints are disabled. If the constraints were linear, one would expect the listed distributions
to follow the chi-square distribution~\eqref{eq:chi-square-pdf} with $m - l$ degrees of freedom. Despite the fact
that the constraints are nonlinear, the distributions in figures~\ref{fig:chi2-vertex-and-momentum-gsim-etapipi},
\ref{fig:chi2-vertex-and-xy-momentum-gsim-etapipi},
\ref{fig:chi2-vertex-and-z-momentum-gsim-etapipi}
and~\ref{fig:chi2-vertex-gsim-etapipi} are consistent with the chi-squared
probability density function~\eqref{eq:chi-square-pdf}. The numbers of degrees of freedom in these cases are $7$,
$6$, $5$ and $4$, respectively.

As mentioned above, a certain single event of the $e^+e^-\rightarrow\pi^+\pi^-\eta,\;\eta\rightarrow\gamma\gamma$
Monte Carlo simulation was chosen as the initial event. However, it should be noted that the fitting results given
in this subsection do not depend on the choice of the initial event. However, in
the cases where the nonlinearity of the
constraints can manifest itself strongly, the results of the Gaussian simulation can significantly depend on the choice
of the initial event. Such a case is described in detail in the next subsection.

\subsection{Hypotheses $e^+e^-\rightarrow X K^{\pm}\pi^{\mp},\;X\rightarrow\pi^+\pi^-$ \label{sec:kskpi-gsim}}
\begin{figure}[tbp]
  \centering
  \begin{subfigure}[t]{0.47\textwidth}
    \centering
    \includegraphics[width=\textwidth]{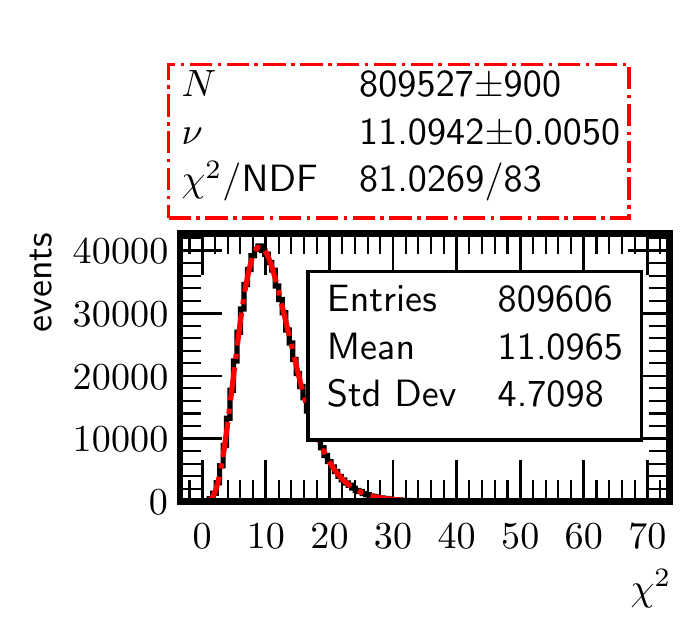}
    \caption{\label{fig:chi2-gsim-kskpi-good}}
  \end{subfigure}
  \hspace{1em}
  \begin{subfigure}[t]{0.47\textwidth}
    \centering
    \includegraphics[width=\textwidth]{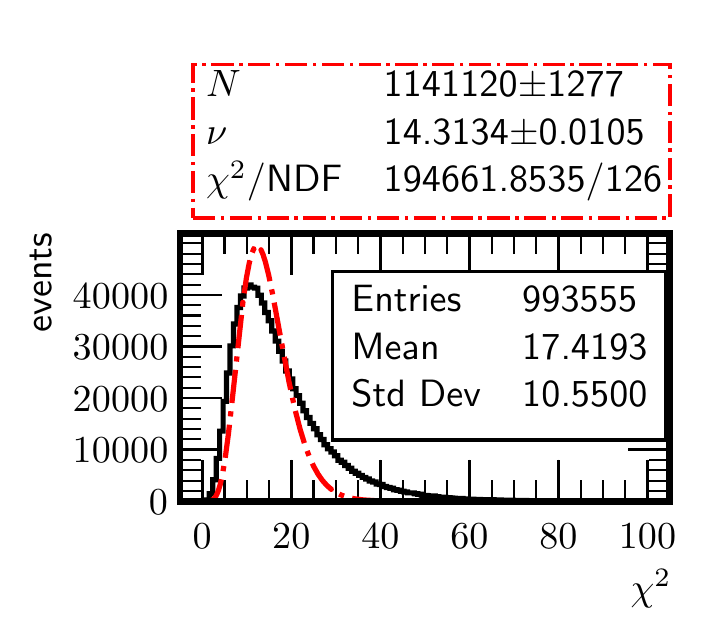}
    \caption{\label{fig:chi2-gsim-kskpi-bad1}}
  \end{subfigure}
  \begin{subfigure}[t]{0.47\textwidth}
    \centering
    \includegraphics[width=\textwidth]{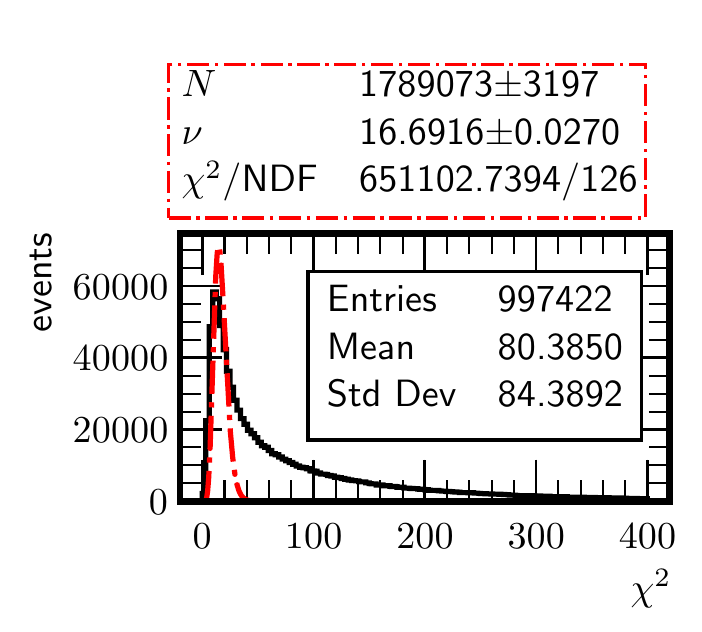}
    \caption{\label{fig:chi2-gsim-kskpi-bad2}}
  \end{subfigure}
  \hspace{1em}
  \begin{subfigure}[t]{0.47\textwidth}
    \centering
    \includegraphics[width=\textwidth]{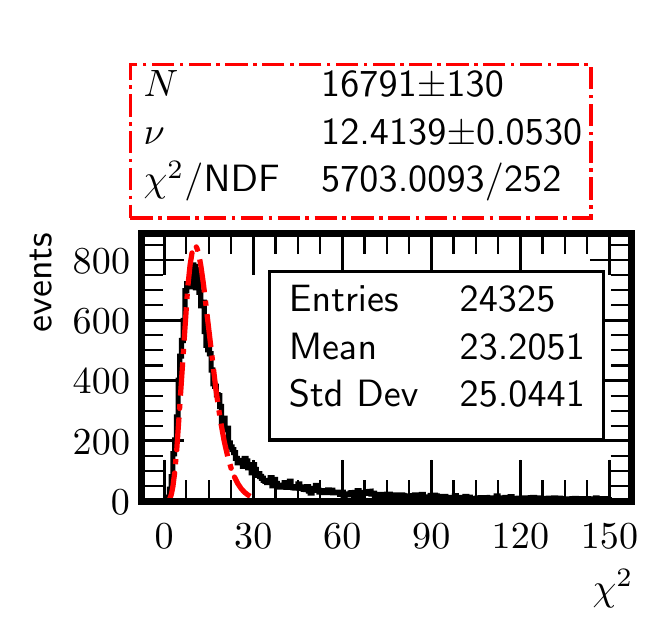}
    \caption{\label{fig:chi2-gsim-kskpi-all-events}}
  \end{subfigure}
  \caption{Chi-square distributions corresponding to the events of the
  $e^+e^-\rightarrow K_S K^{\pm}\pi^{\mp},\;K_S\rightarrow\pi^+\pi^-$ Gaussian
    simulations. These events were fitted under the correct signal hypothesis:
  $e^+e^-\rightarrow X K^+\pi^-,\;X\rightarrow\pi^+\pi^-$ or
    $e^+e^-\rightarrow X K^-\pi^+,\;X\rightarrow\pi^+\pi^-$. The $\chi^2$ histogram is
    indicated by a solid line, while the dash-dotted line is the fitting function. This
    fitting function is proportional to the chi-squared probability density
    function~\eqref{eq:chi-square-pdf}.
    Figure~(a) corresponds to the case of the Gaussian based on one of the Monte Carlo
    simulation events. Figures~(b) corresponds to the case of the Gaussian simulation
    based on another Monte Carlo simulation event. Figure~(c) corresponds to the case
    of the Gaussian simulation based on the third event of the Monte Carlo simulation.
    Finally, figure~(d) corresponds to the case of the Gaussian simulation based on
    all Monte Carlo simulation events. The fitting function describes the histogram
    well only in figure~$\text{(a)}$.}
\end{figure}
This subsection discuss the fitting results corresponding to a number of the
$e^+e^-\rightarrow K_S K^{\pm}\pi^{\mp}$, $K_S\rightarrow\pi^+\pi^-$ Gaussian simulations.
Each Gaussian simulation corresponds to a different initial event. As in the previous
subsection, the events of the corresponding Monte Carlo simulation~(see section~\ref{sec:kskpi-xkpi})
were used as these initial events. Further, to obtain the events of each Gaussian simulation,
a procedure similar to that described in the first paragraph of the previous subsection was
performed. In the case of the examples presented below, one of the
$e^+e^-\rightarrow X K^{\pm}\pi^{\mp}$, $X\rightarrow\pi^+\pi^-$ hypotheses is used, while
fitting the simulated events. Moreover, when fitting each simulation event, only the correct
signal hypothesis is used to avoid misassignments of kaons and pions.

Figure~\ref{fig:chi2-gsim-kskpi-good} shows the chi-square distribution corresponding to the
events of the first Gaussian simulation. The parameters of the particles and their covariance
matrices in the corresponding initial event are such that the effects of the constraints
nonlinearity are not large enough to lead to a significant
distortion of the chi-square distribution. It is
seen from the figure that the chi-square histogram is well described by the probability density
function~\eqref{eq:chi-square-pdf}. Moreover, the number of degrees of freedom is equal to $11$,
as expected.

In the case of the Gaussian simulation obtained using another initial event, the parameters of the
particles in this initial event turned out to be such that they led to significant nonlinearity
effects. The corresponding chi-square histogram is shown in figure~\ref{fig:chi2-gsim-kskpi-bad1}.
It can be seen from the figure that this histogram is not described by the probability density
function~\eqref{eq:chi-square-pdf}.

Figure~\ref{fig:chi2-gsim-kskpi-bad2} shows the chi-square histogram for an even more dramatic case than
in figure~\ref{fig:chi2-gsim-kskpi-bad1}. This figure corresponds to the events of the third Gaussian
simulation obtained using a third initial event different from the first two. In this case, the
distribution is much wider than in figures~\ref{fig:chi2-gsim-kskpi-good}
and~\ref{fig:chi2-gsim-kskpi-bad1}. This distribution is not described by the probability density
function~\eqref{eq:chi-square-pdf}.

Although the nonlinearity of the constraints tends to a wider chi-square distribution, even in the worst
case shown in figure~\ref{fig:chi2-gsim-kskpi-bad2}, the chi-square distribution peaks at $\chi^2\sim 10$.
Comparing this distribution with the chi-square distribution shown in figure~\ref{fig:kf-chi2-4pi-xkpi} and
obtained for background events, one can conclude that even in the worst case, signal and background events
can be separated using a chi-square selection criterion. However, it is clear that the separation power is
the smaller, the wider the signal chi-square distribution.

In connection with the above, the question arises, what is the ratio between the number of events in the
Monte Carlo simulation, leading to the distributions in figure~\ref{fig:chi2-gsim-kskpi-good} and
figures~\ref{fig:chi2-gsim-kskpi-bad1}, \ref{fig:chi2-gsim-kskpi-bad2}. This question can be answered using a
slightly different kind of Gaussian simulation than the one described above. Instead of doing a Gaussian
simulation based on a single Monte Carlo simulation event, let us draw one Gaussian simulation event for
each event of the Monte Carlo simulation. The chi-square distribution corresponding to such a Gaussian
simulation is shown in figure~\ref{fig:chi2-gsim-kskpi-all-events}. Although this distribution is not
consistent with the probability density function~\eqref{eq:chi-square-pdf}, it can be concluded that the
Monte Carlo simulation is dominated by events that correspond to distributions in
figures~\ref{fig:chi2-gsim-kskpi-good} and~\ref{fig:chi2-gsim-kskpi-bad1}.

\begin{figure}[tbp]
  \centering
  \begin{subfigure}[t]{0.47\textwidth}
    \centering
    \includegraphics[width=\textwidth]{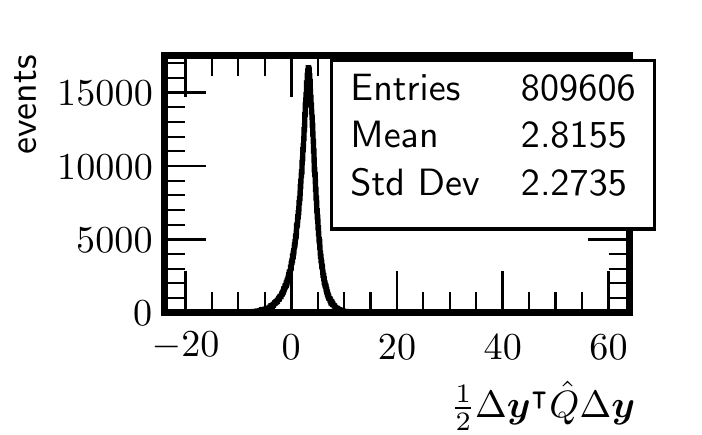}
    \caption{\label{fig:qhalf-gsim-kskpi-good}}
  \end{subfigure}
  \hspace{1em}
  \begin{subfigure}[t]{0.47\textwidth}
    \centering
    \includegraphics[width=\textwidth]{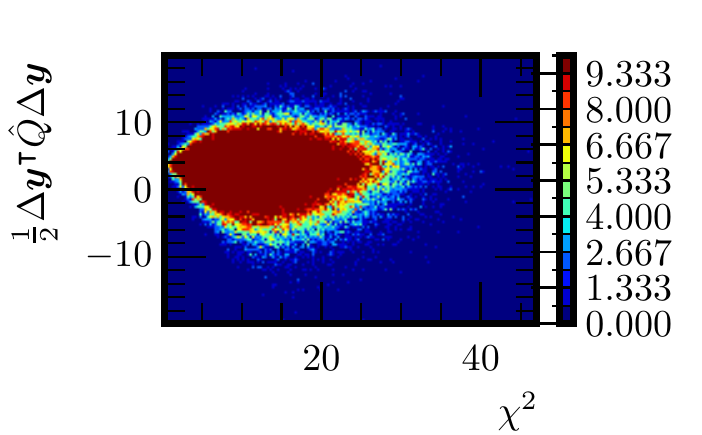}
    \caption{\label{fig:qhalf-vs-chi2-gsim-kskpi-good}}
  \end{subfigure}
  \caption{Distribution~(a) of the quadratic form
    $\frac{1}{2}\Delta\bm{y}^{\intercal}\hat{Q}\Delta\bm{y}$ and two-dimensional distribution~(b):
    $\frac{1}{2}\Delta\bm{y}^{\intercal}\hat{Q}\Delta\bm{y}$ versus $\chi^2$. These distributions
    correspond to the
    events of the $e^+e^-\rightarrow K_S K^{\pm}\pi^{\mp},\;K_S\rightarrow\pi^+\pi^-$
    Gaussian simulation. This Gaussian simulation corresponds to the case of the
    chi-square distribution shown in figure~\ref{fig:chi2-gsim-kskpi-good}.\label{fig:q-form-kskpi-good}}
\end{figure}
Figure~\ref{fig:qhalf-gsim-kskpi-good} shows the
$\frac{1}{2}\Delta\bm{y}^{\intercal}\hat{Q}\Delta\bm{y}$ distribution.
This distribution corresponds to the Gaussian simulation in the case of
figure~\ref{fig:chi2-gsim-kskpi-good}. The mean value of this distribution is approximately equal
to~$2.8$, which is significantly less than the mean value ($\nu=11$) of the corresponding chi-square
distribution. The standard deviation of the considered $\frac{1}{2}\Delta\bm{y}^{\intercal}\hat{Q}\Delta\bm{y}$
distribution is approximately equal to $2.3$. Figure~\ref{fig:qhalf-vs-chi2-gsim-kskpi-good} shows the two-dimensional
distribution of the $\frac{1}{2}\Delta\bm{y}^{\intercal}\hat{Q}\Delta\bm{y}$ versus the $\chi^2$.

\begin{figure}[tbp]
  \centering
  \begin{subfigure}[t]{0.47\textwidth}
    \centering
    \includegraphics[width=\textwidth]{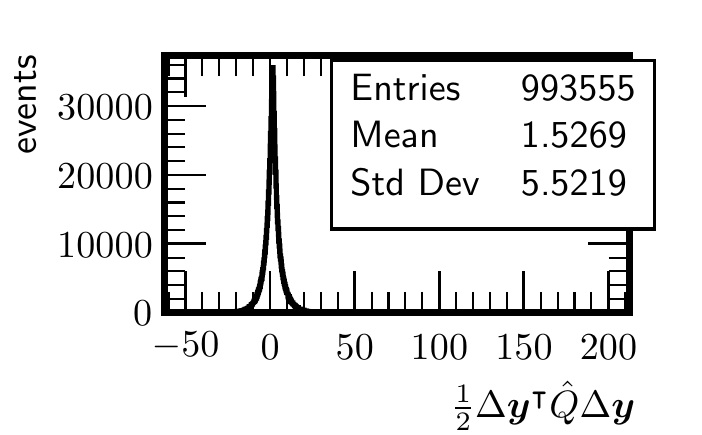}
    \caption{\label{fig:qhalf-gsim-kskpi-bad1}}
  \end{subfigure}
  \hspace{1em}
  \begin{subfigure}[t]{0.47\textwidth}
    \centering
    \includegraphics[width=\textwidth]{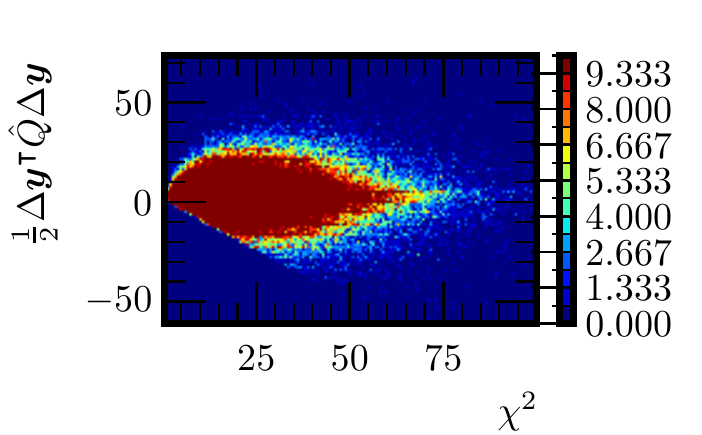}
    \caption{\label{fig:qhalf-vs-chi2-gsim-kskpi-bad1}}
  \end{subfigure}
  \begin{subfigure}[t]{0.47\textwidth}
    \centering
    \includegraphics[width=\textwidth]{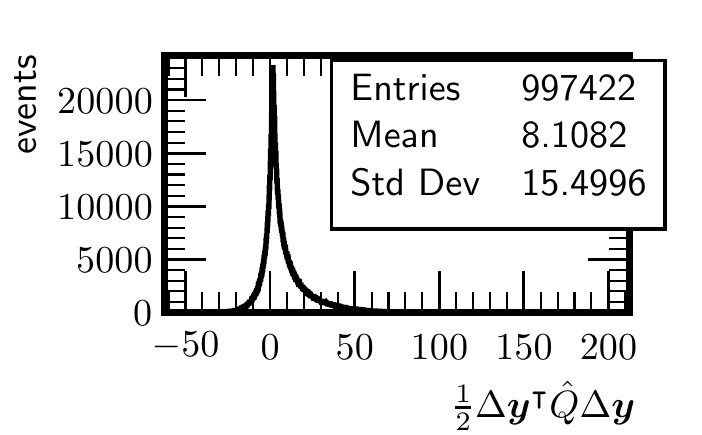}
    \caption{\label{fig:qhalf-gsim-kskpi-bad2}}
  \end{subfigure}
  \hspace{1em}
  \begin{subfigure}[t]{0.47\textwidth}
    \centering
    \includegraphics[width=\textwidth]{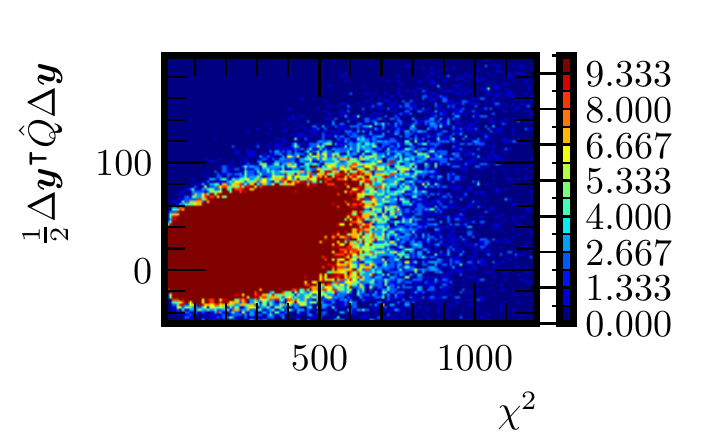}
    \caption{\label{fig:qhalf-vs-chi2-gsim-kskpi-bad2}}
  \end{subfigure}
  \caption{Distributions~(a),~(c) of the quadratic form
    $\frac{1}{2}\Delta\bm{y}^{\intercal}\hat{Q}\Delta\bm{y}$ and two-dimensional distributions~(b),~(d):
    $\frac{1}{2}\Delta\bm{y}^{\intercal}\hat{Q}\Delta\bm{y}$ versus $\chi^2$.
    These figures correspond to the
    events of the two different $e^+e^-\rightarrow K_S K^{\pm}\pi^{\mp},\;K_S\rightarrow\pi^+\pi^-$
    Gaussian simulations (second and third). Figures~$\text{(a)}$ and~$\text{(b)}$ correspond to
    the events of the second simulation, while figures~$\text{(c)}$ and~$\text{(d)}$ correspond
    to the events of the third simulation. The second Gaussian simulation corresponds to the
    chi-square distribution shown in figure~\ref{fig:chi2-gsim-kskpi-bad1}, while the third Gaussian
    simulation corresponds to the distribution shown in
    figure~\ref{fig:chi2-gsim-kskpi-bad2}.\label{fig:q-form-kskpi-bad}}
\end{figure}
Figure~\ref{fig:q-form-kskpi-bad} shows distributions similar to those shown in
figure~\ref{fig:q-form-kskpi-good}, but corresponds to the case of the second and
third Gaussian simulations, i.e. figures~\ref{fig:chi2-gsim-kskpi-bad1}
and~\ref{fig:chi2-gsim-kskpi-bad2}, respectively. Figure~\ref{fig:qhalf-gsim-kskpi-bad1}
shows the $\frac{1}{2}\Delta\bm{y}^{\intercal}\hat{Q}\Delta\bm{y}$ distribution
for the events of the second Gaussian simulation. The standard deviation of this
distribution is about two times larger than in the case of the distribution shown
in figure~\ref{fig:qhalf-gsim-kskpi-good}. As shown in figure~\ref{fig:chi2-gsim-kskpi-bad1},
such a deviation of the $\frac{1}{2}\Delta\bm{y}^{\intercal}\hat{Q}\Delta\bm{y}$ is sufficient
to skew the corresponding chi-square distribution. Figure~\ref{fig:qhalf-vs-chi2-gsim-kskpi-bad1}
shows the two-dimensional distribution of the
$\frac{1}{2}\Delta\bm{y}^{\intercal}\hat{Q}\Delta\bm{y}$ versus the $\chi^2$ for the case of the
second Gaussian simulation. Figures~\ref{fig:qhalf-gsim-kskpi-bad2} and~\ref{fig:qhalf-vs-chi2-gsim-kskpi-bad2}
show distributions similar to those shown in figures~\ref{fig:qhalf-gsim-kskpi-bad1}
and~\ref{fig:chi2-gsim-kskpi-bad1}, respectively. Figures~\ref{fig:qhalf-gsim-kskpi-bad2}
and~\ref{fig:qhalf-vs-chi2-gsim-kskpi-bad2} correspond to the case of the third Gaussian simulation.
Figure~\ref{fig:qhalf-gsim-kskpi-bad2} shows that the distribution in this case is much wider than
in the case of the first and second Gaussian simulations. In addition, this distribution is strongly
asymmetric with respect to zero. Moreover, the long tail on the right slope of this distribution
corresponds to large chi-square values. This can be seen from the two-dimensional distribution
shown in figure~\ref{fig:qhalf-vs-chi2-gsim-kskpi-bad2}.

We can conclude from the examples presented in section~\ref{sec:gaussian-simulation-examples} that if the
uncertainties of the parameters are of Gaussian nature and the nonlinearity of the constraints can be
neglected, then the chi-square distributions are described by the corresponding probability density
function with the expected number of degrees of freedom.

\begin{figure}[tbp]
  \centering
  \begin{subfigure}[t]{0.47\textwidth}
    \centering
    \includegraphics[width=\textwidth]{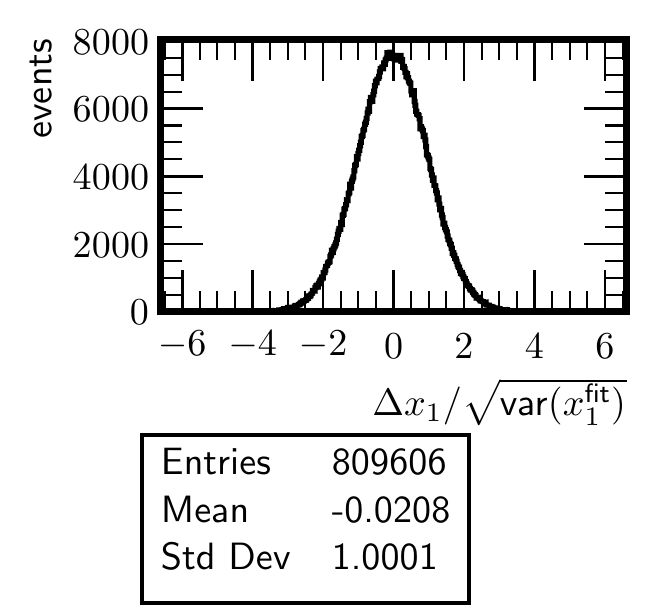}
    \caption{\label{fig:pull_x1_simhypo_xkpi_gauss_sim_good}}
  \end{subfigure}
  \hspace{1em}
  \begin{subfigure}[t]{0.47\textwidth}
    \centering
    \includegraphics[width=\textwidth]{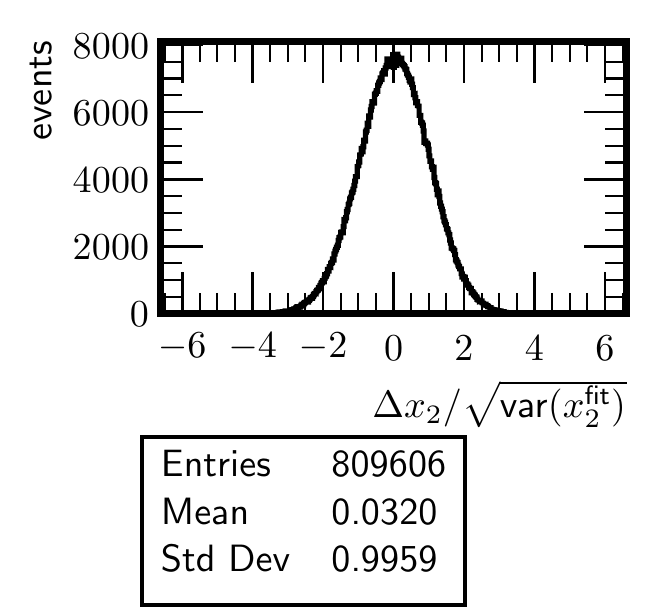}
    \caption{\label{fig:pull_x2_simhypo_xkpi_gauss_sim_good}}
  \end{subfigure}
  \caption{Pull distributions for vertex coordinates. The distributions are
    obtained for the events of the same Gaussian simulation as the distribution
    shown in figure~\ref{fig:chi2-gsim-kskpi-good}. The distribution shown in
    figure~(a) corresponds to the
    $x$-coordinate of the $e^+e^-$ interaction vertex. The distribution shown in
    figure~(b) corresponds to the
    $x$-coordinate of the $X\rightarrow\pi^+\pi^-$ decay vertex.\label{fig:pull_x_simhypo_xkpi_gauss_sim_good}}
\end{figure}

\begin{figure}[tbp]
  \centering
  \begin{subfigure}[t]{0.47\textwidth}
    \centering
    \includegraphics[width=\textwidth]{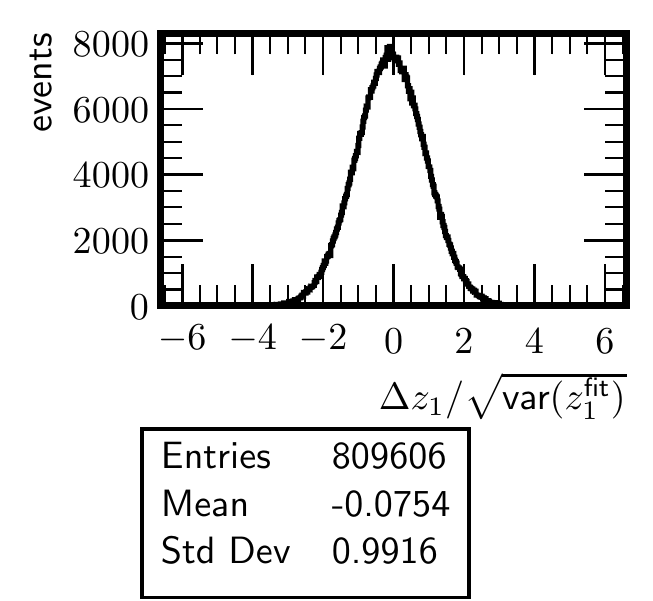}
    \caption{\label{fig:pull_z1_simhypo_xkpi_gauss_sim_good}}
  \end{subfigure}
  \hspace{1em}
  \begin{subfigure}[t]{0.47\textwidth}
    \centering
    \includegraphics[width=\textwidth]{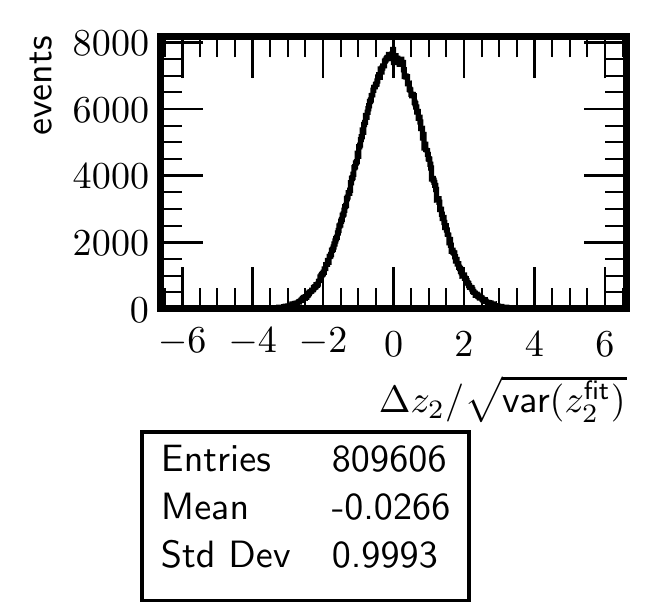}
    \caption{\label{fig:pull_z2_simhypo_xkpi_gauss_sim_good}}
  \end{subfigure}
  \caption{Pull distributions for vertex coordinates. The distributions are
    obtained for the events of the same Gaussian simulation as the distribution
    shown in figure~\ref{fig:chi2-gsim-kskpi-good}. The distribution shown in
    figure~(a) corresponds to the
    $z$-coordinate of the $e^+e^-$ interaction vertex. The distribution shown in
    figure~(b) corresponds to the
    $z$-coordinate of the $X\rightarrow\pi^+\pi^-$ decay vertex.\label{fig:pull_z_simhypo_xkpi_gauss_sim_good}}
\end{figure}
Another interesting question is what the pull distributions for vertex
coordinates look like if the detector response is Gaussian and the covariance
matrix $\hat{\tilde{C}}$ is well calibrated. This question can be answered by
applying the kinematic and vertex fitting procedure to Gaussian simulation
events. Examples of such distributions are shown in
figures~\ref{fig:pull_x_simhypo_xkpi_gauss_sim_good}
and~\ref{fig:pull_z_simhypo_xkpi_gauss_sim_good}.
Figure~\ref{fig:pull_x1_simhypo_xkpi_gauss_sim_good}
shows the pull distribution obtained for the $x$-coordinate of the interaction vertex.
Figure~\ref{fig:pull_x2_simhypo_xkpi_gauss_sim_good} shows the pull distribution
obtained for
the $x$-coordinate of the $X\rightarrow\pi^+\pi^-$ decay vertex. Figures~\ref{fig:pull_z1_simhypo_xkpi_gauss_sim_good}
and~\ref{fig:pull_z2_simhypo_xkpi_gauss_sim_good} show
similar pull distributions corresponding to the $z$-coordinates of these
vertices. The pull distributions shown in
figures~\ref{fig:pull_x_simhypo_xkpi_gauss_sim_good}
and~\ref{fig:pull_z_simhypo_xkpi_gauss_sim_good} were
obtained for the same Gaussian simulation events as the distribution shown in
figure~\ref{fig:chi2-gsim-kskpi-good}. It is seen from
figures~\ref{fig:pull_x_simhypo_xkpi_gauss_sim_good}
and~\ref{fig:pull_z_simhypo_xkpi_gauss_sim_good} that the standard deviations
corresponding to the distributions from these figures are close to one, as
expected. This fact confirms that the kinematic and vertex fitting procedure
correctly reconstructs the coordinates of the vertices.

Section~\ref{sec:kskpi-xkpi} shows that the standard deviations in the case of pull distributions
corresponding to the events of simulation~(that takes into account the CMD-3
response) are significantly greater than one. This is mainly due to the non-Gaussian
response of the
detector and the imperfect calibration of the covariance
matrix~$\hat{\tilde{C}}$. In section~\ref{sec:kskpi-xkpi},
this statement is demonstrated for the simulated events of the
$e^+e^-\rightarrow{K_S}K^{\pm}\pi^{\mp}$, $K_S\rightarrow\pi^+\pi^-$ process fitted to the
$e^+e^-\rightarrow{X}K^{\pm}\pi^{\mp}$, $X\rightarrow\pi^+\pi^-$ hypotheses. However,
this statement is also
true for events of other processes fitted to other hypotheses. Moreover,
in the case of Gaussian simulations, the standard deviations corresponding
to the pull distributions obtained for the events of these processes are
always close to one.

\section{Fitting package \label{sec:fitting-package}}
The kinematic and vertex fitting package consists of two parts. The first part
is the base fitting package KFBase.  The KFBase package contains the optimizer
class and an abstract base hypothesis class. This package also contains classes for all
constraints, vertices and particles, except for charged particles and photons. The second
part is the package of kinematic and vertex fitting for the $\text{CMD-}3$ experiment.
This package is called KFCmd and depends on the base fitting package. The KFCmd
package contains classes of charged particles and photons. It also contains a
class responsible for reading input data, as well as classes
with the implementation of some hypotheses. The packages KFBase and KFCmd, as well as their
documentation, can be found in the repositories~\url{https://github.com/sergeigribanov/KFBase}
and~\url{https://github.com/sergeigribanov/KFCmd}, respectively.

Both packages
are written in the C++ (ISO/IEC 14882:2017) programming language and depend on a number of other packages such as
ROOT~\cite{Brun:1997pa} (version 6.26) and Eigen~3~\cite{eigenweb} (version 3.4.0). The Eigen~3 header
only library is used for matrix computations, while the ROOT
framework is mainly used for file manipulations. The authors also provide a docker image that
contains a number of examples of using the fitting package. The examples can be run in Jupyter
notebooks based on the Python~3 kernel. The fitting package does not currently have a Python API.
Instead of such an API, the PyROOT~\cite{Galli:2020boj} functionality is currently used. The
repository with the docker image can be found at the
link~\url{https://github.com/sergeigribanov/kinfit-cmd3-docker}.

\section{Summary}
In this paper, the authors review in detail the algorithm of operation and
the capabilities of the kinematic and vertex fitting package for the
$\text{CMD-}3$ experiment. The paper provides a description of the constraints
available in the fitting package, as well as the parameterizations of particles
and vertices. The work provides a number of examples that show the operation of
the fitting package using most of the constraints and particle parametrizations.
These examples consider hypotheses containing charged particles and photons in
the final state, hypotheses with intermediate and lost particles. It should be
noted that the fitting package can be easily extended with additional constraints,
particle and vertex parameterizations. Despite the fact that this package was
developed for the $\text{CMD-}3$ experiment, the authors believe that after some
adaptation the package can be used in similar experiments.

In addition,
the fitting package was tested using the Gaussian simulation technique. This kind
of testing is very useful, as it helps to eliminate some issues at the stage of
package development. Possible mistakes in the gradients and Hessians of
constraints, as well
as of the particle four-momentum and trajectories, can be examples of such issues.
Since some gradients and Hessians can be very cumbersome, testing for this kind of
issues requires special attention.

The authors consider the discussed kinematic and vertex fitting package as their first
step towards a more rigorous and universal fitting package. In the future, it is planned
to pay special attention to the optimization algorithm. The fitting package currently
uses Newton's method. This requires inverting the Hessian matrix at each iteration of the
optimization procedure. The more optimization parameters the hypothesis contains, the
more resource-intensive the Hessian inversion becomes. The authors plan to consider
the possibility of using quasi-Newtonian methods~\cite{Gill1982,Izmailov2014},
in which the Hessian inversion does
not occur at every iteration. In the case of hypotheses with a large number of vertices,
the authors also plan to study optimization algorithms based on the Kalman filtering,
similar to the one discussed in the papers~\cite{HULSBERGEN2005566,KROHN2020164269}. In addition, the authors
plan to also explore the possibility of using
inequality constraints~\cite{Gill1982,Birgin2014,Eiselt2019} in some kinematic and
vertex fitting cases.

\section{Acknowledgments}
The authors are grateful to the members of the $\text{CMD-}3$ collaboration for discussion of the
considered topic and numerous suggestions. The authors also express special gratitude to the members
of the collaboration who are already actively using and testing the fitting
package. The work described in
sections~\ref{sec:minimization-algorithm}-\ref{sec:constraints} is supported by
Russian Science Foundation grant
No.\ $19\text{-}72\text{-}20114$. The work related to the testing of the fitting
package is supported by the Russian Foundation for Basic Research grant
No.\ $20\text{-}02\text{-}00496$~A.

\bibliography{article-kinfit-cmd3}

\end{document}